%% file: 000-paper.tex
\documentclass[acmsmall]{acmart}

\usepackage{soul}
\usepackage{enumitem}
\usepackage{rotating}
\usepackage{multirow}
\usepackage{pdflscape}
\usepackage{lineno}
\usepackage{longtable, longfbox}
\usepackage{xcolor, colortbl}
\usepackage{tikz}
\usepackage{tcolorbox}
\usepackage{diagbox}
\usepackage{wrapfig}
\usepackage{siunitx}
\usepackage{graphicx,marvosym}
\usepackage{caption, subcaption}
\usepackage{array}
\usepackage{textcomp}

\include{z-commands}

\AtBeginDocument{%
  \providecommand\BibTeX{{%
    \normalfont B\kern-0.5em{\scshape i\kern-0.25em b}\kern-0.8em\TeX}}}

\setcopyright{acmcopyright}
\copyrightyear{9999}
\acmYear{9999}
\acmDOI{99.9999/9999999.9999999}

\acmJournal{TIIS}
\acmVolume{99}
\acmNumber{99}
\acmArticle{99}
\acmMonth{99}

\begin{document}
\title{Measuring User Experience Inclusivity in Human-AI Interaction via Five User Problem-Solving Styles
}

\author{Andrew Anderson}
\email{anderan2@oregonstate.edu}
\orcid{0000-0003-4964-6059}
\affiliation{\institution{Oregon State University}}

\author{Jimena Noa Guevara}
\email{noaguevg@oregonstate.edu}
\affiliation{\institution{Oregon State University}}

\author{Fatima Moussaoui}
\email{moussaof@oregonstate.edu}
\affiliation{\institution{Oregon State University}}

\author{Tianyi Li}
\email{li4251@purdue.edu}
\affiliation{\institution{Purdue University}}

\author{Mihaela Vorvoreanu}
\email{Mihaela.Vorvoreanu@microsoft.com}
\affiliation{\institution{Microsoft}}

\author{Margaret Burnett}
\email{burnett@oregonstate.edu}
\affiliation{\institution{Oregon State University}}

\renewcommand{\shortauthors}{Anderson et al.}
\renewcommand{\shorttitle}{Measuring User Experience Inclusivity in Human-AI Interaction}

\begin{abstract}
\textbf{Motivations:} 
Recent research has emerged on generally how to improve AI product user experiences, but relatively little is known about an AI product's inclusivity. 
For example, what kinds of users does it support well, and who does it leave out?
And what changes in the product would make it more inclusive?

%
\textbf{Objectives:} 
Our overall objective is to help fill this gap, investigating what kinds of diverse users an AI product leaves out, and how to act upon that knowledge.
To bring actionability to our findings, we focus on users' diversity of problem-solving attributes. 
Thus, our specific objectives were:
(1)~to reveal whether participants with diverse problem-solving styles were left behind in a set of AI products;
and (2)~to relate participants' problem-solving diversity to their demographic diversity, specifically, gender and age.

\textbf{Methods:} 
We performed 18 experiments, discarding two that failed manipulation checks.
Each experiment was a 2x2 factorial experiment with online participants. 
Each experiment compared two AI products: one deliberately violating an HAI guideline and the other applying the guideline.
For our first objective, we analyzed how much each AI product gained/lost inclusivity compared to its counterpart, where inclusivity was supportiveness to participants with particular problem-solving styles.
For our second objective, we analyzed how participants' problem-solving styles aligned with their demographics, namely their genders and ages. 

\textbf{Results \& Implications:} 
Participants' diverse problem-solving styles revealed six types of inclusivity results:
(1)~the AI products that followed an HAI guideline were almost always more inclusive across \textit{diversity} of problem-solving styles than the products that did not follow that guideline---but the ``who'' that got most of the inclusivity varied widely by guideline and by problem-solving style; 
\FIXED{
(2)~when an AI product had risk implications, four variables' values varied in tandem: participants' feelings of control, their (lack of) suspicion, their trust in the product, and their certainty while using the product; 
} 
\FIXED{
(3)~the more control an AI product offered users, the more inclusive it was; 
}
\FIXED{
(4)~whether an AI product was learning from ``my'' data or other people's affected how inclusive that product was; 
} 
\FIXED{
(5)~participants' problem-solving styles skewed differently by gender and age group; 
}  
and
(6)~almost all of the results suggested actions that HAI practitioners could take to improve their products' inclusivity further. 
Together, these results suggest that a key to improving the demographic inclusivity of an AI product (e.g., across a wide range of genders, ages, etc.) can often be obtained by improving the product's support across diverse problem-solving styles.


\end{abstract}

\ccsdesc[500]{Human-centered computing~User studies}
\ccsdesc[300]{Computing methodologies~Intelligent agents}

\keywords{Intelligent User Interfaces, Human-Computer Interaction
}



\maketitle

\input{doc/01-Introduction}

\input{doc/02-background}
\input{doc/03-Methodology}
\input{doc/04-Results-a-Inclusivity}

\input{doc/04-Results-Everything-Else}

\input{doc/06-Gender}

\input{doc/05-Discussion-Inclusiveness-Equity}

\input{doc/08-Conclusion}


\bibliographystyle{ACM-Reference-Format}
\bibliography{000-paper.bib}

\appendix
\section{Online Appendices}

Note to reviewers: These appendices are intended to be ACM TiiS-hosted online appendices, but for reviewing purposes we temporarily provide them on the lead author's homepage.

\FIXME{MMB@AAA: Appendix B fixes needed: (1) Seems to be a typo in G13's Applic vignette: "not displays" probably should be "now displays"? Please chk to see if the participants saw "not" or "now". (this is a potentially baa-a-a-aad typo )\\
(2) Makes no sense that these tables are numbered from 10 up.  If easy, make them something like "B1", "B2", ... Or if you can't, then just start numbering at 1.}

\begin{itemize}
    \item Appendix~{\color{blue}~\href{https://web.engr.oregonstate.edu/~anderan2/MSR-Appendices/Appendix-Facet-Survey-Rules.pdf}{{A}}} provides the problem-solving style survey, along with the rules for discerning participants' problem-solving style values for each of the problem-solving style types. 
    \item Appendix~\href{https://web.engr.oregonstate.edu/~anderan2/MSR-Appendices/Appendix-Vignettes.pdf}{{\color{blue}{B}}} shows the vignettes for both the \vioProduct{} and \appProduct{} for all 16 experiments.
    \item Appendix~{\color{blue}~\href{https://web.engr.oregonstate.edu/~anderan2/MSR-Appendices/Appendix-Demographics.pdf}{C}} provides the demographic data for all participants.
    \item Appendix~{\color{blue}\href{https://web.engr.oregonstate.edu/~anderan2/MSR-Appendices/Appendix-Statistics.pdf}{{D}}} shows \textit{all} of the statistical tests for \textit{all} experiments and \textit{all} problem-solving style types, along with whether they were significant under Holm-Bonferroni correction.
\end{itemize}

\end{document}

%% file: z-commands.tex
\definecolor{AbiOrange}{RGB}{255,255,255}
\definecolor{TimBlue}{RGB}{255,255,255}
\definecolor{AbiOrangeQuote}{RGB}{251,212,180}
\definecolor{TimBlueQuote}{RGB}{219,229,241}
\definecolor{NegativeAbi}{RGB}{255, 115, 0}
\definecolor{NegativeTim}{RGB}{60, 120, 216}
\definecolor{Eggshell}{RGB}{215,198,186}
\definecolor{WomanOrange}{RGB}{237,125,49}
\definecolor{ManBlue}{RGB}{31,78,121}
\definecolor{FigureMaroon}{RGB}{191,10,16}

\definecolor{InitiallyBlue}{RGB}{33,116,194} 
\definecolor{DuringInteractionRed}{RGB}{168,62,27}  
\definecolor{WhenWrongYellow}{RGB}{235,150,60}
\definecolor{OverTimeGreen}{RGB}{79,139,69}

\definecolor{LightGray}{RGB}{211,211,211}

\usepackage{adjustbox}
\newcommand{\colorboxBackgroundForegroundText}[3]
{\protect\adjustbox{padding=1pt 1pt, bgcolor=#1}{\color{#2}#3}}

\newcommand{\user}[3]{{%
~#1#2}}

\newcommand{\redact}[1]{X University}

\newcommand{\quotateInset}[4]{%
    \vspace{0pt}%
    \begin{quote}%
    \user{#2}{#3}{}: ``\textit{#1}''%
    \end{quote}%
    \vspace{-2pt}%
}

\newcommand{\vioProduct}[0]{
    {\fontfamily{cmtt}\selectfont Violation AI product}%
    }%

\newcommand{\appProduct}[0]{
    {\fontfamily{cmtt}\selectfont Application AI product}%
    }%

\newcommand{\tTestResult}[4]{
    (t(#1) = #2, \textit{p} = #3, \textit{d} = #4)%
}%

\newcounter{boldifyCounter}
\newcounter{fixmeSectionCounter}
\newcounter{fixmeTotalCounter}

\newcounter{resultCounter}

\makeatletter
\@addtoreset{fixmeSectionCounter}{section}
\@addtoreset{fixmeSectionCounter}{subsection}
\@addtoreset{boldifyCounter}{section}
\@addtoreset{boldifyCounter}{subsection}

\makeatother

\newcommand{\boldify}[1]{}
\ifdefined\boldifyON
	\renewcommand{\boldify}[1]{\par\noindent%
		\textbf{{**}%
		 #1**\\}
	}
\fi

\newcommand{\fakeResult}[1]{}
\ifdefined\fakeResultON
    \renewcommand{\fakeResult}[1]{\par\noindent%
        \stepcounter{boldifyCounter}%
        \textbf{{\color{blue}**FAKE FAKE FAKE**}%
        ~\arabic{section}.\arabic{subsection}.\arabic{boldifyCounter}%
        : #1\\}
    }
\fi

\newcommand{\FIXED}[1]{{\color{black}#1}}
\newcommand{\FIXME}[1]{}
\ifdefined\fixmeON
	\renewcommand{\FIXME}[1]{\par\noindent%
		\stepcounter{fixmeSectionCounter}\stepcounter{fixmeTotalCounter}%
		{\color{red}\fbox{\color{black}%
			\parbox{.97\linewidth}{%
				\textbf{FIXME \arabic{section}.\arabic{subsection}.%
        		\arabic{fixmeSectionCounter} (\color{red}%
        		\#\arabic{fixmeTotalCounter}):} #1
        		}
        		}%
        }
	}
\fi

\newcommand{\Result}[2]{\par\noindent%
    \stepcounter{resultCounter}
    \color{black}\fbox{\color{black}%
		\parbox{.97\linewidth}{%
             \textbf{Result \##1:} #2
        }
    }
}

    {\par%
    \setlength{\hangindent}{0.05\columnwidth}%
    \setlength{\parindent}{0.05\columnwidth}%
    \textit{Insight-G#1-#2: } #3\\%
    }%

\ifdefined\specialFooterON
\usepackage{fancyhdr, datetime}
\fi

\newcommand{\draftStatus}[1]{}
\ifdefined\draftStatusON
	\renewcommand{\draftStatus}[1]{
        \hfill **D#1%
	}%
\fi 

%% file: doc/01-Introduction.tex
\section{Introduction\draftStatus{ MMB's opinion: D2.5?}
}
\label{sec:intro}


\input{figure/Li-resultG3only}
Suppose the product owner of an AI Product named ``G3'' ran a before/after user study, to find out whether potential customers had better user experiences with G3's new version than an older, not very good version of G3.
Also suppose the results came out like those of Figure~\ref{fig:Li-resultG3only}.

The product owner should be at least somewhat pleased with these results---the product clearly improved. 
Of the 13 user-experience (UX) outcome variables measured (y-axis), 11 were positive.
In fact, as the *'s indicate, the differences were significant, with the new version's UX outcomes  significantly better than the old version's.
Still, the effect sizes of G3's changes were fairly small (yellow bars), with only two of moderate size (blue bars).

What the product owner would now like to know is: who was included in those positive effects, and \textit{who was left out}?
And what further changes are needed to enable G3 to better support more of the potential customers?

These kinds of questions are human-AI interactions (HAI) questions about the user experience (UX) quality that AI products offer their customers.
In this paper, we abbreviate the concept of HAI user experiences as HAI-UX.

\boldify{One way that the HAI community is currently tackling the UX needs of people is through trying to apply design guidelines. Li et al. did a first investigation on a set of guidelines, finding that }

One method that the HAI community currently uses to improve AI products' user experiences is to develop and apply guidelines for human-AI interaction.
At least three major companies---Apple, Google, and Microsoft---have each proposed guidelines, providing high-level advice for how to improve human-AI interaction, such as ``Consider offering multiple options when requesting explicit feedback''~\cite{apple2019guidelines}, ``Let users give feedback''~\cite{google_2019}, and ``Support efficient correction''~\cite{amershi2019guidelines}.
In fact, G3's product owner followed a guideline from the Microsoft set, and doing so did indeed improve the product, as Figure~\ref{fig:Li-resultG3only} showed. 

In this paper, we consider how to measure beyond just \textit{whether} products like G3 improve their HAI-UX.
We investigate how to know \textit{who}, of all the diverse humans who could be using products like G3, is \textit{included} in our HAI-UX improvements and who has been left out.
%
%
Applying the concept of inclusivity to human interactions with AI products, we will say that AI product A is more \textit{inclusive} to some particular group of people than product B is, if product A provides those people with measurably better user experience outcomes than product B does.

\FIXME{MMB to AAA: why are we defining problem-solving styles twice in the same parag below? (I wordsmithed the 2nd instance, but then noticed the 1st...)\\
AAA@MMB: Fixed. First should be problem-solving, the second problem-solving styles.}

The primary groups of interest in this paper are those who are diverse in the ways they go about \textit{problem-solving}.
\FIXED{We use the term \textit{problem-solving} to mean any time that people are engaged in solving problems, such as whether and how to accept/reject an AI's recommendations.}
We consider participants' problem-solving diversity via the set of five problem-solving style spectra from the inclusive design method known as GenderMag~\cite{burnett2010gender}.
\FIXED{We use the term \textit{problem-solving styles} to refer to the approaches that individuals take to go about trying to solve a problem.}
These five problem-solving style spectra are people's diverse attitudes toward risk, their diverse levels of computer self-efficacy, their diverse motivations, their diverse information processing style, and their diverse styles of learning technology.

\boldify{For example...}

For example, ``risk-averse'' is one endpoint of the risk attitude spectrum.
Applying risk-aversion to technology, risk-averse users may be hesitant to invoke a new feature for fear that it may have undesirable side-effects (e.g., privacy), may waste their time, may not be worth learning about, etc. 
At the other end of the spectrum, ``risk-tolerant'' users may be more willing to take such risks, even if the feature has not been proven to work yet and requires additional time to understand~\cite{guizani2022debug, padala2020gender, vorvoreanu2019gender}.

\boldify{To investigate whether this approach can help us understand human-AI data more, we looked at problem-solving diversity inclusivity, which is different from Li et al.}

In this investigation,  we consider how user experiences of people with diverse problem-solving styles were impacted by design differences in AI-powered systems like G3 above.
Specifically, we gathered 1,016 participants' five GenderMag problem-solving styles, and investigated inclusivity differences in 16 pairs of AI products.
Each AI product had controlled differences: one AI product applied an HAI guideline from the Amershi et al. guidelines set~\cite{amershi2019guidelines}, and its counterpart violated that guideline.
All AI products were productivity software (e.g., Microsoft PowerPoint, etc.) that had added AI features.
An earlier investigation on the same data, reported in Li et al.~\cite{li-MSR-work}, investigated the ``whether'' questions of these data, i.e., whether HAI-UX differences between each pair AI products occurred.
That investigation found that participants's HAI-UX outcomes were generally better when the guidelines were applied; Figure~\ref{fig:Li-resultG3only} is in fact one example of their findings.
Our investigation instead focuses on ``who'' questions, i.e, who were included (and who were not) in the HAI-UX outcome changes, from the perspective of participants' diverse problem-solving styles.


\FIXED{
\boldify{We're going to present Risk in detail, and a summary of the rest. We chose Risk to highlight because... see the appendices for the rest in detail.}

To show how analyzing HAI-UX data by these five problem-solving styles can reveal actionable insights into how to improve an AI product's inclusiveness, this paper presents a detailed analysis of one of these problem-solving style types, namely attitudes toward risk. 
However, space constraints prevent providing detailed analyses for all five of these problem-solving style types, so this paper summarizes the remaining four problem-style types' results with an eye toward generality; 
we also provide detailed analyses of all five problem-solving style types in the Appendices.
We selected attitudes toward risk as the problem-solving type to present in detail, because of the preponderance of recent research literature and popular perception focusing on risks with AI, such as risks of inaccuracies, of privacy loss, of excessive or insufficient trust, of job loss, and more (e.g., \cite{tchaikovsky2023nytimes, hu2023dark, harrison2020empirical, gillath2021attachment, dodge-fairness-2019, hulse2018perceptions, curry2023aijoblosses, kelly2023aijoblosses}).
We investigate:
} 


\begin{enumerate}[label = {}, rightmargin = 30px]
    \item \textbf{RQ1-Risk:} \textit{When the HAI guidelines are violated vs. applied to AI products, how inclusive are the resulting AI products to users with diverse attitudes toward risk?}
    \item \textbf{RQ2-AllStyles:} \textit{How inclusive are such products to users with diverse values of GenderMag's other four problem-solving styles?}
\end{enumerate}


We also investigate the relationship between participants'  problem-solving style diversity and their demographic diversity.
Our reason for doing so is that, even in the presence of statistical significance, knowing the existence of demographic disparities in who a product serves may not suggest actionable ways to address that disparity; for example, if one gender is left out of high-quality user experiences with an AI product, how to fix it?
In contrast, problem-solving style disparities often do suggest actionable ways forward; for example, if risk-averse participants are left out of high-quality user experiences, perhaps the product should be clearer about risks of using it (e.g., its privacy impacts). 
Thus, we investigate:
\begin{enumerate}[ label = {}, rightmargin = 30px]
\item \textbf{RQ3-DemographicDiversity:} \textit{How does AI product users' problem-solving diversity align with their demographic diversity?}
\end{enumerate}

Thus, the new contributions of our research are:
\begin{itemize}
    \item \textit{Risk-inclusivity in HAI-UX}: Reveals whether and how a set of AI products that followed a widely used set of HAI guidelines~\cite{amershi2019guidelines} are inclusive to participants with diverse attitudes toward risk.
    \item {\textit{Beyond risk-inclusivity in HAI-UX}}: Generalizes the above results to the other four GenderMag problem-solving style spectra. 
    \item \textit{Actionable inclusivity in HAI-UX}: Reveals whether results to the above suggest actionable steps an HAI practitioner can take to make an AI product more inclusive.
    \item \textit{Problem-solving diversity and demographic diversity}: Reveals relationships between participants' problem-solving styles and their intersectional gender-and-age demographic diversity, to enable HAI practitioners to bring actionable results from problem-solving diversity investigations to bear on demographic disparities.
\end{itemize}

%% file: figure/Li-resultG3only.tex
\begin{wrapfigure}{i}{0.3\linewidth}
        \centering
        \includegraphics[width = 0.8\linewidth]{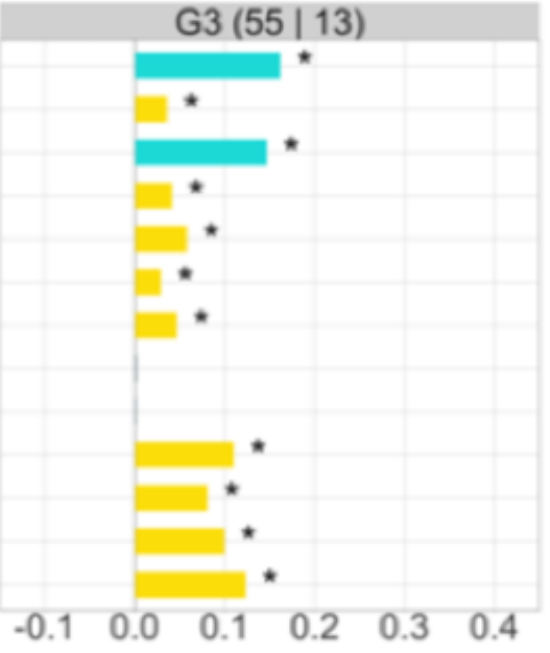}
        \caption{An outcome of an experiment comparing two versions of ``G3'' AI products~\cite{li-MSR-work}.  X-axis shows the amount of improvement resulting, for the 13 variables (not labeled here) on the y-axis. *:  difference was statistically significant.}
        \label{fig:Li-resultG3only}
\end{wrapfigure}

%% file: doc/02-background.tex
\FIXED{
\section{Background \& Related Work 
\draftStatus{0}}
\label{sec:background}

\subsection{Background}
\label{subsec:background}


\subsubsection{The Gender Inclusiveness Magnifier (GenderMag)}

\boldify{The GM styles come from GM, and have been repeatedly shown to be about problem-solving and gender. }

\FIXME{\\} 
\FIXME{AE-4---"In particular, R4 and R8 both pointed out that more explanation is needed to address how GenderMag has been validated specifically for problem-solving styles (other than gender inclusivity).''}

The GenderMag problem-solving styles are foundational underpinnings to this investigation (Table~\ref{table:personas}).
The GenderMag \textit{method} is an inclusive design and evaluation method based on these five problem-solving style types~\cite{burnett2016gendermag}; software professionals use the method to detect ``(gender\nobreakdash-)inclusivity bugs.''%
\footnote{\FIXED{GenderMag finds the issues not by using people's gender identity, but rather by the five problem-solving style types. These problem-solving styles' values statistically cluster around genders.} 
} 
GenderMag's problem-solving styles are particularly well-suited to our investigation into diverse users' experiences with AI features, because GenderMag was developed to improve technology's \textit{inclusiveness} for \textit{problem-solving technology}---such as spreadsheet development, software debugging, and any other domains where problems can arise with which users must grapple~\cite{burnett2016gendermag}.
Because the GenderMag method is intended for practical use by developers without social science backgrounds, a set of criteria~\cite{burnett2016gendermag,mendez2019inclusivemag} were applied to the original long list of applicable problem-solving style types~\cite{beckwith2004gender}, to reduce to the five styles in Table~\ref{table:personas}.
These five style types have been repeatedly identified as having strong ties to both problem-solving and gender%
\footnote{\FIXED{Some of these works gathered participants' biological sex rather than their gender identities; others by gender.  Further, most of the literature has been binary, reporting only females/males or women/men, so the upcoming description of the styles is also necessarily binary.  In this discussion, we use gender terminology (e.g., ``woman'' instead of ``female'') simply to avoid switching back and forth for different studies.}}; 
we will summarize some of this research shortly.

\input{tables/2-Facet-to-Persona}

Each of these problem-solving style types have continuous ranges whose endpoints (Table~\ref{table:personas}'s column~2 \& 4) are the only distinguished values.
Values at one end are assigned to a persona named ``Abi,'' those at the other end are assigned to a persona named ``Tim,'' and a mix of values are assigned to a persona named ``Pat.'' 
For example (Table~\ref{table:personas}'s row~1), Abi and Pat are more risk-averse about technology risks than Tim, so Abi and Pat might be less likely than Tim to use the same password on multiple sites.

\FIXME{For AE-4 and R4-2.2: say that these are all backed by a bunch of research}

To summarize these five styles (see~\cite{burnett2016gendermag} for details): 

\begin{itemize}
    \item Attitudes toward risk: Studies across multiple domains have reported a wide diversity in people's attitudes toward risk (e.g.,~\cite{dohmen2011individual, wagner1993english}) and how they solve problems~\cite{weber2002domain,charness2012strong,hyll2015impact}.
    Gender differences have also been reported in risk and decision in numerous
    problem-solving domains, with women almost always (statistically) less willing to take risks than other people~\cite{weber2002domain,charness2012strong}.
    Note that risks relevant to decision-making include not only obvious risks like privacy and security, but also risk of wasting time and/or of failing~\cite{kim-microsoft2023inclusive}.  

    \item Computer self-efficacy: One specific form of confidence is self-efficacy---people's belief in their ability to succeed in a specific task~\cite{bandura1986explanatory}.
    Self-efficacy matters to problem solving because it influences people's use of cognitive strategies, effort exerted, persistence with a problem, and coping strategies~\cite{bandura1986explanatory}.
    Regarding gender, empirical data have shown that women tend statistically to have lower computer self-efficacy than other people~\cite{stumpf2020gender}.
    Overall, ties between people's self-efficacy and how they approach a variety of problem-solving tasks have been well-documented in many domains (e.g., ~\cite{yaugci2016effect,padala2020gender,yang2023self,voica2020motivation}).
    \item Motivations: In the context of technology, motivations are  the reasons an individual decides to interact with technology, such as using technology mainly to accomplish a task, versus having an interest and enjoyment in using and exploring technology (e.g., ~\cite{stafford2001identifying}).
    Gender differences have been reported in these ``task-oriented'' and ``tech-oriented'' motivations~\cite{stumpf2020gender}.
   An individual's motivations can affect not only which technology features they decide to use, but also how they go about using those features~\cite{grosso2021exploring,stafford2001identifying,bridges2018hedonic,o2010influence,kartal2022preservice}.

   \item Information processing style: Solving problems often requires gathering information. However, individuals vary on how much information they gather and when.
    Some gather information comprehensively---i.e., in sizeable batches---to form an approach first and then carry it out, whereas others gather it selectively, acting upon the first promising information, then possibly gathering a little more information before taking the next action, and so on.
    Regarding gender, women are statistically more likely to process information comprehensively and men are statistically more likely to process it selectively~\cite{stumpf2020gender}.
    In both research and practice, much attention has been given to how technology can enable different individuals to obtain the right amount of information at the right time (e.g., \cite{chapman2022USDSguidelines, pirollicard1995IFT, torrens2020lacking, whittaker2011personalInfoMgt}).

    \item Learning Style for Technology (By process vs. by tinkering): Learning style considers how people go about solving problems by how they structure their approach.
    For example, some people prefer to learn new technology by an organized process, like a recipe.
    Others prefer to tinker, exploring options and experimenting in a ``what if I did this'' way.
    Regarding gender, women have been shown to be statistically less likely than other people to use the latter approach when encountering features new to them~\cite{stumpf2020gender}.
    Because of such differences, technology organizations have begun to stress the importance of supporting the entire range of Learning Style values; one example is in Microsoft's design guidelines~\cite{kim-microsoft2023inclusive}.
    
\end{itemize}

\FIXME{R9-6: The fact that the study was limited to people living in the USA (which is never justified by the way). In particular, I doubt it can be assumed that the gender/age study for RQ3, or even people scoring on the GenderMag items, are similar in all countries around the world. }

\FIXME{R4-1---"conclude 2.1.1 with a discussion of how this paper builds upon / extends / etc. this body of literature.'' \\
R4-1-fix---"End this section with “all these papers do X, but ours adds Y.''}

In GenderMag evaluations across the world, these five problem-solving style types have repeatedly shown impacts on which technology features diverse people decide to use and/or how they use them, as evidenced ~\cite{agarwal2023MOSIP, cunningham2016supporting,marsden2016evaluation,russian2017gender,shekhar2018cognitive,gralha2020genderDiffs,carver2018gender, padala2020gender,vorvoreanu2019gender}. 
However, most such evaluations have been outside the context of AI.
This paper is not only within AI contexts, it isolates the human-AI interactions from all other interactions.


\subsubsection{Guidelines for Human-AI Interaction}


\boldify{Another key portion of our work deals with the 18 guidelines for human-AI interaction, a set of general advice for researchers and practitioners developing AI-powered systems}

A second key component of this paper is Amershi et al.'s 18 guidelines for human-AI interaction~\cite{amershi2019guidelines}.
This set of 18 guidelines for human-AI interaction, depicted in Figure~\ref{fig:human-AI-guidelines}, provided high-level advice for designers about what the user should expect from AI-powered systems.
Each guideline had three components (1)~a number, (2)~a name which provides high-level advice for designers (e.g., ``Make clear what the system can do''), and (3)~a brief description of what the guideline means (e.g., ``Help the user understand what the AI system is capable of doing'').
Amershi et al. also ran an initial study to investigate if the designers of AI-powered systems could find examples of the guidelines and if the guidelines were clear~\cite{amershi2019guidelines}.

\input{tables/8-Guidelines}


}

\FIXED{


\subsection{Related Works }
\label{subsec:related-works}


\FIXME{AE-5.5 **FM: DONE 2/10/23** In fact, given that GenderMag plays such a central role in the paper, I wonder if the other related work can put more focus on prior studies related to GenderMag’s problem-solving styles and perhaps put less emphases on other cognitive styles that were not ...\\
AE-7---"The reviews of the list of studies, while comprehensive, do not seem to be converging to a set of ideas or findings that directly support the current method/analysis. Perhaps a more careful reorganization and selection can help.''
}
 \FIXME{MMB@AAA: I added "in Human-AI contexts" to make really clear the scope.\\
 AAA - Elevating this comment because I think that it matters.
 }

\boldify{Some works in inclusivity focus on algorithmic inclusivity, and here's a host of citations for them. However, we must stress that our paper does NOT FOCUS ON ALGORITHMIC INCLUSIVITY}

Inclusivity and related concepts like fairness in human-AI interaction can be thought of in two broad categories: 1) ``under-the-hood'' algorithmic inclusivity (i.e., how to detect and fix when \textit{algorithms} behave inappropiately or unfairly for some groups of people) and 2) ``over-the-hood'' inclusivity of diverse users' \textit{differing experiences} when they interact directly with AI products)
There has been a host of literature for the former category (e.g.,~\cite{bird2020fairlearn,katzman2023taxonomizing,propublica-compas,harrison2020empirical,green2019disparate,buolamwini2018gender,yang2020towards}), but this paper instead focuses  on the latter category.

\subsubsection{Investigations into individuals' problem-solving styles in Human-AI contexts}
\label{subsubsec:problem-solving-AI}

\boldify{Esteemed reader, here's why we need this subsubsection---we cover problem-solving styles in Sections~\ref{sec:Results-Insights} \&~\ref{sec:beyond-risk}}

Because this paper addresses five problem-solving style ranges---risk, computer self-efficacy, motivations, information processing style, and learning style with technology (e.g., by process or by tinkering), we focus on those five styles here.



Risk is the focus of \textbf{RQ1-Risk}.
There is a preponderance of research identifying  risks associated with AI products (e.g., risks of inaccuracies, privacy loss, misaligned trust, etc.)
In light of AI's risks, Cohen et al.~\cite{cohen2020sensitivity} 
has posited 
that AI's explanations should consider both risk-averse and risk-tolerant users.
Schmidt \& Biessmann~\cite{schmidt2020calibrating} also considered people's attitudes toward risk, classifying participants as either risk-averse or risk-tolerant using an incentivized gambling task.
They found that participants who were more risk-averse exhibited a more pronounced algorithmic bias (trusting an AI too much) than risk-tolerant participants with improved transparency, which they suggested was ``a sign of blind trust in ML predictions that can be attributed to increased transparency.''~\cite{schmidt2020calibrating}


\boldify{More closely related to our work are those works that used the GM facet values in AI:}

\textbf{RQ2-AllStyles} considers the remaining four problem-solving style types we consider in this paper, one of which is computer self-efficacy.
Kulesza et al.~\cite{kulesza2012tell} explicitly measured their participants' computer self-efficacy.
Their work measured the change in their participants' computer self-efficacy as an outcome of explaining ``why''-oriented eXplainable AI (XAI) approaches.
They showed that scaffolding participants' experiences with ``behind the scenes'' training led to higher self-efficacy improvements and increased mental model soundness  when compared to the participants without the scaffolding.
Additionally, Jiang et al.~\cite{jiang2000persuasive} found that when their participants used an AI system, those with elevated self-confidence were less likely to accept the system's proposed solution, preventing them from being persuaded by the system in the presence of system discrepancies.


\boldify{Our paper deals with motivation\underline{s}, regarding why people interact with AI technologies. Here's some work that deals with this.}

Another problem-solving style type considered in this paper is motivations.
In this paper, motivation\underline{s} refers to reasons why people are interacting with the technology.
Other researchers have also considered motivations in this sense, in the context of AI systems.
For example, Shao \& Kwon~\cite{shao2021hello} identified four such motivations: people using AI products for the purposes of entertainment, companionship, functional utility, and dynamic control.
They found that when participants interacted with AI products for functional utility and dynamic control, there was a positive relationship between their participants' satisfaction and these two motivations.
Skjuve et al.~\cite{skjuve2024people} contributed six user motivations among 
respondents' responses for why they interacted with ChatGPT, including productivity, novelty, creative work, learning/development, and as a means of social interaction/support.
Li et al. investigated motivations and user satisfaction in AI settings~\cite{li2020understanding}, and found a significant interaction between explanation type and motivations with respect to an AI product's persuasiveness (i.e., neighbor-rating explanation and hedonic motivation) for three types of explanations in a movie recommender system.

\boldify{Although seemingly related to our own work, there have been works that work on motivation (no s). We acknowledge these works exist, but they aren't related to our paper.}

However, other researchers have used a different meaning of motivation for users of AI products, namely how motivated a user was (i.e., how great their desire to succeed or win).
For example, Eisbach et al.~\cite{eisbach2023optimizing} found that when participants were more motivated to succeed at a task, they were more likely to intentionally process AI recommendations and explanations.
Visser~\cite{vissergender} found that when participants perceived an AI as masculine, they were more motivated to play games with more intensity to win than when the AI was perceived as feminine.
Some researchers have also investigated how motivated (i.e., likely) users are to interact with an AI product in the future.
For example, Baek \& Kim~\cite{baek2023chatgpt} found that when ChatGPT was perceived as creepy, participants were less motivated to interact with it in the future.


\boldify{Some of these works even tangentially relate to GenderMag, like those that study need for cognition (information processing style)}

Another problem-solving style type considered in this paper is information processing style.
As explained in Section~\ref{subsec:background}, information processing style refers to the diverse ways that people gather information to solve problems.
A closely-related concept is need for cognition~\cite{rudolph2018understanding,coutinho2005metacognition,khan2023psychological,jaquay2023exploring}, which refers to the extent to which individuals are inclined towards effortful cognitive activities~\cite{cacioppo1984efficient}, ranging from those with a lower need for cognition to those with higher.
In AI contexts, when considering need for cognition among participants, Dodge et al.~\cite{dodge-fairness-2019} found that diverse needs for cognition came with different needs for explanation types and amount of explanation.
Millecamp et al.~\cite{millecamp2020cogito} found that their participants with a higher need for cognition put in more effort to find the ``best'' AI recommendation, and Riefle et al.~\cite{riefle2022influence} found that their participants with higher need for cognition felt they understood an AI's explanations more than their counterparts.

\boldify{With one particularly pertinent result from Bucinca et al
}

One particularly pertinent inclusivity result while considering need for cognition in AI contexts was that of Bu{\c{c}}inca et al.~\cite{buccinca2021trust}, researching how to reduce over-reliance on AI explanations.
They found that adding cognitive forcing functions benefited only those with higher need for cognition, creating intervention-generated inequalities%
\footnote{According to Veinot et al., intervention-generated inequalities occur when a technological intervention disproportionately benefits a group of people who are already privileged in a particular context~\cite{veinot2018good}.} 
 because those with higher need for cognition have historically been a more advantaged group.


The fifth problem-solving style type considered in this paper is information processing style, and some researchers have considered process-oriented versus tinkering-oriented learning styles into their analyses of AI products and development.
For example, Nam et al.~\cite{nam2023ide} used the GenderMag problem-solving style survey to investigate how 32 developers' information processing styles and learning styles affected their ways of using Large Language Model (LLM) developer tools.
They leveraged three linear regression models, one for each of the investigated LLM features, and used participants' information processing style and learning style as explanatory variables for feature usage.
For learning style, they found that process-oriented learners were significantly more likely to probe the LLMs with follow-up queries, whereas tinkering-oriented learners tended to jump directly into tinkering with the code after getting minimal direction from the LLM.

\boldify{However, here's how we're different, pulled straight from contributions and conclusion }

This paper differs from the above works by considering all five of these user problem-solving styles (as opposed to subsets of them), and comparing their effects on the same products.
This paper also differs by showing how investigating these styles can help reveal actionable steps toward improving AI-powered technologies' inclusivity across  diverse users.



\subsubsection{Investigations of HAI inclusivity to diverse humans, from a demographic perspective}
\label{subsubsec:demographics-AI}

\FIXME{If I feel I can write a lot about gender/age, then insert ``Because this paper considers gender/age in Section..., we focus on works related to gender/age inclusivity in AI''}

\boldify{For example, here's some works who deal with gender in human-AI interaction, where they're trying to capture potential differences.}


The first component relevant to \textbf{RQ3-DemographicDiversity} are works in AI contexts that investigate gender differences while analyzing human data.
van Berkel et al.~\cite{van2021effect} studied perceived fairness in AI recidivism and loan predictions and found that their participants who identified as men were significantly more likely to say that both systems were fairer than those who identified as women.
Such results have also been found across domains, such as de Graaf et al.~\cite{de2015sharing}, who found that gender influenced participants' willingness to accept robotic technologies.
Derrick \& Ligon~\cite{derrick2014affective} also found gender differences on how likable the AI was, depending on how it behaved.
Similarly, Joseph et al.~\cite{joseph2024artificial} utilized a regression model to report on how awareness of AI impacted perception and utilization of AI tools.
They found that increases in male students' awareness of AI resulted in an increase in their utilization of AI tools.
However, increases in female students' awareness resulted in a \textit{decrease} in their utilization of AI.
Of particular interest to this paper, Hu \& Min~\cite{hu2023dark} found privacy concern gender effects in AI products.
Although both men and women were concerned about the ``watching eye'' of AI, the participants who identified as women were more concerned about privacy violations than the men.

The second component relevant to \textbf{RQ3-DemographicDiversity} are works investigating age differences.
Researchers have also found age differences in human-AI interaction.
Gillath et al. considered and found that older participants were significantly less likely to trust AI.
Similarly, both Shahid et al.~\cite{shahid2014child} and Martinez-Miranda~\cite{martinez2018age} found that age impacted their participants' perceptions of AI-powered robots.
Their participants were much younger than ours (i.e., under 18 years old).
However, other works have identified that regarding user experience, utilizing technologies like augmented reality and affective computing can help social robots become better companions for older adults~\cite{anjum2021augmented} or that aging populations have been empowered by using artificial intelligence to personalize smart home interfaces~\cite{giakoumis2019smart}.
Additionally, Zhou et al.~\cite{zhou2023empirical} found that considering participants' age while adapting human-facing AI interfaces in the smart home domain led to improvements in usability for elderly participants. 
Other works related to user experience have similarly found how people's age might influence attitudes towards AI, impacting things like trust~\cite{gillath2021attachment} and acceptance~\cite{jonsson2021ai}.
As with gender, some works have also discovered age differences in risky situations, such as Shandilya et al.~\cite{shandilya2022understanding}, who interviewed 15 participants, all aged 60 or over, and their findings included user experience themes which may cluster by attitudes toward risk, such as the perceived annoyance when AI-enabled products deviated from expected behavior or data privacy threats. 

%

\boldify{There are works out there that deal with implications for human-AI interaction that deal with specific kinds of people, like women being concerned about privacy, etc.}

%

\boldify{These works are differnt because they do not draw a comparison or align problem-solving style diversity with demographic diversity}

Although these works considered participants' demographic diversity to find differences in user experience while interacting with AI via demographic dimensions such as gender, our work differs by instead providing actionable avenues for HAI practitioners through the alignment with participants' five GenderMag problem-solving values.
We establish this tie in Section~\ref{sec:gender} to demonstrate that designing to improve inclusivity for problem-solving style diversity also improves inclusivity for demographic diversity.


\subsubsection{Actionable recommendations for human-AI interaction}
\label{subsubsec:actionable-recommendations-AI}

\FIXME{MMB@AAA: not quite sure about this one.  
We aren't proposing guidelines, so this title seems weird.  
But the content of the section seems to be on the right track.  
Maybe we want the title to be something like "Actionable recommendations for human-AI interaction"?
And then the difference is that those are guidelines for GENERIC human-AI, whereas ours START with such guidelines, then empirically produce Actionable results from an INCLUSIVITY perspective?  }

\boldify{Although our investigation is within the context of Amershi et al.'s guidelines, some have identified even since then that designing for human-AI interaction remains a challenge.}

This investigation occurred within the context of Amershi et al.'s guidelines for human-AI interaction, but there are other ongoing efforts to support human-AI interaction.
In January, 2022, Xu et al.~\cite{xu2021human} suggested that the set of design standards and guidelines supporting Human Computer AI-based systems was quite sparse, corroborating Yang et al.'s~\cite{yang2020re} observations that designing for quality HAI experiences remains a challenge for researchers and designers.
Some of the challenges for the user experience Yang et al. identified included assisting users in understanding AI capabilities, how to craft thoughtful interactions, and even collaborating with AI engineers throughout the design process.

\boldify{To address these challenges, other researchers and companies have also developed their own set of design principles and guidelines for human-AI interaction.}

To address these challenges, other works have also proposed (and evaluated) sets of design principles for human-AI interaction.
In 1999, Horvitz~\cite{horvitz1999principles} identified 12 critical factors for mixed-initiative user interfaces, since humans would transition towards performing collaborative tasks with intelligent agents\footnote{Amershi et al. acknowledge that 8 of the guidelines map to principles outlined in Horvitz's work.}.
Some of the critical factors pointed towards the need to consider things like the uncertainty of a user's goals, as well as how to empower the user to infer ideal actions in light of costs, benefits, and uncertainties.
Since then, researchers have proposed multiple principles towards aspects of human-AI interaction, such as Kulesza et al.'s~\cite{kulesza2015principles} principles of explanatory debugging, with situational considerations like principles for explaining how an AI made its decisions in the event that is wrong.
Other proposed principles focus on specific technologies, such as Ahmad et al.'s~\cite{ahmad2022designing} focus on personality-adaptive conversational agents.
Ahmad et al.'s work produced six principles, some of which suggest a need to design agents in such a way that they can support diverse users in a mental health setting.

\boldify{The last kind of advice are in the forms of guidelines. There have been different attempts to develop a set of guidelines for human-AI interaction, and Wright et al.~\cite{wright2020comparative} conducted a survey of all the guidelines from Apple, Google, and Microsoft}

Similar to our own research, others have investigated methods of informing the design of human-AI interaction through guidelines.
Wright et al.~\cite{wright2020comparative} survey guidelines from three major companies---Apple, Google, and Microsoft---and unify more than 200 guidelines into multiple categories.
In their work, they classify the guidelines into categories such as initial considerations of AI, curating the models themselves, the deployment of the AI-powered system, and the human-AI interface.
As Wright et al. point out, both Apple's~\cite{apple2019guidelines} and Google's~\cite{google_2019} guidelines are developed with the \textit{developer} in mind, whereas Amershi et al.'s guidelines focus on how the design pertains to the \textit{user}.
The closest work to our own that does an empirical investigation of guidelines for human-AI interaction comes from the first investigation, reported in Li et al.~\cite{li-MSR-work}.
The results, discussed in more detail in Section~\ref{subsec:methods-phase-one-results}, found that in almost all of the experiments, participants preferred products which applied the guidelines, and applying the guidelines positively impacted participants' user experience.

These works' outcomes produced guidelines for generic human-AI interaction, whereas our own investigation not begins within a specific AI context (productivity software) and empirically produces actionable results from an inclusivity perspective.

}

%% file: tables/2-Facet-to-Persona.tex
\begin{table}[h]
\centering

\includegraphics[width = \columnwidth]{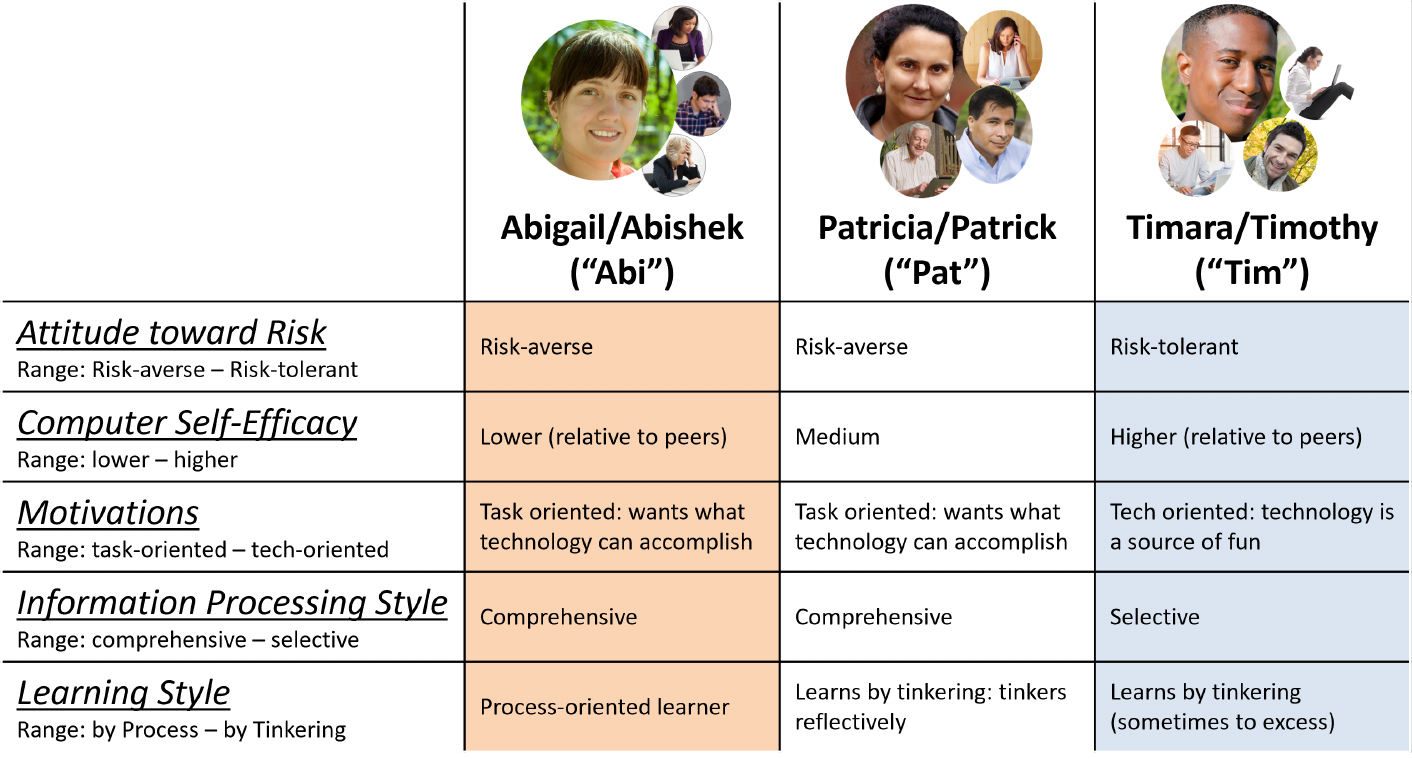}

\vspace{5pt}
    \caption{The five GenderMag problem solving style types (rows), each type's range of possible values, and the set of values for each.
    The \colorboxBackgroundForegroundText{AbiOrangeQuote}{black}{``Abi''} values (left) are the values at one end of each type, and the \colorboxBackgroundForegroundText{TimBlueQuote}{black}{``Tim''} values (right) are at the other end.
    Any individual can have any combination of values within these types, but in aggregate, the results have statistically clustered by people's self-identified gender~(e.g., \cite{burnett2016gendermag,stumpf2020gender,vorvoreanu2019gender}).}
    
    
    \label{table:personas}
    \vspace{-10pt}
\end{table}


%% file: tables/8-Guidelines.tex
\begin{landscape}
\begin{figure}[h]
    \centering
    \includegraphics[width = .95\linewidth]{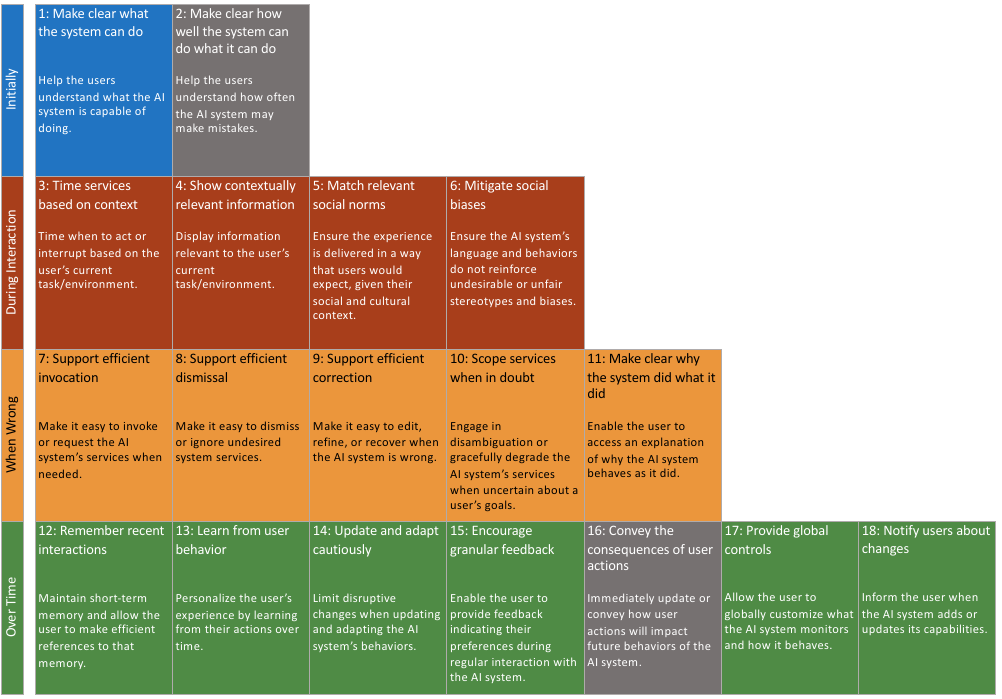}
    \caption{Amershi et al.'s 18 guidelines for human-AI interaction~\cite{amershi2019guidelines}.
    For the 4 phases (left column), each guideline has a number, title, and brief description.
    Our analyses exclude the two guidelines' experiments (Guidelines 2 \& 16, greyed out) which did not pass the manipulation check, as Li et al. did~\cite{li-MSR-work}.
    }
    \label{fig:human-AI-guidelines}
\end{figure}
\end{landscape}

%% file: doc/03-Methodology.tex
\section{Methodology 
}
\label{sec:methodology}

\boldify{This investigation was composed of 18 independent experiments, one for each of the guidelines in Figure~\ref{fig:human-AI-guidelines}} 

To investigate our research questions, we performed 18 independent experiments, one for each of Amershi et al.'s 18 HAI guidelines~\cite{amershi2019guidelines}.
We used these experiments to perform two investigations.
Investigation One, reported in Li et al.~\cite{li-MSR-work}, sought to understand the impacts of violating/applying these guidelines (Figure~\ref{fig:human-AI-guidelines}).
Investigation Two, which is the one we report in this paper, sought to better understand  potential disparities in the user experiences of participants with diverse problem-solving style values.
Our investigation used five problem-solving style spectra---the spectrum of participants' attitudes toward risk, of their computer self-efficacy, of their motivations, of their information processing styles, and of their learning styles (by process vs. by tinkering). 

To answer these research questions, we generated the following statistical hypotheses before data collection.
For any dependent variable and any of the five problem-solving styles, our statistical hypotheses between applications (app) and violations (vio) of any guideline were:
{
\begin{center}
    $H_0: \mu_{app} - \mu_{vio} = 0$

    $H_A: \mu_{app} - \mu_{vio} \neq 0$
\end{center}
} 






\subsection{Study Design
}
\label{subsec:study-design}

\boldify{The expts were about productivity software as a medium-risk domain. Participants were assigned to only one of the experiments. Each experiment was a 2x2 factorial experiment, and here were the factors. The guideline adherence factor was within-subjects, and the other was between subjects.}

The experiments' context was productivity software, such as document editors, slide editors, search engines, email applications, and spreadsheet applications.
Each experiment was a 2x2 factorial experiment, where each factor had two levels.

The first factor, the ``guideline adherence'' factor, was within-subjects, and the factor's levels were ``guideline violation'' and ``guideline application''.
For any one experiment's guideline, the ``guideline violation'' condition violated that particular HAI guideline;
for example, in Guideline 1's experiment (make clear what the system can do), the ``guideline violation'' did \textit{not} make clear what the system can do.
Similarly, in Guideline 11's experiment (make clear why the system did what it did), the ``guideline violation'' did \textit{not} make clear why the system did what it did.
In contrast, the ``guideline application'' level applied each guideline; 
for example, in Guideline 1's experiment (make clear what the system can do), the ``guideline application'' condition added clarifying information about what the system can do.

The second factor, the ``AI performance'' factor, was between-subjects.
This factor's levels were ``AI optimal'' and ``AI sub-optimal''.
In the ``AI optimal'' level, the AI sometimes made mistakes but worked well most of the time, whereas in the ``AI sub-optimal'' level, the AI sometimes made mistakes and sometimes worked well.

\boldify{Each level in guideline adherence factor was represented by a vignette, which several other works have done too. I'll summarise what Li et al. wrote.}

In each experiment, both the product that violated the guideline and the product that applied it were represented by vignettes, as in several other works in human-AI interaction~\cite{lima2021human,abendschein2021influence,mckee2021understanding,davis2022gender,lee-efficientAnnotations}.
The vignettes were developed in two phases:
in the first phase, two researchers went through an iterative brainstorming process, where they independently thought about how the 18 guidelines might show up in productivity software, drafting between 5--8 interaction scenarios for each guideline.
Then, the researchers collaborated to review, rewrite, and sometimes replace the scenarios.
In the second phase, the researchers adhered to Auspurg et al.'s~\cite{auspurg2009complexity} best practices to make the vignettes simple, clear, and realistic.
In cases where the interaction description was not understandable through text, images were used to promote understandability.
\FIXME{\textbf{R9-4.5: **MMB on 1/28/24 says DONE:} Produced by authors without involving end-users...}
\FIXED{Before deploying the study, each vignette went through two rounds of piloting.
In the first round, each vignette received feedback from 7 HCI researchers not familiar with the project, and changes were made based on that feedback.
In the second round, we piloted the updated vignettes on Amazon MTurk with five participants per vignette; no issues were identified from this second pilot.
} 
\FIXED{Each of the final
} 
vignettes was composed of three parts: (1)~a product/feature introduction; (2)~a description of what the AI feature did; and (3)~a summary of how well the AI performed.

%

\boldify{Figure~\ref{fig:G1-vignettes} provides an example of the two vignettes participants saw. Notice that they saw only a generic name for each vignette, and they had no concept of the guidelinesFor readability, we use the terms \vioProduct{} and \appProduct{} }

\FIXME{\textbf{R9-5: **MMB on 1/28/24 says DONE} ...it seems possible that a significant effect is due to the underlining application and the familiarity of the participants with them.}

Figure~\ref{fig:G1-vignettes} provides an example of the two vignettes from the experiment for Guideline 1 (``Make clear what the system can do'').
In first part, the only difference between the two conditions' vignettes was in the name (Ione and Kelso), generic names given to each product to distinguish them from each other 
\FIXED{and to avoid the influence of prior familiarity with a real product.
} 
The second part manipulates the ``guideline adherence'' factor.
In Figure~\ref{fig:G1-violation-vignette}, part 2 states: 
\begin{center}
``\textit{We will help you improve your presentation style}''
\end{center} 
without giving specific examples or details, thus violating the guideline by \textit{not} making clear what the system can do.
In contrast, Figure~\ref{fig:G1-application-vignette}'s part 2 applies the guideline to make clear exactly what the system can do, stating:
\begin{center}``\textit{As you practice your presentation, we will give you feedback about your presentation style:
how fast you speak,
use of filler words (such as `um' and `like'),
use of inappropriate words (such as `damn').}''
\end{center}
We will refer to the vignette that violated the guideline as the \vioProduct{}.
Similarly, we will refer to the vignette that applied the guideline as the \appProduct{}.

\input{figure/03-Vignettes}

\boldify{The dependent variables were drawn from multiple reputable sources to assess the user experience with AI products. Table~\ref{table:step1-dependent_variables} shows the ten dependent variables we analysed. (PARTICIPANTS ONLY SAW THE WORDING AND RESPONDED, R5!)}

Table~\ref{table:step1-dependent_variables} lists the questions that the participants responded to for both the \vioProduct{} and the \appProduct{}.
These dependent variables gather information about different dimensions of participants' user experiences.
The first five questions (control, secure, inadequate, uncertain, and productive) follow Benedek \& Miner's~\cite{benedek2002measuring} approach of measuring end users' feelings in user experience.
Perceived usefulness was taken from Fred Reichheld~\cite{reichheld2011ultimate} and has been known to relate to acceptance and use of AI-infused systems.
The last five questions (perceived usefulness, suspicious, harmful, reliable, and trust) came directly from Jian et al.~\cite{jian2000foundations}, who focused on scales for trust in automated systems.
The answer to each question was provided on an agreement scale, ranging from ``extremely unlikely'' (encoded as a 1) to ``extremely likely'' (encoded as a 7).

\input{tables/1-DependentVariables}


\subsection{Participants \& Procedures 
}
\label{subsec:methods-participants-procedures}

\boldify{1300 participants were recruited from MTurk, and they had to have at least 100 human-intelligence tasks (HITs) with above a 95\% acceptance rate on the platform.}

1,300 participants were recruited from Amazon Mechanical Turk (MTurk), a popular crowdsourcing platform.
To ensure quality data, participants had to meet certain performance criteria on MTurk before they could participate in the study, such as having at least 100 approved human intelligence tasks (HITs) and having above a 95\% acceptance rate on the platform.
Additionally, participants had to be located in the USA and be at least 18 years old.
After workers accepted the HIT, they were presented with an IRB consent form, and then  answered three screening questions.
The first two asked about their familiarity with productivity software and the last confirmed that they were above the minimum age requirements.
Upon completion of the screening survey, participants were provided \$0.20.




\boldify{Participants were randomly assigned to only one of the 18 experiments, one for each of the guidelines for human-AI interaction (Figure~\ref{fig:human-AI-guidelines}), and they were shown the two vignettes in a random order.}

Once participants had completed the screening survey, they were randomly assigned to one (and only one) of the 18 experiments, one for each of the guidelines for human-AI interaction (Figure~\ref{fig:human-AI-guidelines}).
First, participants randomly saw either the \vioProduct{} or \appProduct{}, such as the example provided in Figure~\ref{fig:G1-vignettes}\footnote{Participants were not told that one product violated/applied a guideline and one did the opposite.}.
Participants then responded to the user experience questions shown in Table~\ref{table:step1-dependent_variables}, which were asked in a random order.
Once participants completed their responses for the first AI product, they saw the second product and answered the same user experience questions in another random order.



\boldify{Once participants had seen both of the products, participants were asked to select which of the two products they preferred and why they preferred it, and they answered two manipulation check questions.}

Once participants had seen both the \vioProduct{} and \appProduct{} and answered the user experience questions for each, they were asked to select which of the two products they preferred and explain why they preferred it.
As mentioned in Li et al.~\cite{li-MSR-work}, one of the authors read the open-ended answers provided in each factorial survey repeatedly, until codes began to emerge. 
Then, the codes were recorded and each comment was coded.
Other team members conducted spot checks to verify the qualitative coding.
Participants were then asked two manipulation check questions%
\footnote{In experimental design, a manipulation check is a test used to determine the effectiveness of a manipulation in an experimental design.
Passing manipulation checks indicates that the manipulation in an experimental design was effective, whereas failing manipulation checks indicates that it was not.}, one closed- and one open-ended.
The closed-ended manipulation check asked participants whether or not they agreed with text that mirrored the guidelines themselves (e.g., ``make clear what the system can do'', ``make clear why the system did what it did'', etc.).
For example, if participants in Guideline 1's experiment agreed that the \appProduct{} made clear what the system can do,  and they disagreed with the statement for the \vioProduct{}, then they passed the manipulation check.
The open-ended manipulation check asked participants to ``...briefly describe the differences between Kelso and Ione'' (the fictitious names randomly assigned to the \vioProduct{} and \appProduct{}).
The open-ended answers were qualitatively coded to check whether or not each participant had successfully perceived the experimental manipulation.

\boldify{Last, participants filled out demographic information and the problem-solving style survey (Section~\ref{subsec:methods-phase-two-analysis}), and they were paid \$5 for completing the experiment.}

Participants then filled out a questionnaire with their demographic data, including their age, self-identified gender, race, highest education level, and field of employment.
They also filled out the problem-solving style questionnaire (Section~\ref{subsec:methods-phase-two-analysis}) and were paid a bonus of \$5 for completing the experiment.

\subsection{Investigation One Results Summary
}
\label{subsec:methods-phase-one-results}

\boldify{As mentioned in Section~\ref{sec:intro}, Investigation One (Li et al.) gauged the efficacy of Amershi et al.'s~\cite{amershi2019guidelines} guidelines, and found that...}

As mentioned in Section~\ref{sec:intro}, Investigation One, reported in Li et al.~\cite{li-MSR-work}, compared user experience outcomes of AI products that had applied the guidelines against AI products that had not.
That investigation's measures were generalized eta-squared ($\eta^2$) effect sizes for each of the dependent variables in each of the experiments.

The primary takeaway from Investigation One was that, for most of the guidelines, participants perceived the \appProduct{}s as more useful and as providing better user experiences than the \vioProduct{}s did.
Figure~\ref{fig:Li-resultThumbnails} shows thumbnails of their results for each guideline's experiment.
The more color-filled each thumbnail, the larger the  positive effect sizes were for that guideline's experiment.
For example, G3's thumbnail shows significant differences with small or medium effect sizes on most of the HAI-UX aspects measured.
G6's experiment produced particularly strong results.
Its thumbnail is almost filled with color, indicating that G6's experiment produced significant differences on all HAI-UX measures, with medium or large effect sizes for all but one.

\input{figure/Li-resultThumbnails}


\boldify{A few things inform our second phase: 1) two experiments failed the manipulation checks, so those experiments were dropped, leaving 1,043 participants total, and 2) the AI optimality factor was significant in only one experiment (G13), so this factor was dropped}

In addition, Investigation One's analysis informed a few aspects of Investigation Two's analysis.
It did so in two ways.
First, Investigation One's analysis revealed that 2 of the 18 experiments failed the manipulation checks (Section~\ref{subsec:methods-participants-procedures})---the experiments for Guideline 2 and Guideline 16---and as such were dropped from Investigation One.
Thus, our investigation also drops those two experiments, which leaves a total of 1,043 participants across the remaining 16 experiments.
Second, Investigation One's analysis of these remaining 16 experiments revealed that the AI optimality factor (Section~\ref{subsec:study-design}) was significant in only \textit{one} of these experiments.
This resulted in Investigation One dropping this experimental factor, and we do the same for Investigation Two.

\subsection{Investigation Two (Current Investigation) Data Analysis 
}
\label{subsec:methods-phase-two-analysis}

\boldify{We performed Investigation Two on this data from the perspective of participants' diverse problem-solving styles to uncover fairness issues (=inclusiveness and equity, but we just do INCLUSIVENESS, not EQUITY).
}

This paper's investigation analyzes the same independent experiments' data from a new perspective: the inclusivity that the violation vs. application AI products afforded diverse participants.
Specifically, we consider diversity in terms of participants' diverse problem-solving styles (\textbf{RQ1-Risk} and \textbf{RQ2-AllStyles}) and their diverse gender/age demographics (\textbf{RQ3-DemographicDiversity}).  

\boldify{We collected their problem-solving styles using the validated problem-solving style survey; because some people failed the attention checks, we ended up with 1,016 participants.}

To collect demographics, we used a questionnaire asking participants their gender identity and age group. 
To collect participants' diverse problem-solving styles, we used the GenderMag facets survey~\cite{hamid-2023}, a validated survey that measures participants' values of the five GenderMag problem-solving style types (enumerated in Table~\ref{table:personas}), termed ``facets'' in GenderMag publications.
Each problem-solving style type has multiple Likert-style questions that run from \textit{Disagree Completely} (encoded as a 1) to \textit{Agree Completely} (encoded as a 9), as few examples of which are shown in Table~\ref{tab:questionnaire-examples}.
For example, using this instrument, if one participant answers the first question (top row) closer to \textit{Agree Completely} than a second participant, the first participant is considered to be more risk-averse than the second participant.
27 of the 1043 participants failed at least one attention check in the problem-solving style survey, leaving 1,016 participants for this  investigation.
Appendix A lists the full questionnaire, including the attention checks.


\input{tables/18-Questionnaire-Examples}

\FIXME{AE-4 and R4-2.2 \textcolor{red}{MMB says this fix is DONE on 1/28/24}: how survey validated?}
\boldify{Here's how the survey was validated}
\FIXED{The GenderMag survey has previously been validated in multiple ways.
Hamid et al.~\cite{hamid-2023} summarize the six-step validation process; 
among the steps were literature searches, multiple statistical analyses, demographic validation, and problem-solving style validation.
Particularly relevant to this paper was Guizani et al.'s~\cite{guizani2022debug} participant validation of the problem-solving styles the survey captures.
In that study, participants took the survey, then spoke aloud throughout problem-solving tasks.  
Participants' in-the-moment verbalizations when problem-solving validated their own questionnaire responses 78\% of the time, a reasonably good measure of consistency~\cite{graham2011measuring}.
} 

\boldify{To score them, we sum them up and compare relative to their peers. }

To score a participant's problem-solving style values, we summed up that participant's responses to the risk questions; then the self-efficacy questions, and so on. 
Each sum is the participant's ``score'' for that problem-solving style.
Comparing these scores reveals a participant's placement in that problem-solving style type compared to others in the same peer group, such as among computer science professors, or among residents of  eldercare facilities, etc.; in our case, the peer group is the adult productivity software users who participated in the study.

\boldify{Then we chop the distribution into halves using the median}
These scores formed 16 distributions, one for each experiment (e.g., see Figure~\ref{fig:risk-scores-by-experiment} for the risk score distributions).
Using each experiment's own median\footnote{This approach has also been used in other measures of problem-solving styles, such as \textit{need for cognition}~\cite{buccinca2021trust}.}, which is robust against outliers, we then defined participants as being either more risk-averse than their peers (i.e., above the median) or not, and similarly for the other four problem solving styles.

\FIXME{MMB@AAA: I commented out the below "FIXED" becuase (1) I can't tell what comment it's trying to fix; (2) it seems to me that we've said this better in other places already in this paper; (3) it seems  weak; and (4) it's not part of the topic of this paragraph. If you agree that it can go, you can delete this FIXME.\\
AAA@MMB: It was supposed to be for R3-2.5. Part of our fix said ``(2)  in Methodology, talk about participants as a tuple of facetValues (Risk, SE, Motiv, Learn, Info, ''allothercharacteristicsintheworld). We doing the 1st five,  analyzing each facet separately.
}

\input{figure/01-Risk-Scores-by-Experiment}

\FIXME{R4-8: huh? Central Limit Theorem?}

\boldify{For each experiment, we used t-tests because... }

To analyze the dependent variables for each of the 16 experiments, we used t-tests after ensuring that the assumptions held, as follows. 
To investigate inclusivity (Section~\ref{subsec:inclusivity-risk}), we compared within-subjects using paired t-tests, treating each \vioProduct{} as a ``before'' and \appProduct{} as an ``after''.
As Table~\ref{tab:risk-sample-size} shows, each of the 16 experiments had over 30 participants, suggesting normality of every experiment by the Central Limit Theorem.%
\FIXED{
\footnote{``The Central Limit Theorem asserts that averages based on large samples have approximately normal sampling distributions, regardless of the shape of the population distribution''~\cite{ramsey2012statistical}.
By convention, the rule of thumb for a large enough sample is often considered to be >=30~\cite{ramsey2012statistical}.}  
In addition, in cases where the sample size fell beneath 30, we used Shapiro-Wilk tests to validate that the underlying reference distribution was not significantly different than normal (i.e., $p \geq .05$).
} 


\boldify{So, t-tests are the right thing. Also, we don't need stat corrections like Bonferroni because the hyps were pre-planned.}

Satisfying these assumptions indicated that the above t-tests were appropriate analysis techniques for these data.
Each of the 16 experiments were designed with pre-planned hypotheses for each dependent variable, so we did not add statistical corrections.
As other researchers~\cite{armstrong2014use,perneger1998s} point out,
statistical corrections (e.g., Bonferroni, Holm Bonferroni, Benjamini-Hochberg, etc.) are necessary only if ``...a large number of tests are carried out without pre-planned hypotheses''~\cite{armstrong2011statistical,armstrong2014use}.
\FIXME{\textbf{R9-2:} I recommend using one of \{Holm-Bonferroni, Benjamini-Hochberg \} methods of correction and double-checking  all of the results.}
Still, we recognize that not all readers may agree with this choice, 
\FIXED{so we also show all the Holm-Bonferroni corrections~\cite{holm1979simple} in Appendix D.
}

\input{tables/19-sample-size}

%% file: figure/03-Vignettes.tex
\begin{figure}[h]
    \centering
    \begin{subfigure}[]{0.47\linewidth} 
        \centering
        \includegraphics[width = \linewidth]{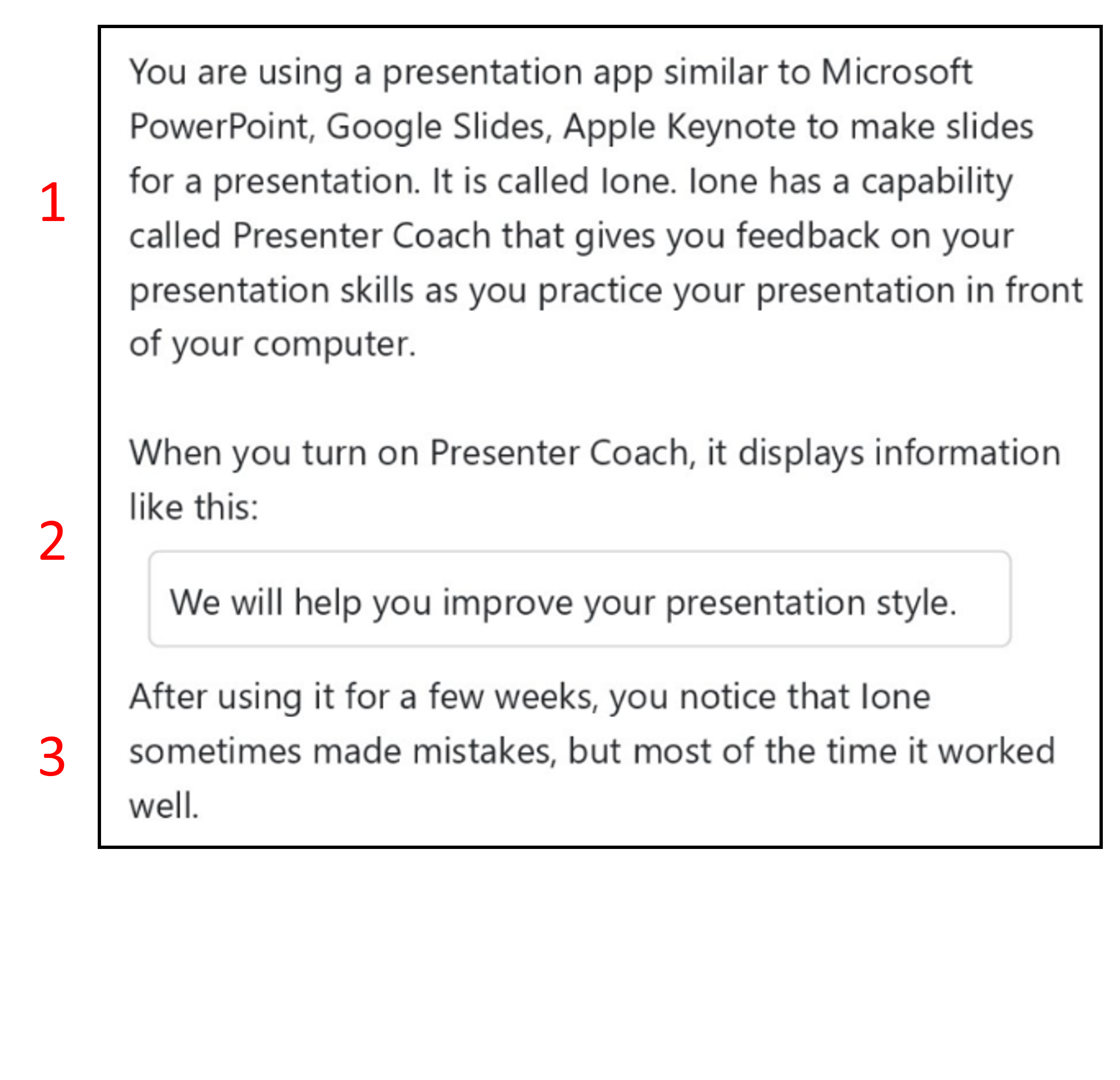}
        \caption{The \vioProduct{}.}
        \label{fig:G1-violation-vignette}
    \end{subfigure}
    \hspace{10px}
    \begin{subfigure}[]{0.47\linewidth} 
        \centering
        \includegraphics[width =\linewidth]{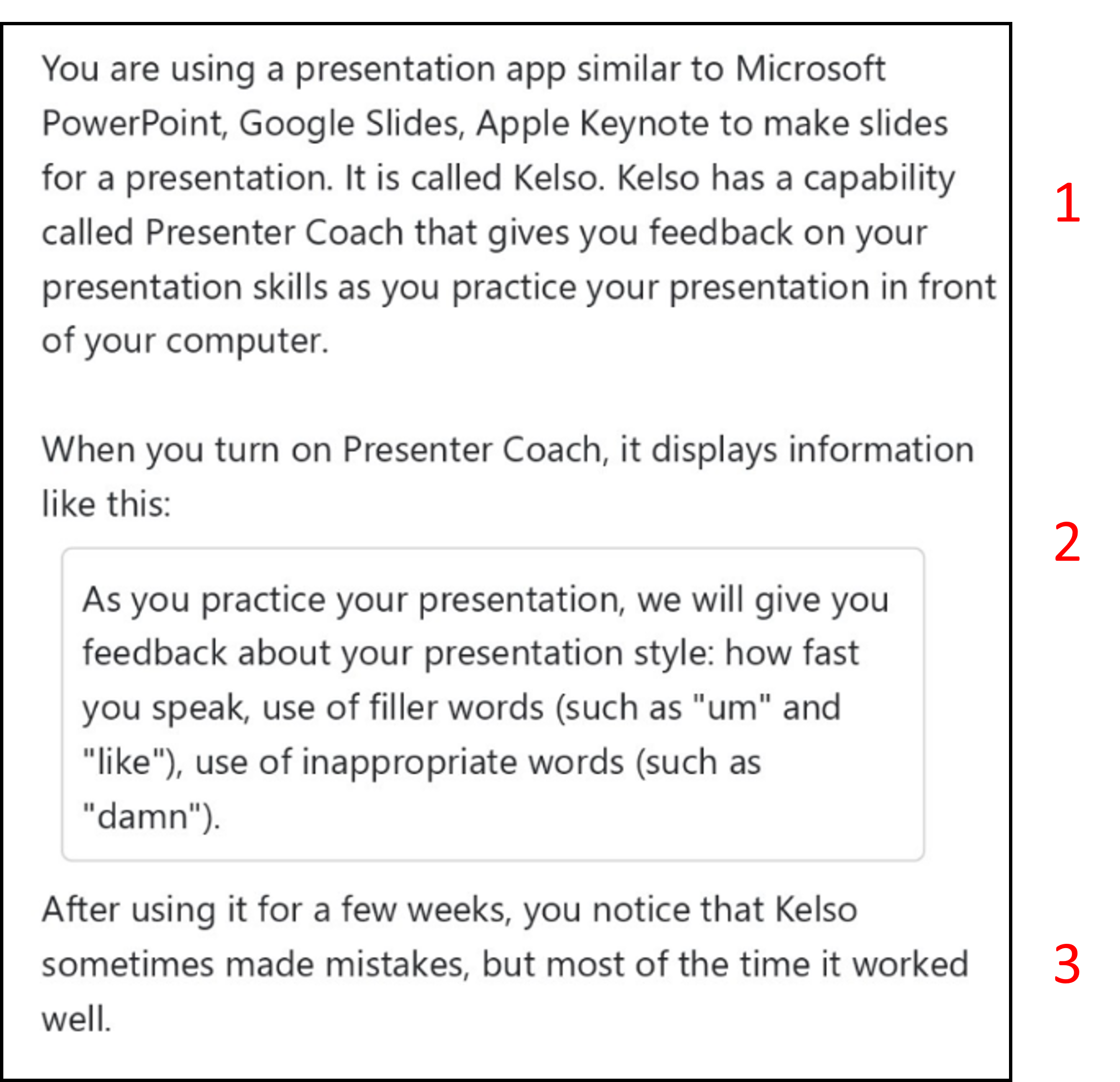}
        \caption{The \appProduct{}.}
        \label{fig:G1-application-vignette}
    \end{subfigure}
    \caption{Guideline 1's (``make clear what the system can do'') two vignettes.
    Each vignette had three components:
    (1) A product and feature introduction, describing what the product was and what it did,
    (2) the behavior description of the manipulated AI feature that differentiated the guideline's violation from its application, and
    (3) the AI performance description.
    Note that participants were never exposed to the concept of guideline violations or applications;
    instead, they saw only generic names (Ione \& Kelso).}
    \label{fig:G1-vignettes}
\end{figure}

%% file: tables/1-DependentVariables.tex
\begin{table}[h]
\centering
\small

\begin{tabular}{p{0.23\linewidth} p{0.6\linewidth} c}

 \textbf{Dependent Variable} & \textbf{Dependent Variable Wording}  & \textbf{Reverse-}\\
 \textbf{Name} & & \textbf{Coded?}  \\
\toprule

I would feel in control & ``\textit{I would feel in control while using the product.}'' & \\
\midrule

I would feel secure & ``\textit{I would feel secure while using the product.}'' & \\
\midrule

{I would feel inadequate} & ``\textit{I would feel inadequate while using the product.}'' & \checkmark   \\
\midrule

{I would feel uncertain} & ``\textit{I would feel uncertain while using the product.}'' & \checkmark \\
\midrule

I would feel productive & ``\textit{I would feel productive while using the product.}'' &\\
\midrule

I perceived it as useful & ``\textit{I would find the product useful.}'' & \\
\midrule

{I would be suspicious} & ``\textit{I would be suspicious of the intent, action, or outputs of the product.}'' & \checkmark  \\
\midrule

{It would be harmful} & ``\textit{I would expect the product to have a harmful or injurious outcome.}''& \checkmark  \\
\midrule

I find the product reliable & ``\textit{I would expect the product to be reliable.}'' & \\
\midrule

I would trust the product & ``\textit{I would trust the product.}'' &\\
\bottomrule

\end{tabular}

    \caption{The 10 dependent variables regarding users' perceived feelings~\cite{benedek2002measuring}, usefulness~\cite{reichheld2011ultimate}, and trust~\cite{jian2000foundations} questions.
    Participants answered these 7-point agreement scale questions for both the Violation product \textit{and} the Application product, which they saw in a random order.
    We indicate the reverse-coded questions ($\checkmark$) -- Feel Inadequate, Feel Uncertain, Suspicious, Harmful -- which Li et al. also did.
    As such, they became: Feel Adequate, Feel Certain, Not Suspicious, Not Harmful.
    Participants saw \textit{only} the wording shown in the ``Dependent Variable Wording'' column.}
    
    \label{table:step1-dependent_variables}
\end{table}

%% file: figure/Li-resultThumbnails.tex
\begin{figure}[]
    \centering
    \includegraphics[width=0.3\columnwidth]{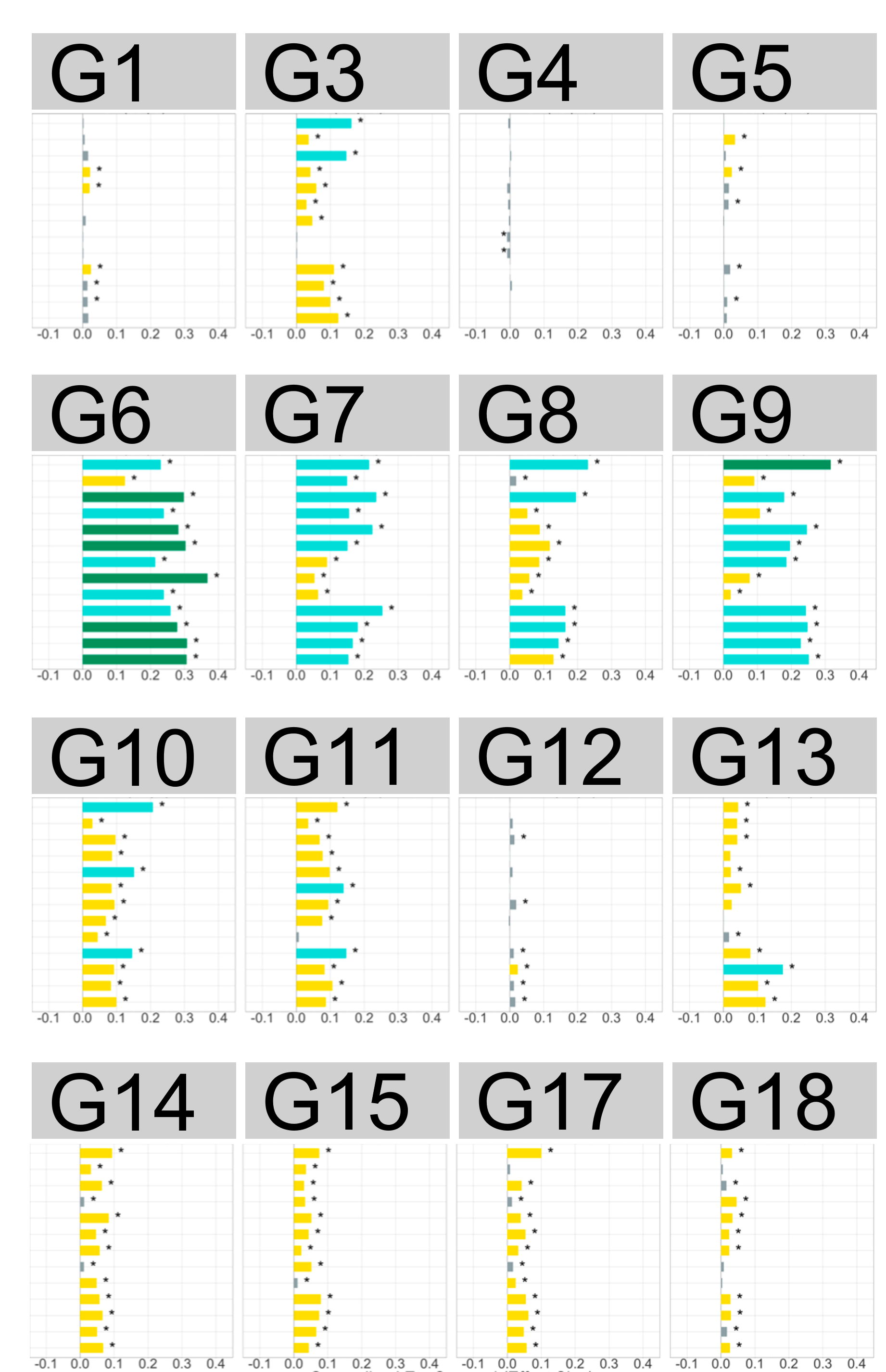}
    \caption{Thumbnails of Investigation One's results for each guideline's experiment. 
    More color indicates larger effect sizes. 
    *: difference was statistically significant.
    See Li et al.~\cite{li-MSR-work} for full details.}
    \label{fig:Li-resultThumbnails}
 
\end{figure}

%% file: tables/18-Questionnaire-Examples.tex
\begin{table}[]
    \centering
    \begin{tabular}{p{0.3\linewidth} p{0.6\linewidth}}
        \textbf{For this Problem Solving Style:} & \textbf{Sample Question:} \\
        \toprule
        Attitude toward risk & ``\textit{I avoid using new apps or technology before they are well-tested}''\\
        \midrule
        
         Computer self-efficacy & ``\textit{I am able to use unfamiliar technology when I have seen someone else using it before trying it.}''\\
         \midrule
         
        Motivations & ``\textit{It's fun to try new technology that is not yet available to everyone, such as being a participant in beta programs to test unfinished technology.}''\\
        \midrule

        Information processing style & ``\textit{I always do extensive research and comparison shopping before making important purchases.}''\\
        \midrule

        Learning style (by process vs. by tinkering) & ``\textit{I enjoy finding the lesser-known features and capabilities of the devices and software I use.}'' \\
        \bottomrule
    
    \end{tabular}
    \caption{Examples of questions from the validated problem-solving style survey (full survey in Appendix~\href{https://web.engr.oregonstate.edu/~anderan2/MSR-Appendices/Appendix-Facet-Survey-Rules.pdf}{{A}}).}
    \label{tab:questionnaire-examples}
\end{table}

%% file: figure/01-Risk-Scores-by-Experiment.tex
\begin{figure}[]
    \centering
    \includegraphics[width = 0.75\linewidth]{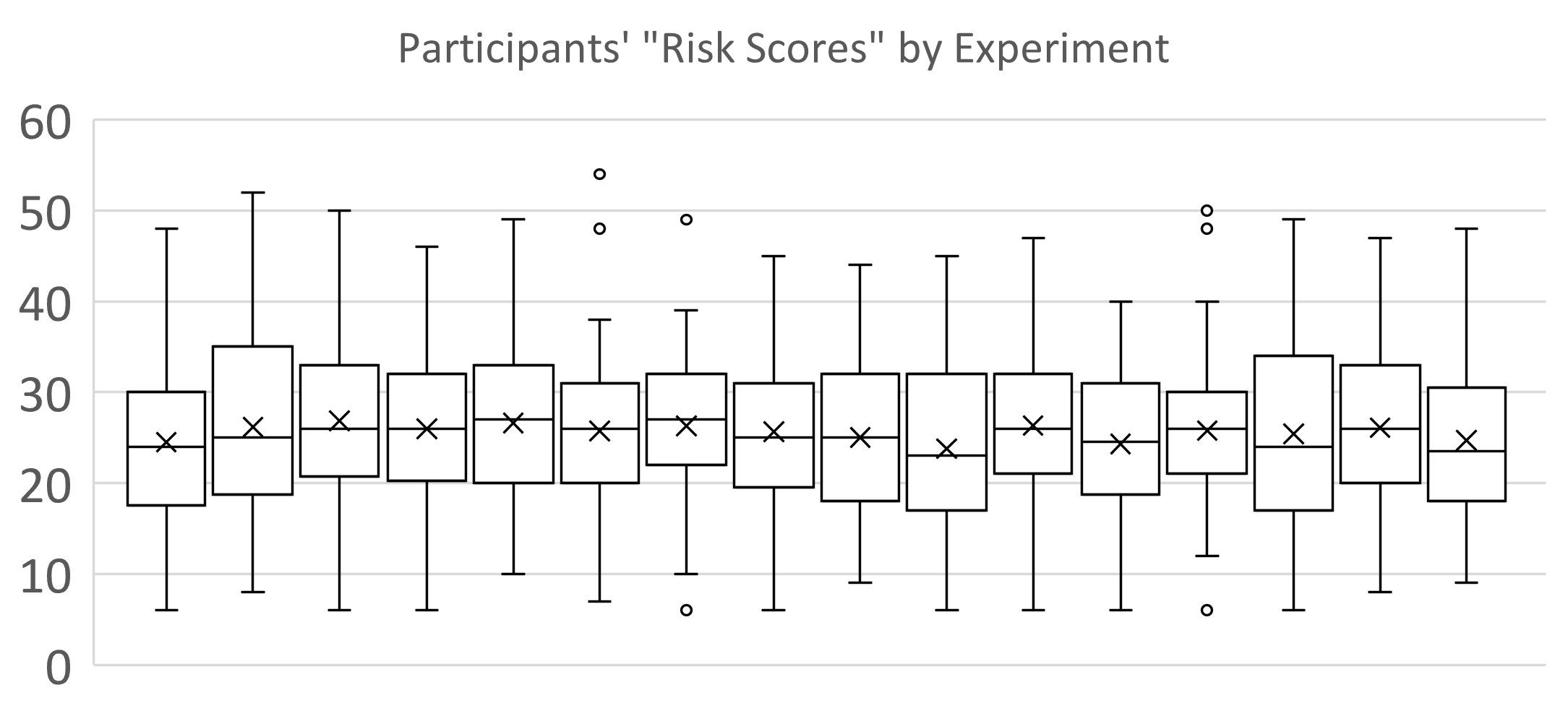}
    \caption{Participants' risk scores (y-axis) for each experiment (x-axis). ``x''s mark the means, horizontal lines mark the medians. 
    Participants above the median are more risk-averse than their peers below the median.}
    \label{fig:risk-scores-by-experiment}
\end{figure}

%% file: tables/19-sample-size.tex
\begin{table}[]
    \centering
    \begin{tabular}{
    p{0.15\linewidth}
    p{0.02\linewidth}
    p{0.02\linewidth}
    p{0.02\linewidth}
    p{0.02\linewidth}
    p{0.02\linewidth}
    p{0.02\linewidth}
    p{0.02\linewidth}
    p{0.02\linewidth}
    p{0.03\linewidth}
    p{0.03\linewidth}
    p{0.03\linewidth}
    p{0.03\linewidth}
    p{0.03\linewidth}
    p{0.03\linewidth}
    p{0.03\linewidth}
    p{0.03\linewidth}}
         \textbf{Experiment} & \textbf{G1} & \textbf{G3} & \textbf{G4} & \textbf{G5} & \textbf{G6} & \textbf{G7} & \textbf{G8} & \textbf{G9} & \textbf{G10} & \textbf{G11} & \textbf{G12} & \textbf{G13} & \textbf{G14} & \textbf{G15} & \textbf{G17} & \textbf{G18}  \\
         \toprule
         Risk-Averse Participants & 26 & 31 & 29  & 32 & 28 & 26 & 26 & 26 & 31 & 31 & 35 & 27 & 31 & 29 & 27 & 32\\
         \midrule
          Risk-Tolerant Participants & 31 & 37 & 36 & 37 & 36 & 32 & 35 & 32 & 34 & 32 & 34 & 30 & 37 & 35 & 36 & 35\\
         \bottomrule
         Total & 57 & 68 & 65 & 69 & 64 & 58 & 61 & 58 & 65 & 63 & 69 & 57 & 68 & 64 & 63 & 67
    \end{tabular}
    \caption{The number of risk-averse vs. risk-tolerant participants (rows) per experiment (columns). The group sizes were similar, with the smaller group at least 43\% of the total in every experiment.}
    \label{tab:risk-sample-size}
\end{table}

%% file: doc/04-Results-a-Inclusivity.tex
\section{Results: What participants' risk styles revealed \draftStatus{top2.9}
}
\label{sec:Results-Insights}
\label{subsec:inclusivity-risk}



\FIXME{\textbf{R4-4}}
\textbf{RQ1-Risk} considers the 16 pairs of AI products described in Section~\ref{sec:methodology}---one violating a guideline and its counterpart applying that guideline---and how the two differed in their inclusivity of risk-diverse participants' HAI experiences.
\FIXED{(We will then generalize beyond risk in Section~\ref{sec:beyond-risk}).
}

\boldify{We measured inclusivity change as follows.}


In this paper, we measure whether/how applying a guideline to an AI product \textit{changed} the product's inclusivity toward some particular group of participants.
For any user experience dependent variable in an AI product, we will say the HAI-UX is \textit{more (less) inclusive} to a group of participants if the Application AI product's result for that variable are \textit{significantly higher (lower)} than the  Violation AI product's for \textit{those} participants.%
\footnote{Recall that, because this inclusivity measure is within-subject, we used paired t-tests to measure HAI-UX in the before (\vioProduct{}) vs. after (\appProduct{}) versions of the AI products.}

\boldify{The outcomes could go any of the following 4 ways, and they did (see table). And in every case it was an improvement or neutral, never negative.}

To answer this question, we performed an in-depth analysis of all HAI-UX measurements' inclusivity by considering participants' attitudes toward risk.
HAI-UX inclusivity could change in only four possible ways: 
(1)~inclusivity changes for both the risk-averse and risk-tolerant participants, 
(2)~inclusivity changes for neither of them, 
(3)~inclusivity changes for the risk-averse only, and 
(4)~inclusivity changes for the risk-tolerant only.
As Table~\ref{tab:Risk-Inclusivity-Frequency} shows, instances of all of these categories occurred.

Table~\ref{tab:Risk-Inclusivity-Frequency} also shows that the risk results fell mainly in categories (1) and (2) above.
Perhaps most important, the table shows that whenever applying a guideline produced a change in inclusivity, it was almost always a \textit{positive} change for at least some risk-attitude group of participants---without loss of inclusivity for the other group.

\Result{1}{\textit{Following the guidelines usually led to inclusivity gains.} Applying these guidelines led to 115 (75+13+27) inclusivity gains for either or both risk groups, and only 1 inclusivity loss.}



\input{tables/04-Inclusivity-Risk-Frequency-Table}


%

\subsection{When everybody gained: More inclusivity for both the risk-averse and the risk-tolerant\draftStatus{MMBD2.8.}\\
}
\label{subsec:both-inclusivity}



\boldify{Why did so many people's includedness go up across the board? Well, there's interesting stuff in risk...}

How did these inclusivity results relate to risk?
At first glance, it may appear that this category of results is a natural consequence of the overall success rates shown by Investigation One.
For example, many of the experiments that produced strong positive results in Investigation One also did so in Investigation Two, as with G6--G9.
Still, even the G6--G9 experiments reveal  relationships between outcomes and perceptions of risk that shed new light on the \textit{why}s of these results. 

\boldify{For example, let's look at G8.}

For example, consider Guideline~8.
In this experiment, the AI-powered feature was a design helper to automatically provide design suggestions for alternative layouts in a presentation application.
In the \vioProduct{}'s vignette, the feature's behavior was:
\textit{``You are working on a slide and Design Helper pops up, showing you some design suggestions. You do not need any design help at this time, but there is no way to hide the design suggestions.''} 
In contrast, the \appProduct{}'s vignette started out the same, but its last sentence was: \textit{``You do not need any design help at this time, so you click on a button visible on screen to hide the design suggestions.''}

\boldify{This DV in G8 went way up in inclusivity.}
The second row of 
Table~\ref{tab:inclusivity-change-both} shows one of Guideline~8's outcomes,%
\footnote{Appendix D provides boxplots of results for all dependent variables in all experiments. 
}        
with the \colorboxBackgroundForegroundText{AbiOrangeQuote}{black}{risk-averse} participants' suspicions of the \vioProduct{} (\colorboxBackgroundForegroundText{AbiOrangeQuote}{black}{left hatched boxplot}) significantly worse than their suspicions of the \appProduct{} (\colorboxBackgroundForegroundText{AbiOrangeQuote}{black}{left clear boxplot})%
\footnote{Recall from Table~\ref{table:step1-dependent_variables} that the ``suspicious'' dependent variable was one of the variables that reverse-coded for presentation clarity, so that more positive outcomes were always higher on the scales.} 
\FIXME{\textbf{R9-1:} Not convinced t-tests are right...report in a footnote whether a non-parametric/ordinal test yields the same results. If not, consider swapping for good.}
\tTestResult{25}{3.2354}{.003}{.648}\footnote{\FIXED{This result was derived using Student's t-test, although these data are not continuous.
We validated all results in this paper using Wilcoxon signed rank test, and the non-parametric results agreed with our own 97\% of the time.
} }.
Likewise, the \colorboxBackgroundForegroundText{TimBlueQuote}{black}{risk-tolerant} participants also were significantly less suspicious of the \appProduct{} than of the \vioProduct{}
\tTestResult{34}{3.0020}{.005}{.507}, as shown in the \colorboxBackgroundForegroundText{TimBlueQuote}{black}{right boxplots}.

\input{tables/04-Inclusivity-Improved-Both}

\boldify{Participants' reasoning often had to do with their risk attitude. Risk is nuanced...}

Yet, despite their agreement on these outcomes, participants' free-text remarks showed that their
reasoning differed with their attitudes toward risk.  
The risk facet is nuanced---it includes aversion/tolerance risks with privacy/security, of producing low-quality work, of wasting too much time, of having trouble with the product, etc. 
In Guideline~8's experiment, about a fourth (14/61) of the participants' comments focused on the second of these, the risk of low-quality work.

\boldify{Wanting to avoid risk of low-quality work was especially true of risk-averse, but risk-tolerant wanted more control too.}

This focus on risk of low-quality work was especially true of risk-averse participants.
31\% (8/26) of this experiment's \colorboxBackgroundForegroundText{AbiOrangeQuote}{black}{risk-averse} participants wrote about preferring the increased control they had over their work quality with the \appProduct{}.



\quotateInset{...very convenient and still make me feel very much in control of my choices.}
{}
{\colorboxBackgroundForegroundText{AbiOrangeQuote}{black}{G08-1921-risk-averse}}
{Whole quote - Kelso would make my job very convenient and still make me feel very much in control of my choices.}

\quotateInset{I don't trust [\appProduct{}] ..., but the fact I can turn the feature off lets me be in more control.}
{}
{\colorboxBackgroundForegroundText{AbiOrangeQuote}{black}{G08-3619-risk-averse}}
{Whole quote - I don't trust lone because of the lower case to be honest, but the fact I can turn the feature off lets me be in more control.}


Even the more \colorboxBackgroundForegroundText{TimBlueQuote}{black}{risk-tolerant} were worried about this type of risk, and 17\% (6/35) of these participants expressed the same sentiments.
However, for these more \colorboxBackgroundForegroundText{TimBlueQuote}{black}{risk-tolerant} participants, annoyance and frustration also figured  prominently in their reasoning (26\%: 9/35), compared to only 1 \colorboxBackgroundForegroundText{AbiOrangeQuote}{black}{risk-averse} participant expressing this sentiment.

\quotateInset{Because I can get rid of the content that might ... influence me to do something stupid. If I am going to do something stupid it will be my idea.}
{}
{\colorboxBackgroundForegroundText{TimBlueQuote}{black}{G08-2831-risk-tolerant}}
{Whole quote - Because I can get rid of the content that might be wrong and might possibly influence me to do something stupid.  If I am going to do something stupid it will be my idea.}


\quotateInset{...[\appProduct{}] would allow me more freedom, and be less annoying with its suggestions, even when they are wrong.}
{}
{\colorboxBackgroundForegroundText{TimBlueQuote}{black}{G08-3681-risk-tolerant}}
{Whole quote - I feel Kelso would allow me more freedom, and be less annoying with its suggestions, even when they are wrong.}

\quotateInset{Without an option to turn off an unnecessary feature, I would be extremely frustrated...as it would be a severe distraction... never would I use [\vioProduct{}]...}
{}
{\colorboxBackgroundForegroundText{TimBlueQuote}{black}{G08-2627-risk-tolerant}}
{Whole quote - Without an option to turn off an unnecessary feature, I would be extremely frustrated while working with Kelso, as it would be a severe distraction, suggesting ideas that I have no need for at appropriate moments. With Ione however, if I ever feel the need to improve the design of a slide, I can enlist its help, but also have the option to hide it in case I no longer needs those services. Kelso does not have that option, thus never would I use it in any circumstance.}

\boldify{This points out interesting relationships between participants seeing a risk of bad output and DV outcomes.}

Comments like these, when coupled with the \colorboxBackgroundForegroundText{AbiOrangeQuote}{black}{risk-averse} and \colorboxBackgroundForegroundText{TimBlueQuote}{black}{risk-tolerant} participants' feeling both significantly more in control and less suspicious of the \appProduct{}, suggest relationships between an expectation of risk and four particular HAI-UX inclusivity outcomes.
As Table~\ref{tab:Risk-Inclusivity-Frequency} shows, across all 7 experiments where the \appProduct{} gained inclusivity for both the risk-averse and risk-tolerant participants' (not)-suspicious outcome (row 7), it also gained inclusivity for their certainty (row 4), control (row 1), and trust outcomes (row 10).
\Result{2}{\textit{Suspicion, control, trust, and certainty changed in tandem, for both risk groups.} In all experiments, every inclusivity gain in (1)~(not-)suspicious for both the risk-averse and risk-tolerant was coupled with an inclusivity gain in all three of (2)~in-control and (3)~trust, and (4)~certainty.}

This result provides insight into why the five experiments that gained the most inclusivity across the risk spectrum participants---G6, G7, G8, G9, and G15---did as well as they did.
What these five \textit{guidelines} have in common is that they all give users more control over the product.
What their five \textit{experiments} have in common (from Table~\ref{tab:Risk-Inclusivity-Frequency}) is that, in all of them, the \appProduct{} gained inclusivity in all four of the above variables:

\Result{3}{\textit{Giving users control mattered for both risk groups.} The five experiments with the most inclusivity gains across risk-diverse participants were those whose guidelines increased users' control over the AI products.}

\subsection{When nobody gained: No inclusivity improvements for either risk group \draftStatus{MMBd2.4.}\\
}
\label{subsec:none-inclusivity}
\boldify{The second-most prevalent category was where nobody's inclusivity got better.}

Not all the results were as positive for diversity. 
Some changes did not change inclusivity outcomes for either of the two groups, measured in this paper as no significant difference in HAI-UX inclusivity for either the \colorboxBackgroundForegroundText{AbiOrangeQuote}{black}{risk-averse} or the \colorboxBackgroundForegroundText{TimBlueQuote}{black}{risk-tolerant} participants between the \vioProduct{} and \appProduct{}.
This was the second-most prevalent category, occurring 44 times across 10 experiments.

\boldify{For example...}
Consider Guideline 4's (``show contextually information'') results, from Table~\ref{tab:no-inclusivity-change-general}'s examples.
Guideline 4's experiment produced 9 instances of this category.
In that experiment, the application was a document editor, and the AI-powered feature was an acronym explainer.
The \vioProduct{} violated the guideline: \textit{``When you highlight an acronym to see what it stands for, [Violation] shows you a standard list of possible definitions taken from a popular acronym dictionary.''}
In contrast, the \appProduct{}: \textit{``When you highlight an acronym to see what it stands for, [Application] shows you definitions that are used in your workplace and pertain to the topic of the current document.''}

\input{tables/04-Inclusivity-No-Change-General}

\boldify{Why? (1) same as before: qual of work}

In some ways, the participants' reasoning for their unchanging responses to the \vioProduct{} vs. the \appProduct{} echoed those of the previous subsection, namely wanting to avoid the risk of low-quality work.
As in the previous section, this reasoning was especially common among the \colorboxBackgroundForegroundText{AbiOrangeQuote}{black}{risk-averse} participants (34\% = 10/29), although 22\% (8/36) of the \colorboxBackgroundForegroundText{TimBlueQuote}{black}{risk-tolerant} also used it.
However, whereas in the previous section participants gave this risk as an \textit{asset} of the \appProduct{}, in this section they gave it as a \textit{liability} of the \appProduct{}.

\quotateInset{[\vioProduct{}] may make mistakes ... but its use of a generic dictionary makes it easier to recognize mistakes...  With [\appProduct{}],  I would be more likely to miss mistakes.}
{}
{\colorboxBackgroundForegroundText{AbiOrangeQuote}{black}{G4-4098-risk-averse}}
{Whole quote -Kelso may make mistakes in context,  but its use of a generic dictionary makes it easier to recognize mistakes,  I'll know that it isn't identifying workplace-specific acronyms.  With Ione,  I would be more likely to miss mistakes.  }

\quotateInset{...if [\appProduct{}] were to make a mistake on me, I would have a hard time trusting it because I did not make any part of the decision.}{}{\colorboxBackgroundForegroundText{AbiOrangeQuote}{black}{G4-3799-risk-averse}}{I would rather use Kelso because Kelso gives me the choice of choosing which definition I would want to go by. Ione would be easier but if Ione were to make a mistake on me, I would have a hard time trusting it because I did not make any part of the decision.}

\boldify{Why? (2) but also privacy.}

Guideline 4 also raised privacy concerns among some participants:

\quotateInset{... I would be nervous that [\appProduct{}] is pulling data from things like my other software and my browsing history.}
{}
{\colorboxBackgroundForegroundText{AbiOrangeQuote}{black}{G4-3905-risk-averse}}
{Whole quote - I would prefer to use Ione because Kelso feels a bit more intrusive. I would be nervous that it is pulling data from things like my other software and my browsing history. This would be unacceptable since I work with sensitive PII.}

\quotateInset{I don't like the idea of [\appProduct{}] taking definitions from my workplace. 
It makes me worry I'm being listened to...}
{}
{\colorboxBackgroundForegroundText{TimBlueQuote}{black}{G4-3947-risk-tolerant}}
{Whole quote -I don't like the idea of Ione taking definitions from my workplace. 
It makes me worry I'm being listened to, or that my network is being monitored, etc.}

\boldify{and guess what, we see that same relationship among trust, control, certainty and (not)-suspicious.}
In the ``everybody gained'' category (previous section), the five most inclusive guidelines across the risk spectrum revealed a relationship among risk-inclusivity and trust, control, certainty, and (not)-suspicious.
The five \textit{least} inclusive guidelines as per Table~\ref{tab:Risk-Inclusivity-Frequency}---G1, G4, G5, G12, and G13---show that the relationship persisted in the ``nobody gained'' category.
None of G1, G4, G5, and G12 produced any inclusivity gains for any of these interrelated variables; and G13 showed only two such gains.
These results not only confirm \textbf{Result \#2}, but also provide a complement to \textbf{Result \#3}:

\Result{4}{\textit{Not having user control mattered, for both risk groups.} None of the five guidelines showing the fewest risk-inclusivity gains offered increased user control over the AI products.}


\subsection{Selective inclusivity: who gained, who did not, and why? \draftStatus{MMBd2.4?}}
\label{subsec:abi-inclusivity}

\boldify{The third and 4th category of HAI-UX inclusivity change were when \textit{only} one or the other won. Abi (only) winning was rare: 13 times in 8 experiments. Tim (only) winning happened about twice as often: 27 times in 13 experiments}

The third and fourth categories of HAI-UX inclusivity changes were inclusivity gains for the \colorboxBackgroundForegroundText{AbiOrangeQuote}{black}{risk-averse} participants only or the \colorboxBackgroundForegroundText{TimBlueQuote}{black}{risk-tolerant} participants only.
Neither category was very large, with 13 and 27 total instances, respectively. 
Table~\ref{tab:inclusivity-only-one} shows a few examples.

\input{tables/04-Inclusivity-Only-One}

\boldify{Altho these weren't big categories,  the Tim category shines a spotlight on three guidelines' expts: G3's, G13's, G18's.}

Despite the relatively small totals, the fourth category, that of bringing gains to the \colorboxBackgroundForegroundText{TimBlueQuote}{black}{risk-tolerant} only, reveals a unique pattern shared by three experiments---Guideline~3's, Guideline~13's, and Guideline~18's.
As a column-wise reading of Table~\ref{tab:Risk-Inclusivity-Frequency} shows, in these three experiments, inclusivity gains for the \colorboxBackgroundForegroundText{TimBlueQuote}{black}{risk-tolerant} participants abounded, but the \colorboxBackgroundForegroundText{AbiOrangeQuote}{black}{risk-averse} participants rarely gained.

\boldify{Consider G3's expt, in which the AI figures out how busy you are to parcel out help it think you need it. Abi's didn't like this, Tim's were fine with it.}

Guideline~3's experiment offers a case in point.
In that experiment, the \appProduct{} provides services only when it decides the user's current task/environment would benefit.
The \appProduct{}'s vignette applied this guideline by stopping  email notifications ``\textit{...when you are busy.}''

As Table ~\ref{tab:Risk-Inclusivity-Frequency} shows, for Guideline~3's \colorboxBackgroundForegroundText{TimBlueQuote}{black}{risk-tolerant} participants, the \appProduct{} showed inclusivity gains on every dependent variable except one for the \colorboxBackgroundForegroundText{TimBlueQuote}{black}{risk-tolerant} participants.
However, only four of these gains extended to the \colorboxBackgroundForegroundText{AbiOrangeQuote}{black}{risk-averse} participants.

\boldify{why risk-tolerant and NOT risk-averse? Although they both liked getting fewer interruptions, For Abi, it came back to work quality and privacy, neither of which bothered Tim.}

Why such differences?
For Guideline~3's experiments, the \colorboxBackgroundForegroundText{AbiOrangeQuote}{black}{risk-averse} participants' concerns about risks to their work or their privacy appeared to outweigh the benefits of fewer notifications, whereas for the \colorboxBackgroundForegroundText{TimBlueQuote}{black}{risk-tolerant}, the balance seemed to tip the other way.
For example, 26\% (8/31) of the risk-averse participants explicitly brought up concerns about these risks, but only 11\% (4/37) of the risk-tolerant did.

\quotateInset{...Also, I wouldn't be sure that [\appProduct{}] would be able to accurately qualify my activities.}
{}
{\colorboxBackgroundForegroundText{AbiOrangeQuote}{black}{G3-3504-risk-averse}}
{Whole quote - Sometimes  I want email notifications even when I'm busy. Also, I wouldn't be sure that Kelso would be able to accurately qualify my activities.}

\quotateInset{[\appProduct{}]... would have to be able to monitor your online activity...that would be a little invasion of privacy...}
{}
{\colorboxBackgroundForegroundText{AbiOrangeQuote}{black}{G3-3054-risk-averse}}
{Whole quote - In order for lone to be able to tell what you are doing online they would have to be able to monitor your online activity. I would think that would be a little invasion of privacy them having access to my online activity.}



\boldify{G3's, G13's, and G18's expts (all favored by the Tim's), all had something in common: the vignettes' AI products were all learning from the USER's data. G3: MY context. G13: MY behavior. G18: MY emails.  Only the risk-tolerant were appreciative of this.}

A risk-oriented commonality among these three experiments lies in what these AI products were actually doing. 
All three of these \appProduct{}s learn from the \textit{user's own data}, as opposed to learning mainly from huge datasets mostly consisting of \textit{other people's} data. 
Specifically, both Guideline~3's and Guideline~13's AI products involved learning from the user's own context and behaviors, and Guideline~18's AI product involved learning from that user's emails.
In the latter case, the AI product also moved that user's emails around, adding the risk that the user might not find some of their emails later.

\boldify{Two others, G4 and G12, also had this attribute.  Even the risk tolerant were not particularly happy with these. This leads us to Result 5.}

Two other \appProduct{}s, Guideline~4's and Guideline~12's, also had this attribute.
Guideline~4 involved learning from the user's contexts, and Guideline~12 learned from the user's recent interactions.
Most of these two products' outcomes were in the ``nobody gained'' category: the \colorboxBackgroundForegroundText{AbiOrangeQuote}{black}{risk-averse} were very uncomfortable with these products, and even the \colorboxBackgroundForegroundText{TimBlueQuote}{black}{risk-tolerant} saw too many risks (e.g., recall the discussion of Guideline~4's experiment in the previous subsection).

These five guidelines' \appProduct{}s were the only ones with this attribute.
And for these five experiments, the \colorboxBackgroundForegroundText{AbiOrangeQuote}{black}{risk-averse} participants hardly ever experienced any inclusivity gains (from Table~\ref{tab:Risk-Inclusivity-Frequency}). 

\Result{5}{\textit{Learning from ``my'' data mattered}. Whenever the \appProduct{}s learned from participants' own data, inclusivity gains for risk-averse participants were rare.}

\subsection{The risk results and actionability \draftStatus{MMBd2.4.}}
\label{subsec:riskActionability}

\boldify{the AI products were designed to isolate applying vs. violating guidelines. But suppose it was a product some company was trying to get good.  }

The AI products in these experiments were designed to isolate effects of applying vs. violating each guideline.
However, if they were real products for sale, the products' owners would probably hope to make each product as well-received by as many of its customers as possible.

\boldify{The risk results gave us actionable ideas for some of the Gn results. Yaay!  But not on some others that we're really like to know about. If we look at the other facets, maybe we're get more ideas.}

The risk results provide actionable ideas for such product owners, for seven of these AI products.
Figure~\ref{fig:Li-resultThumbnails-forResults} points out which products those were.

\input{figure/Li-resultThumbnails-forResults}

The green boxes in the figure mark those products: they are G1's, G3's, G4's, G5's, G12's, G13's, and G18's.
For example, Section~\ref{subsec:both-inclusivity} revealed how lack of user control affected some of the low-performing AI products (e.g., G1's, G4's, G5's, G12's, and G13's).
An actionable implication for these products is that those products would improve by offering users more control.
As another example, Section~\ref{subsec:abi-inclusivity} revealed the sensitivity the \colorboxBackgroundForegroundText{AbiOrangeQuote}{black}{risk-averse} participants had to products that potentially did ``too much'' with \textit{their} data ( e.g., G3's, G4's, G12's, G13's, and G18's products).  
One actionable idea for those products would be to provide information on what else the user's personal data are used for and how long these data are stored.
More generally, results like these suggest that a way to improve AI products favored by only the \colorboxBackgroundForegroundText{AbiOrangeQuote}{black}{risk-averse} participants or only the \colorboxBackgroundForegroundText{TimBlueQuote}{black}{risk-tolerant} participants, is to attend to risk-oriented attributes of the product that were not tolerated well by participants at that end of the risk spectrum.

\Result{6}{\textit{Some risk results were actionable.} For the seven \appProduct{}s associated with G1, G3, G4, G5, G12, G13, and G18, the Risk results provided actionable ideas for further improving the HAI-UX those products offered.}

%% file: tables/04-Inclusivity-Risk-Frequency-Table.tex
\begin{table}[h]
    \centering
    \includegraphics[width = 0.82\linewidth]{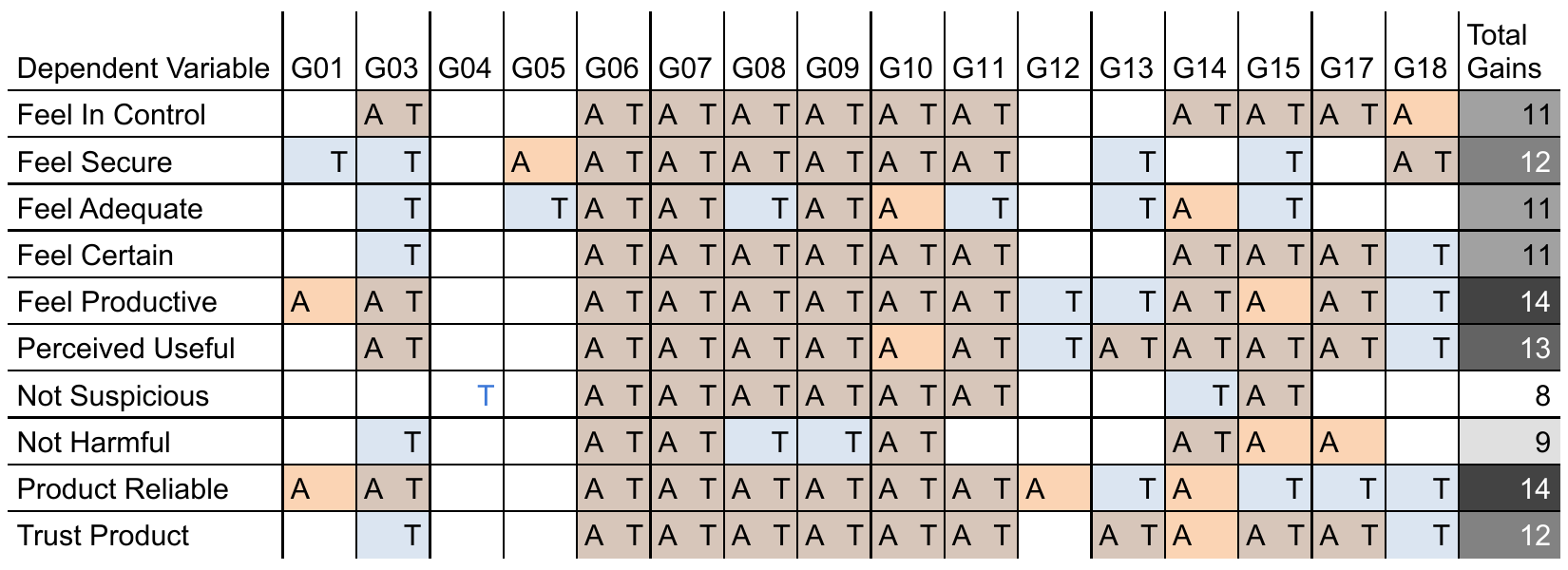}
    \caption{Risk results summary, by  guideline experiment (columns) and dependent variable (rows).
    Reading columnwise reveals the guidelines with the most inclusivity gains, and reading rowwise reveals the dependent variables with the most inclusivity gains (rightmost column).
    There was also one inclusivity loss.\\
    Total occurrences: 75 inclusivity gains for \colorboxBackgroundForegroundText{Eggshell}{black}{both risk-Averse and risk-Tolerant (A T)};
    44 without inclusivity gains/losses (blank); 13 inclusivity gains for  \colorboxBackgroundForegroundText{AbiOrangeQuote}{black}{risk-Averse only (A)};
     27 inclusivity gains for  \colorboxBackgroundForegroundText{TimBlueQuote}{black}{ risk-Tolerant only (T)};
     1 inclusivity loss for \textcolor{NegativeTim}{risk-Tolerant only (T)}. Total possible: 160.
    }
    \label{tab:Risk-Inclusivity-Frequency}
\end{table}

%% file: tables/04-Inclusivity-Improved-Both.tex
\begin{table}[h]
    \centering
    \begin{tabular}{p{0.25\linewidth}|c|c}
         Guideline&  Risk-Averse Inclusivity & Risk-Tolerant Inclusivity \\
         \toprule
         \parbox[c]{\linewidth}{3: Time services based on context }  
         &
         \parbox[c]{0.25\linewidth}{
         \includegraphics[width = \linewidth]{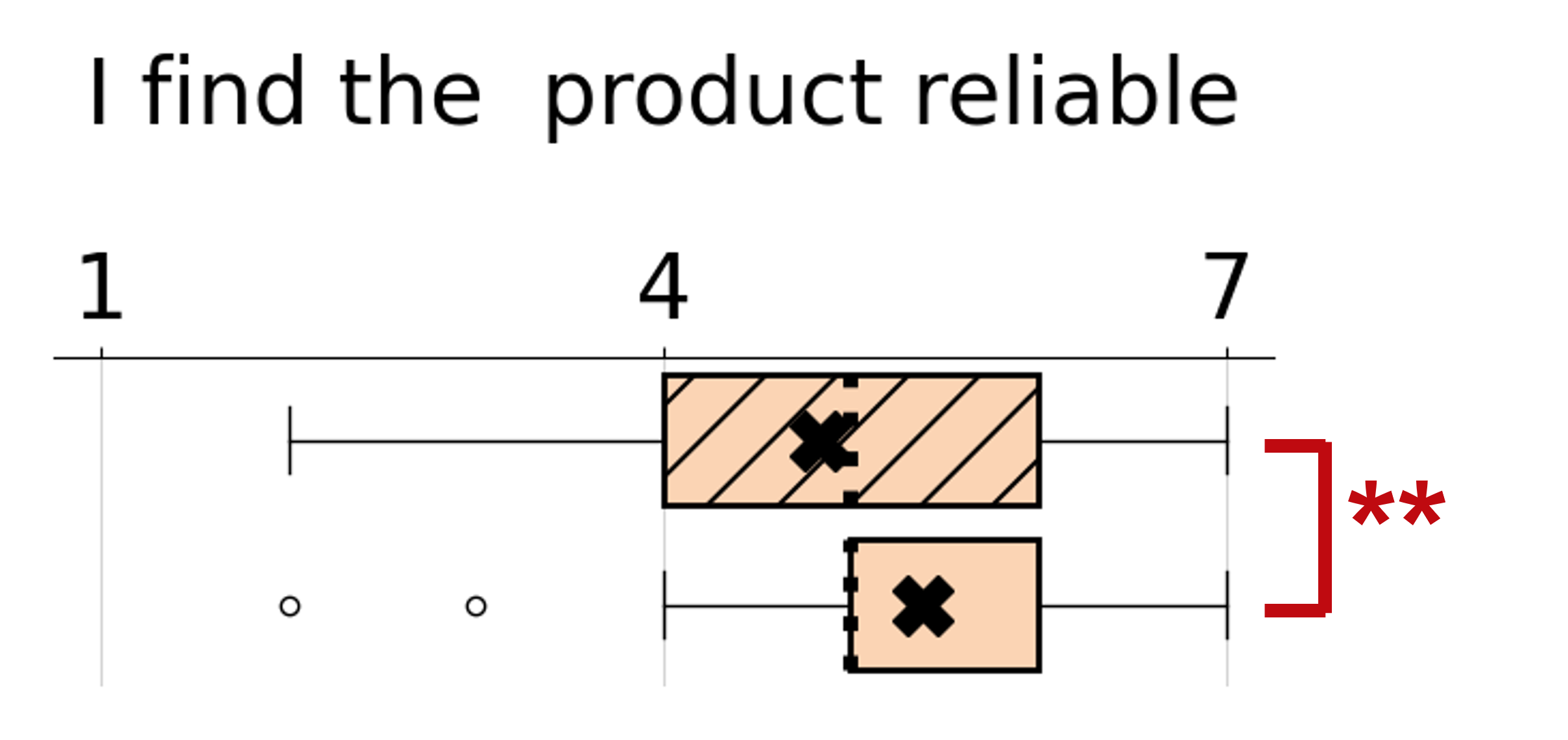}
         }
         & 
         \parbox[c]{0.25\linewidth}{
         \includegraphics[width = \linewidth]{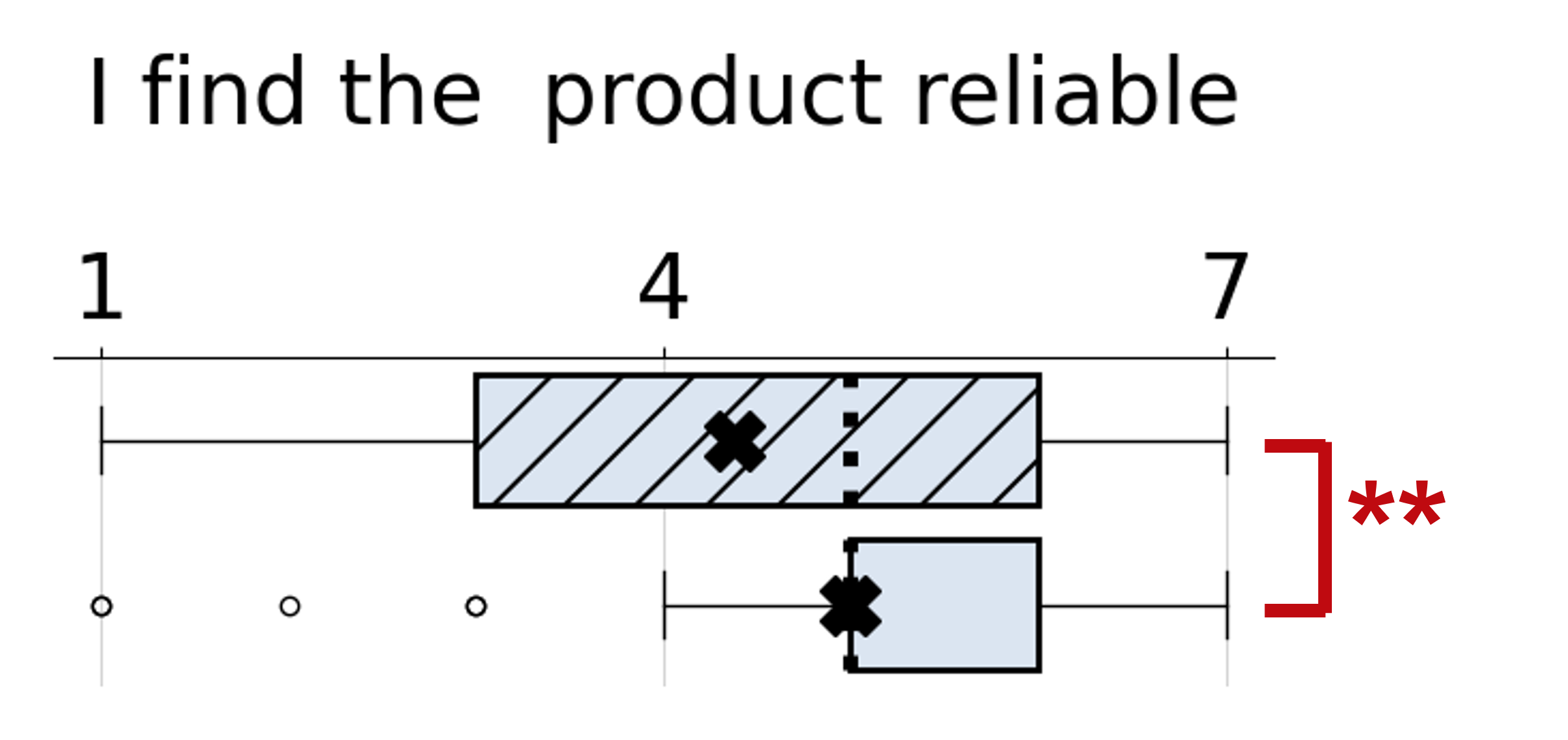} 
         }
         \\
         \midrule
         \parbox[c]{\linewidth}{8: Support efficient dismissal}
         &
         \parbox[c]{0.25\linewidth}{
         \includegraphics[width = \linewidth]{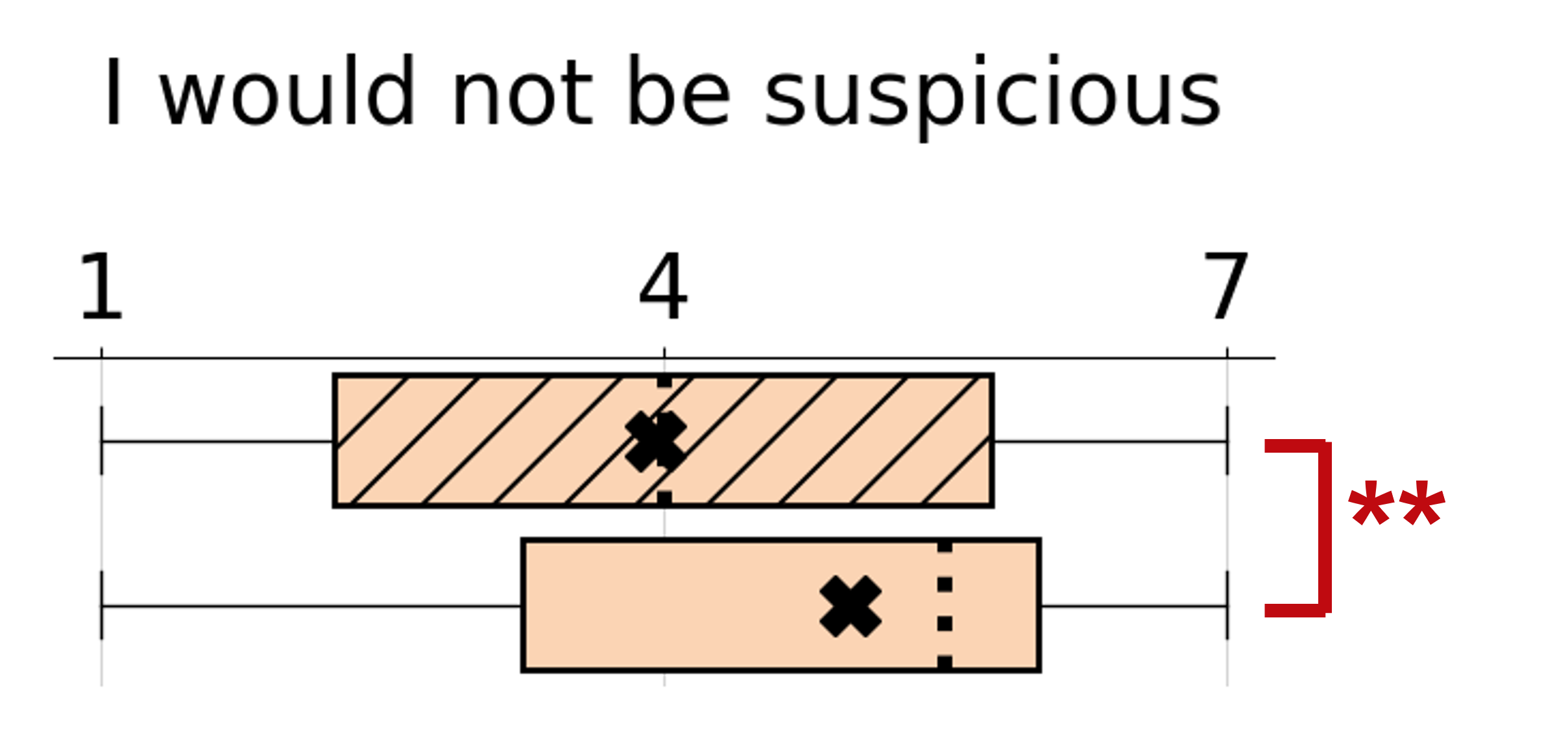} 
         }
         &
         \parbox[c]{0.25\linewidth}{
         \includegraphics[width = \linewidth]{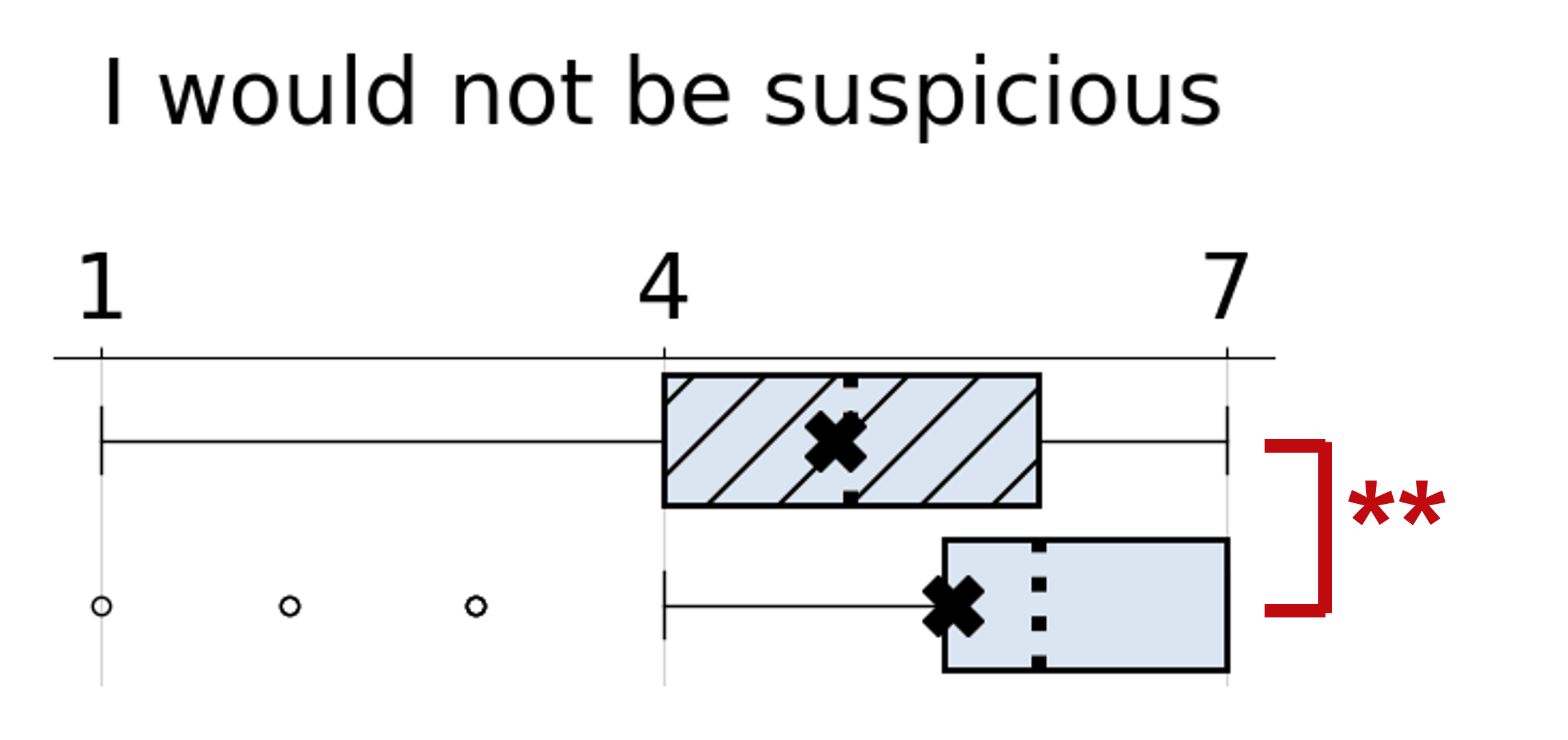}
         }
        \\
         \midrule
         \parbox[c]{\linewidth}{14: Update and adapt cautiously }
         &
         \parbox[c]{0.25\linewidth}{
         \includegraphics[width = \linewidth]{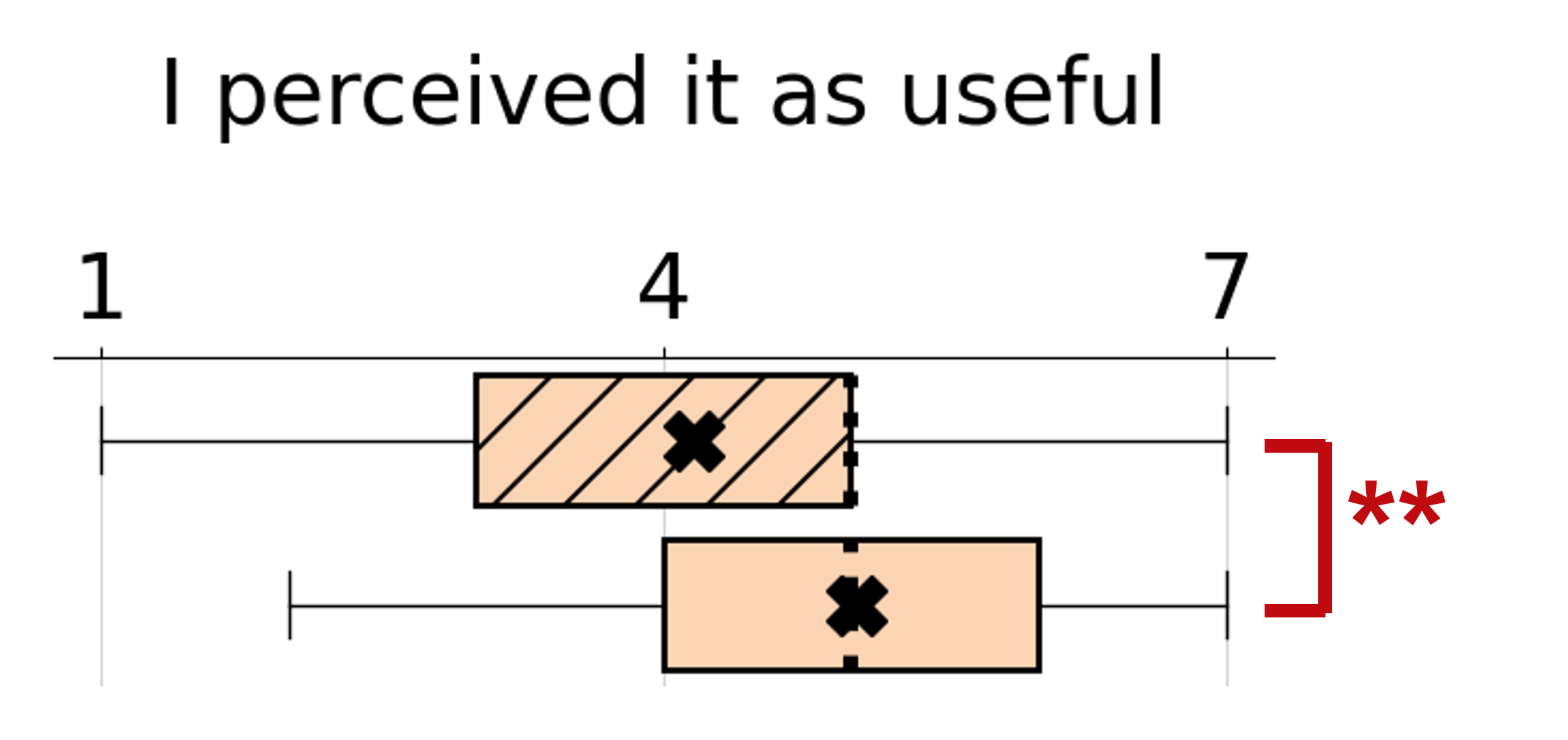}
         }
         & 
         \parbox[c]{0.25\linewidth}{
         \includegraphics[width = \linewidth]{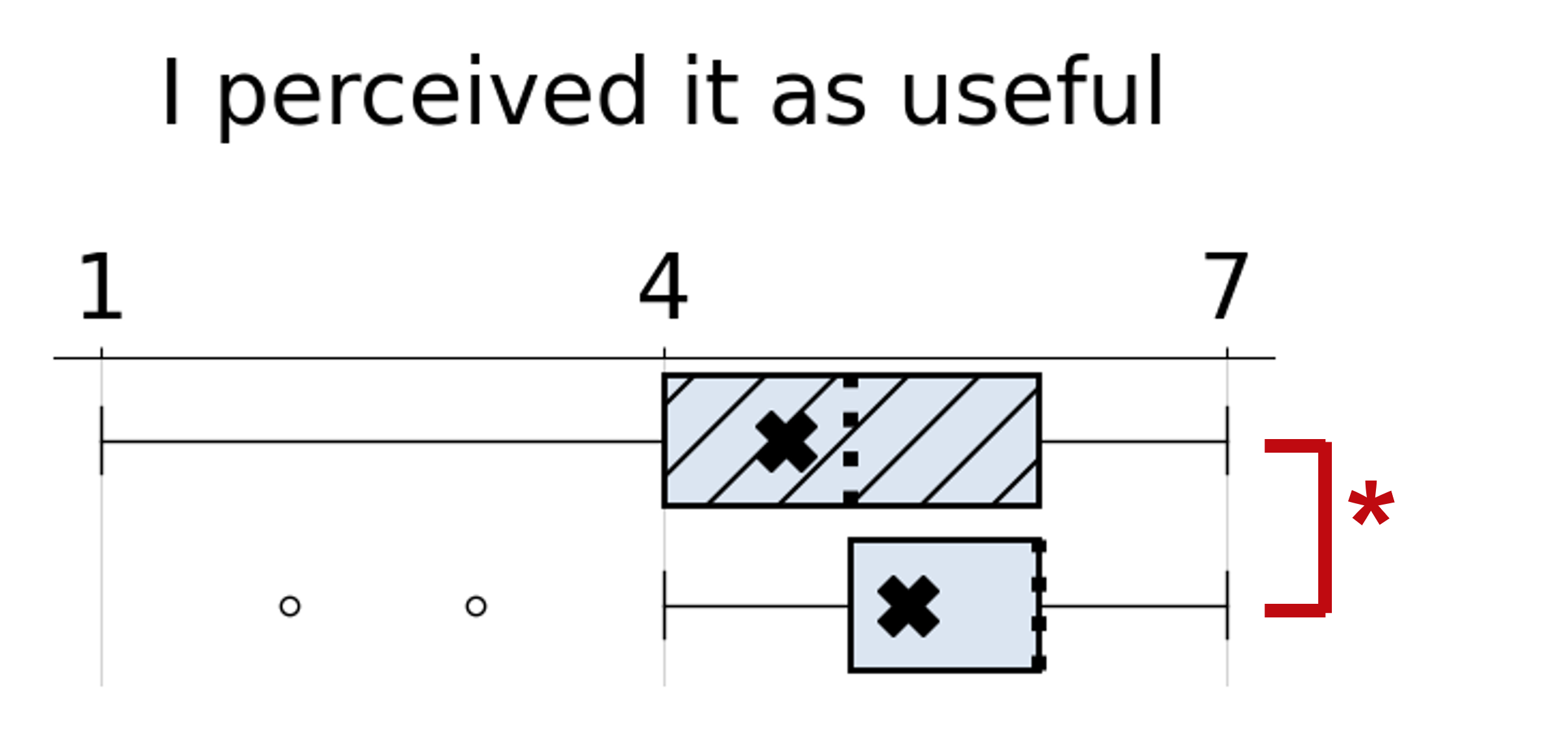}   
         }
         \\

    \end{tabular}
    \caption{Inclusivity gains: A few examples of the 75 times inclusivity improved for both \colorboxBackgroundForegroundText{AbiOrangeQuote}{black}{risk-averse} and \colorboxBackgroundForegroundText{TimBlueQuote}{black}{risk-tolerant} participants.
    E.g., in Guideline~8's experiment (second row), both the \colorboxBackgroundForegroundText{AbiOrangeQuote}{black}{risk-averse} and the \colorboxBackgroundForegroundText{TimBlueQuote}{black}{risk-tolerant} participants were significantly more suspicious of the \vioProduct{} (top, hatched boxplots) than of the \appProduct{} (bottom, clear boxplots). 
    x = average, \textbrokenbar~= median.
    {\color{FigureMaroon}{*}} = p<.05, {\color{FigureMaroon}{**}} = p<.01, {\color{FigureMaroon}{***}} = p<.001, NS=not significant.
    }
    \label{tab:inclusivity-change-both}

\end{table}

%% file: tables/04-Inclusivity-No-Change-General.tex
\begin{table}[h]
    \centering
    \begin{tabular}{p{0.25\linewidth}|c|c}
         Guideline&  Risk-Averse Inclusivity & Risk-Tolerant Inclusivity\\
         \toprule
         \parbox[c]{\linewidth}{1: Make clear what the system can do }   
        & 
        \parbox[c]{0.25\linewidth}{
        \includegraphics[width = \linewidth]{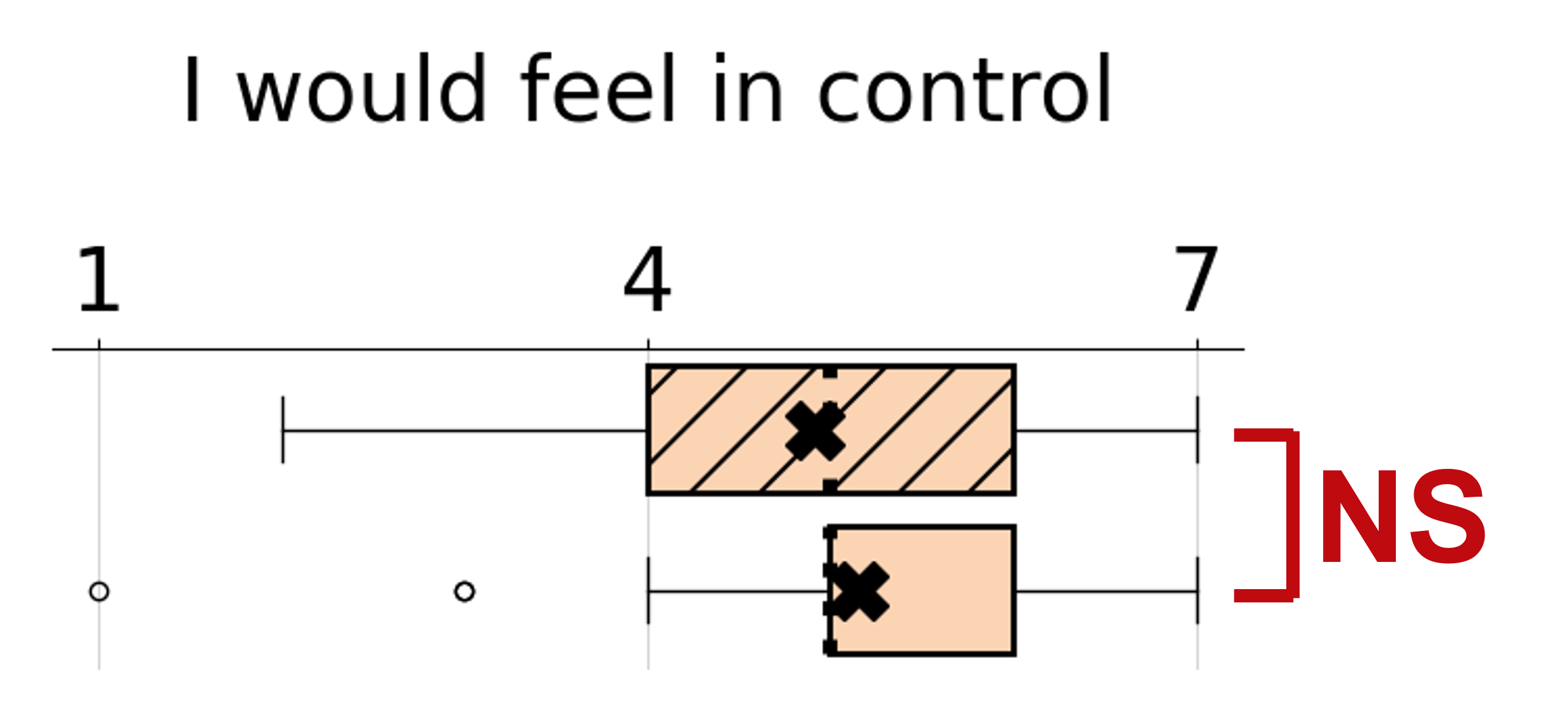}
        }
         & 
         \parbox[c]{0.25\linewidth}{
         \includegraphics[width = \linewidth]{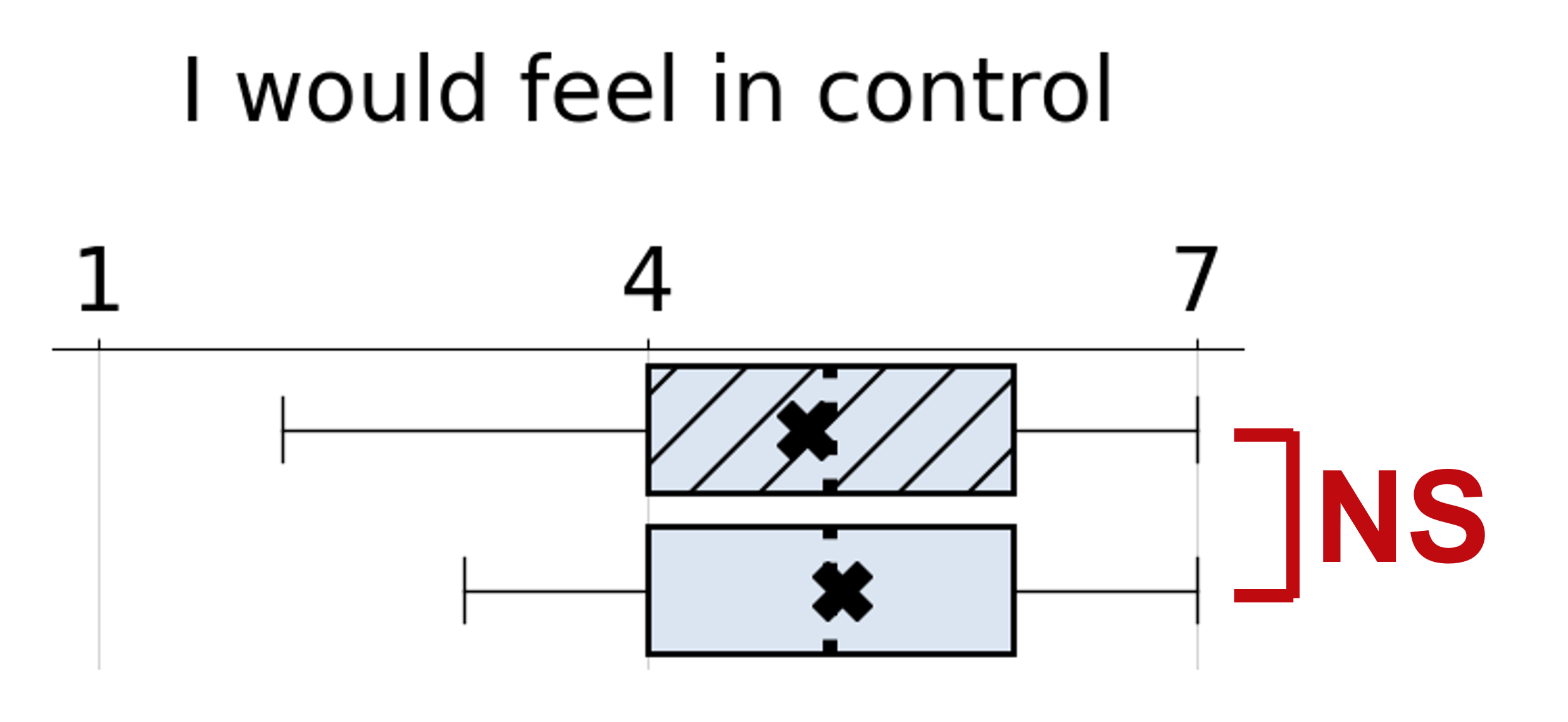}    
         }
         \\

%

         \midrule
         \parbox[c]{\linewidth}{4: Show contextually relevant information} 
         &
         \parbox[c]{0.25\linewidth}{
         \includegraphics[width=\linewidth]{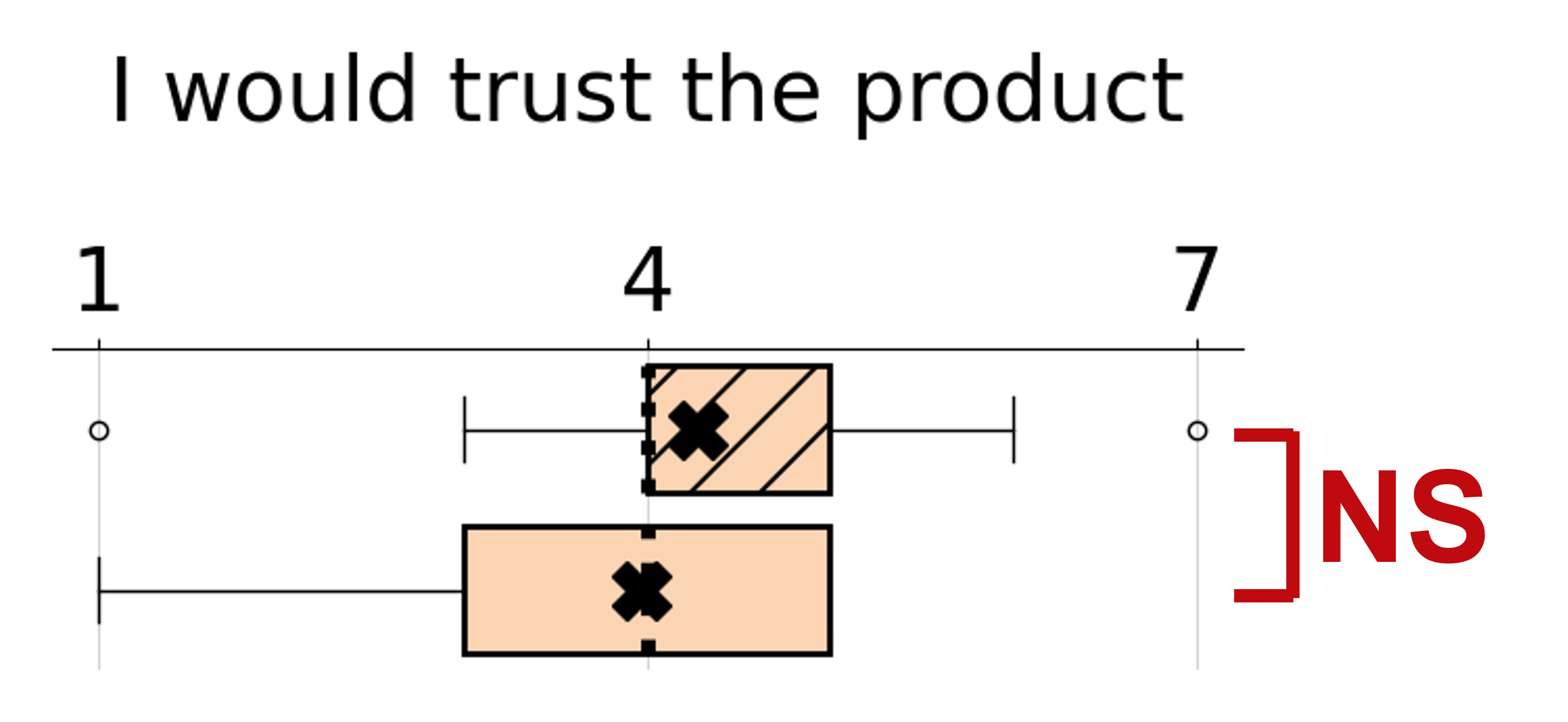}
         }
         & 
         \parbox[c]{0.25\linewidth}{
         \includegraphics[width=\linewidth]{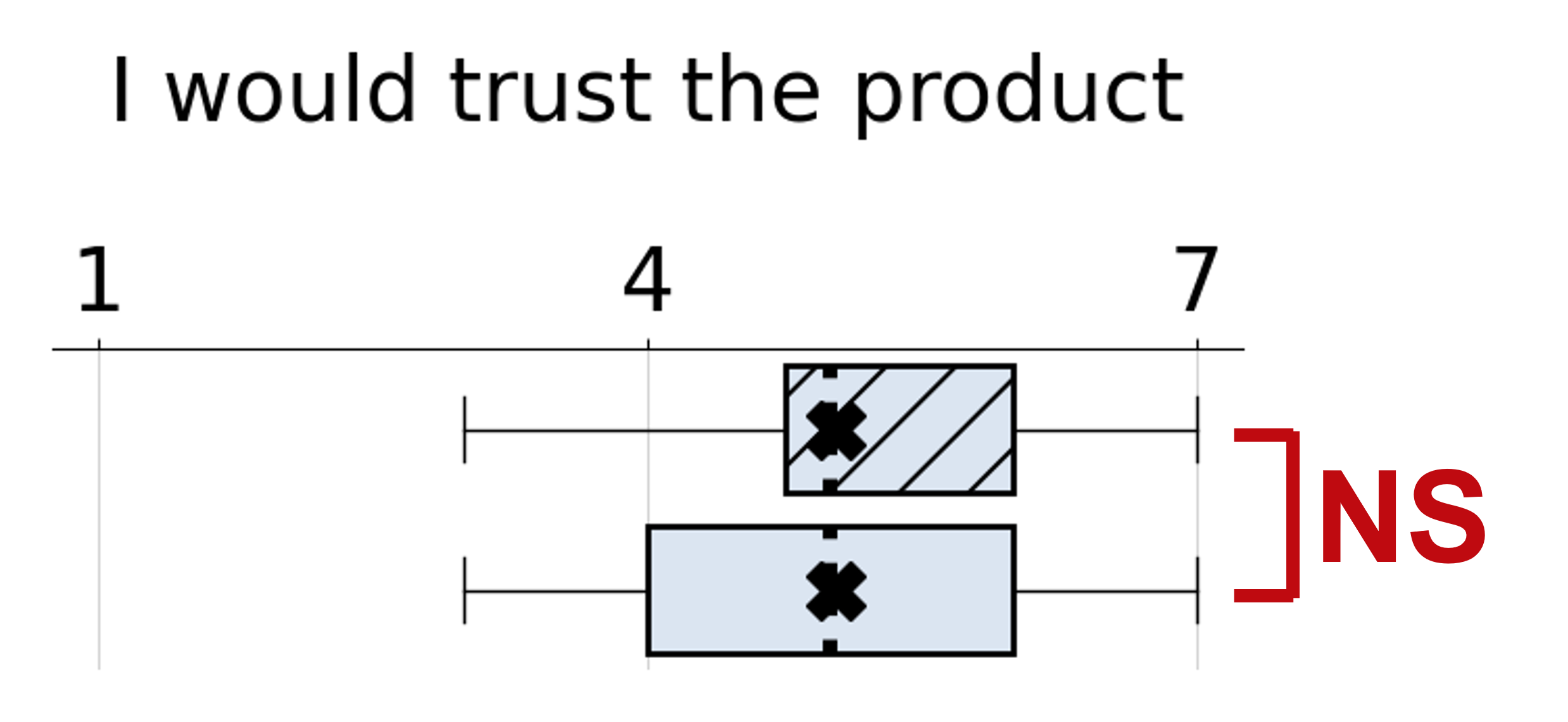} 
         }
         \\

%
         \midrule
         \parbox[c]{\linewidth}{12: Remember recent interactions} 
         &
         \parbox[c]{0.25\linewidth}{
         \includegraphics[width = \linewidth]{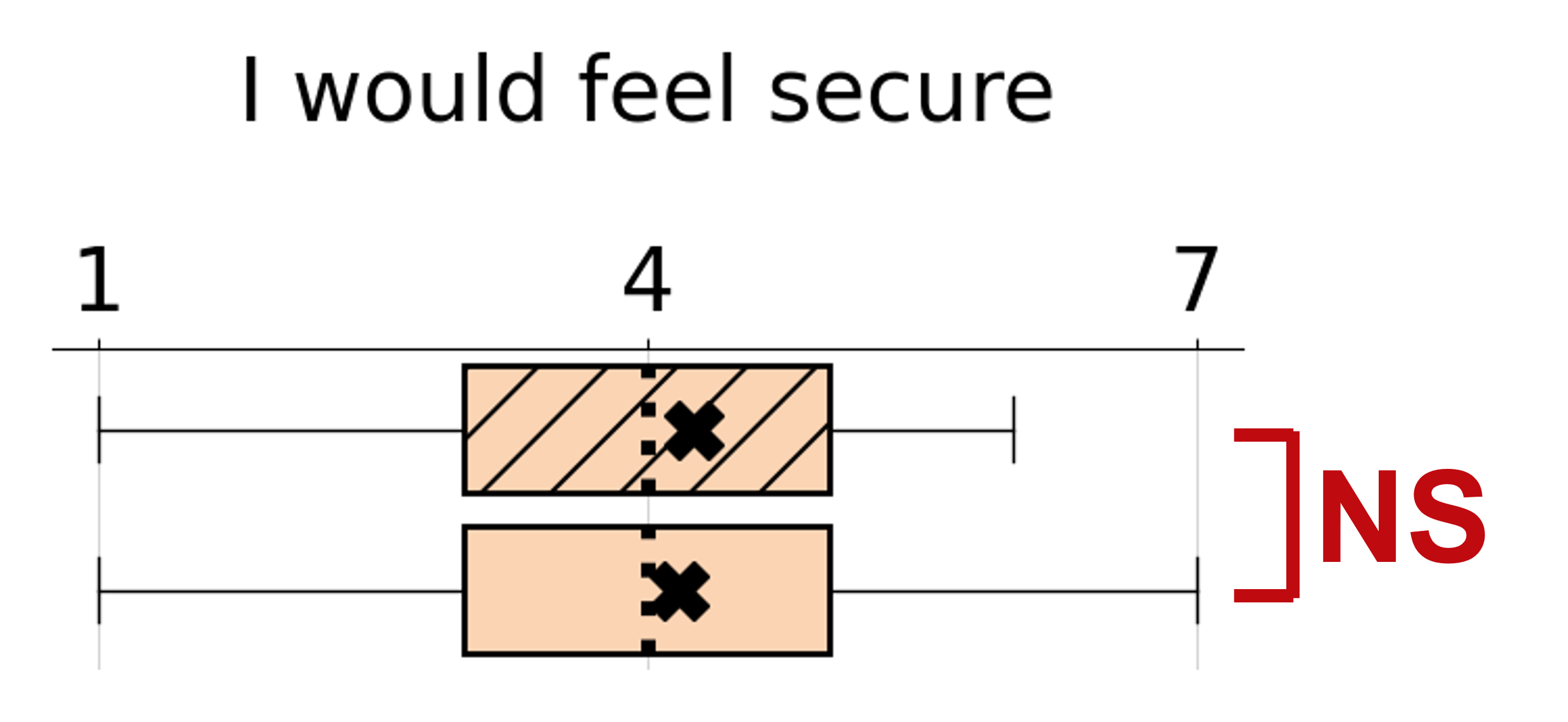}
         }
         & 
         \parbox[c]{0.25\linewidth}{
         \includegraphics[width = \linewidth]{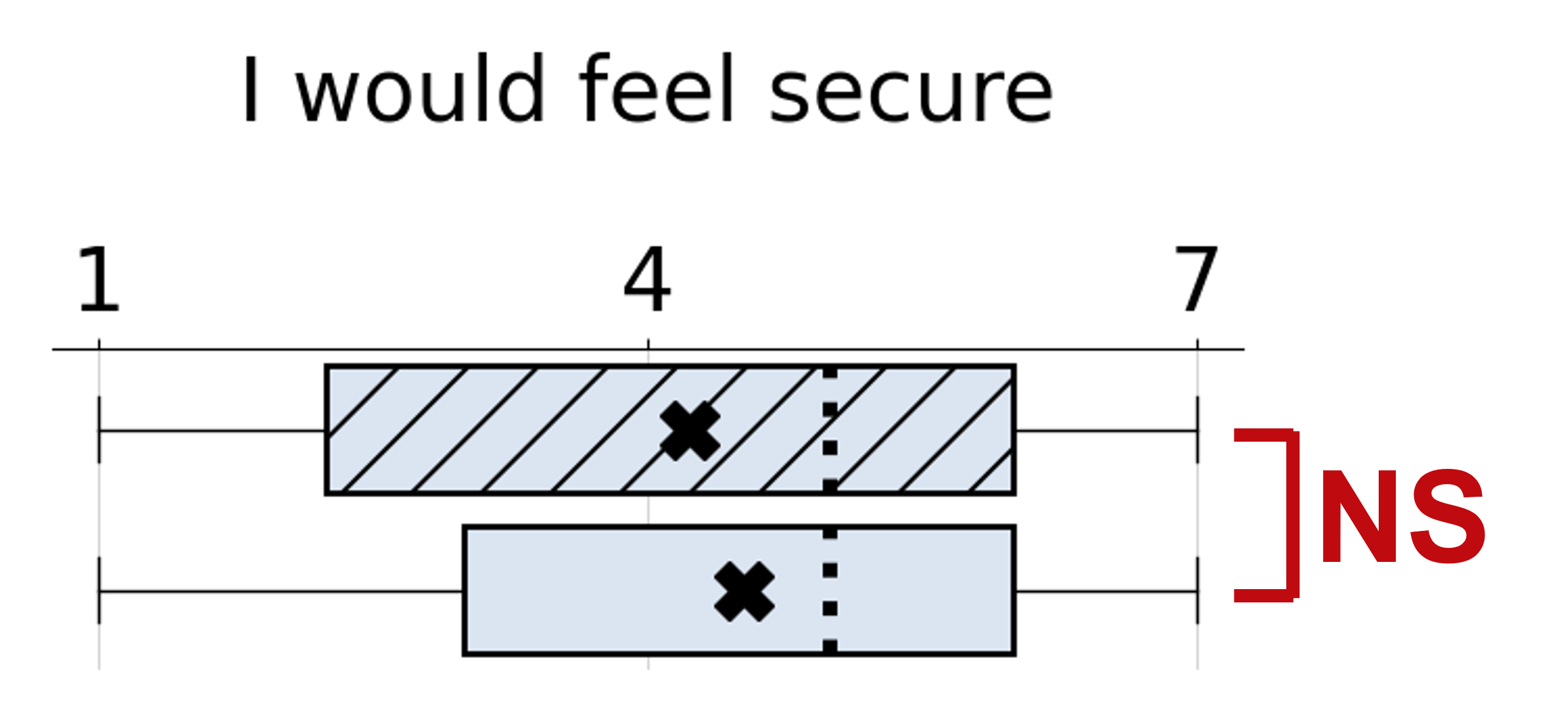} 
         }
         \\

%
%

    \end{tabular}
    \caption{Inclusivity unchanged: Examples from the 44 instances in which HAI-UX inclusivity did not significantly change for either the risk-averse or the risk-tolerant participants.
    }
    \label{tab:no-inclusivity-change-general}
\end{table}

%% file: tables/04-Inclusivity-Only-One.tex
\begin{table}[h]
    \centering
    \begin{tabular}{p{0.25\linewidth}|c|c}
         Guideline&  Risk-Averse Inclusivity & Risk-Tolerant Inclusivity\\
         \toprule
         \parbox[c][]{\linewidth}{3: Time services based on context} 
         &
         \parbox[c]{0.25\linewidth}{
         \includegraphics[width = \linewidth]{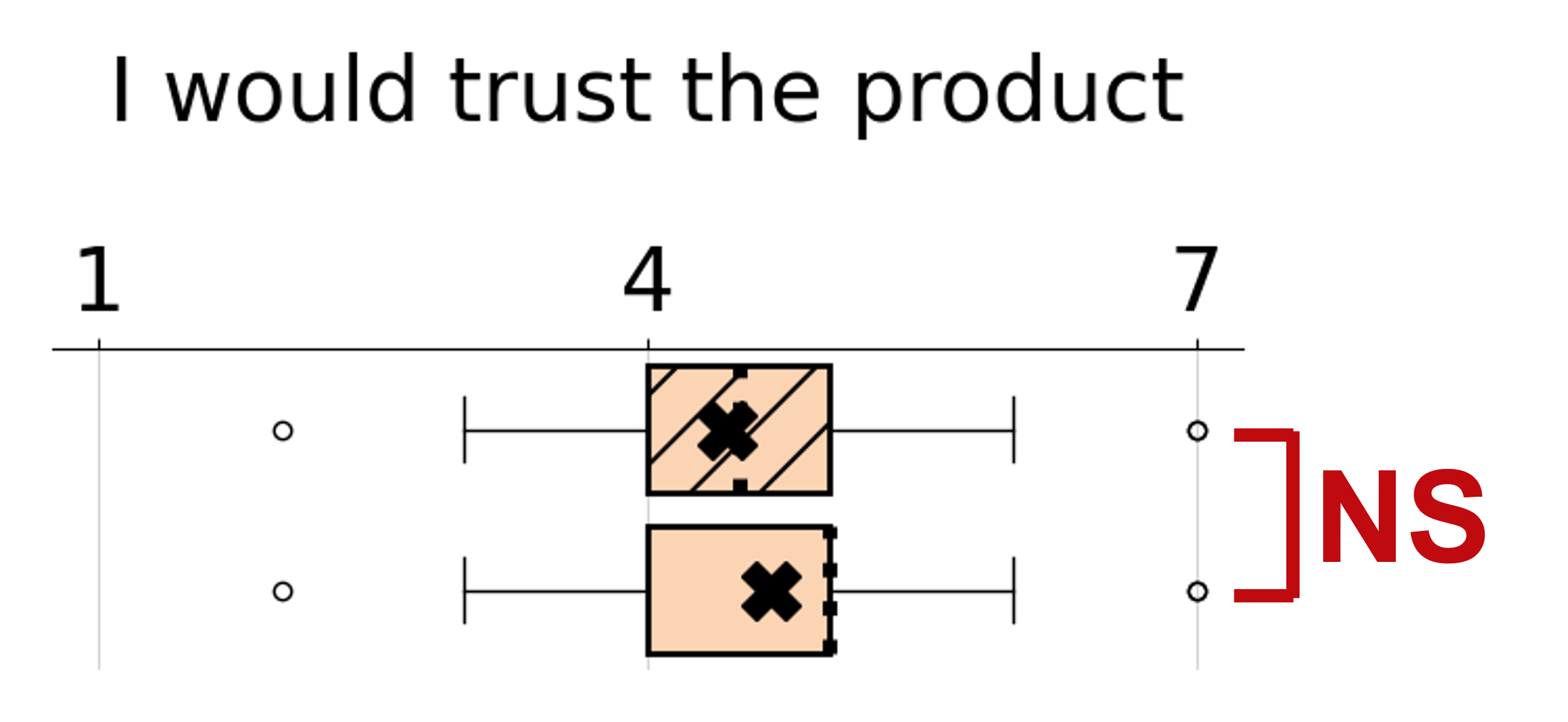}}

         & 
         \parbox[c]{0.25\linewidth}{
        \includegraphics[width = \linewidth]{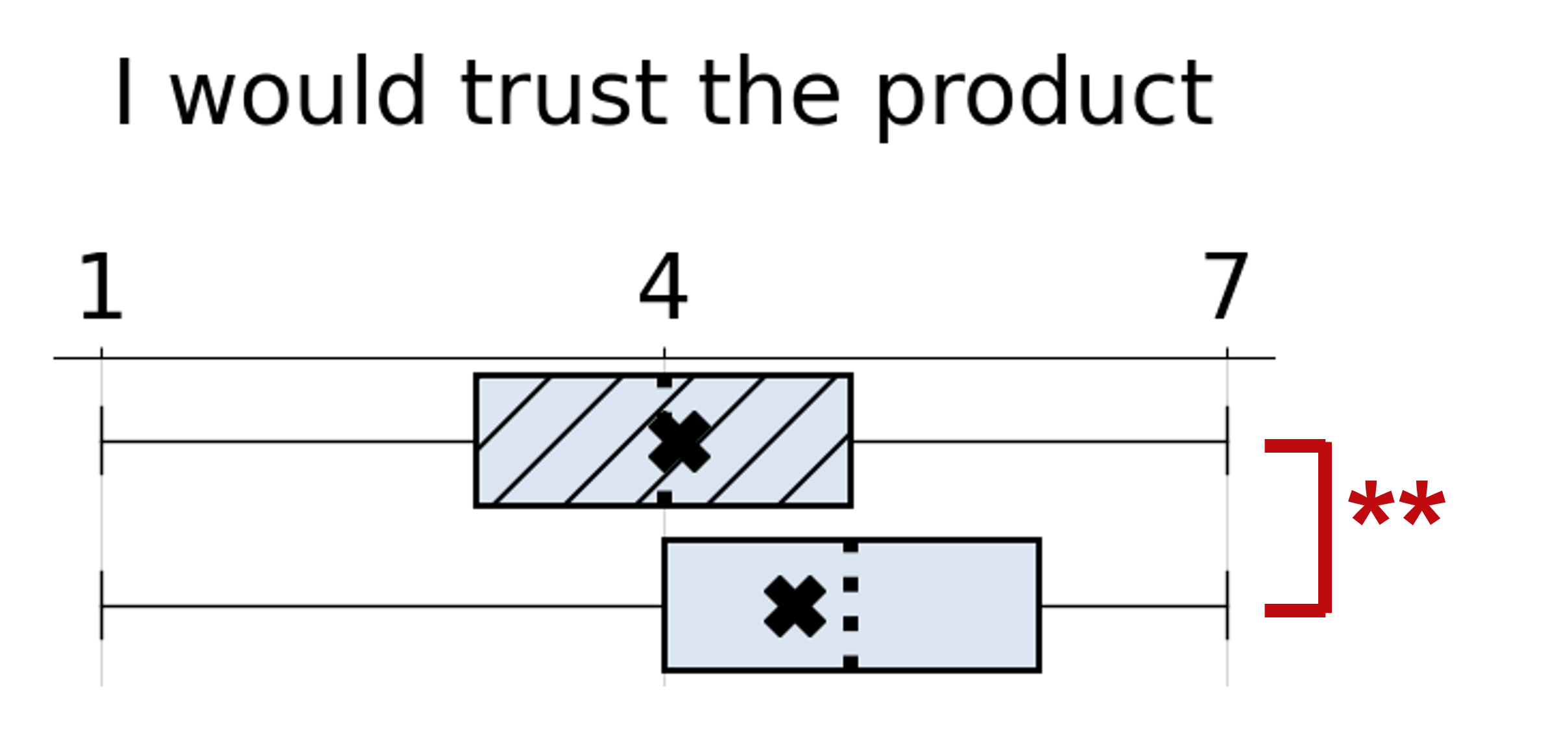}}
         \\
                  
%
%
         \midrule
         \parbox[c]{\linewidth}{14: Update and adapt cautiously}
         &
         \parbox[c]{0.25\linewidth}{
         \includegraphics[width=\linewidth]{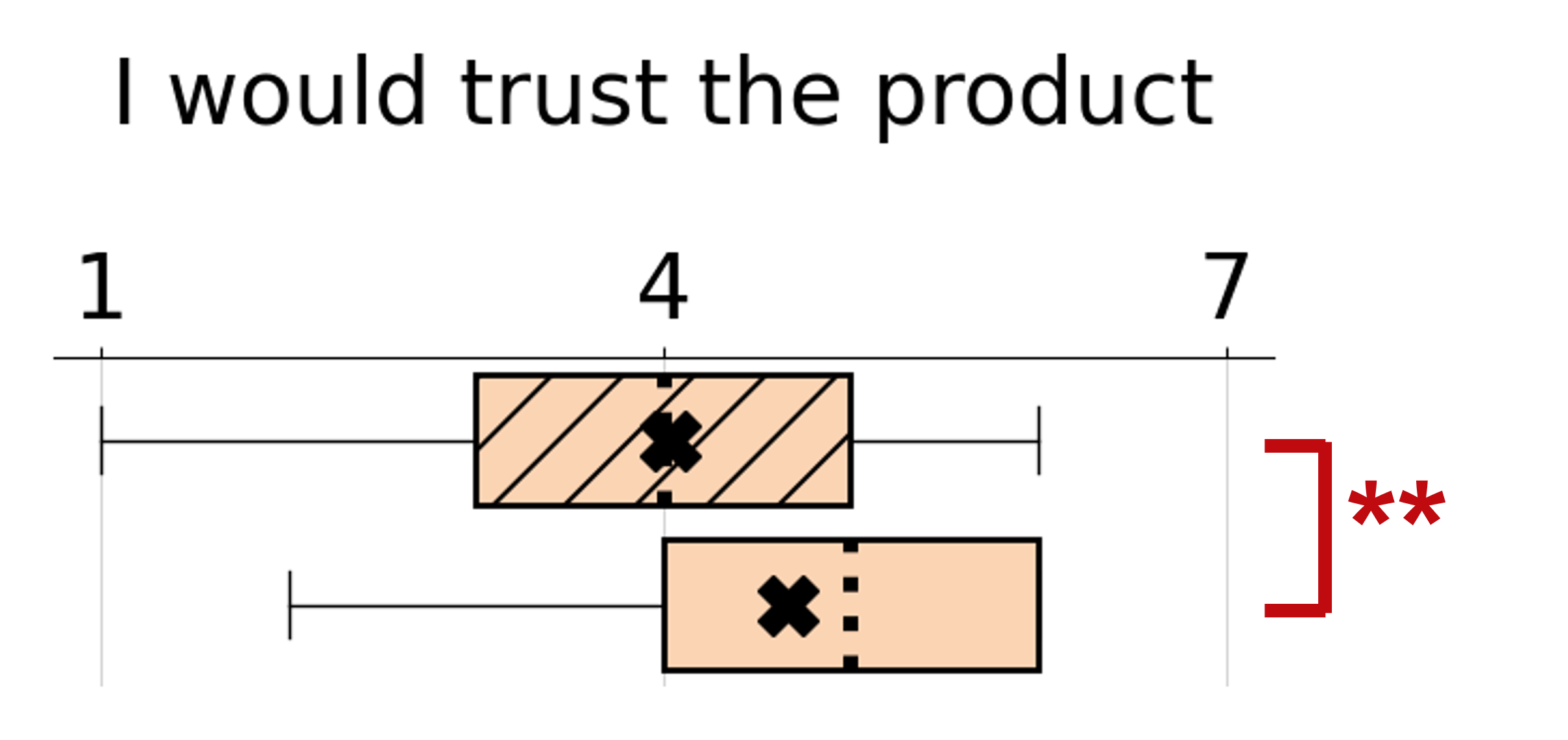}
         }
         & 
         \parbox[c]{0.25\linewidth}{
         \includegraphics[width=\linewidth]{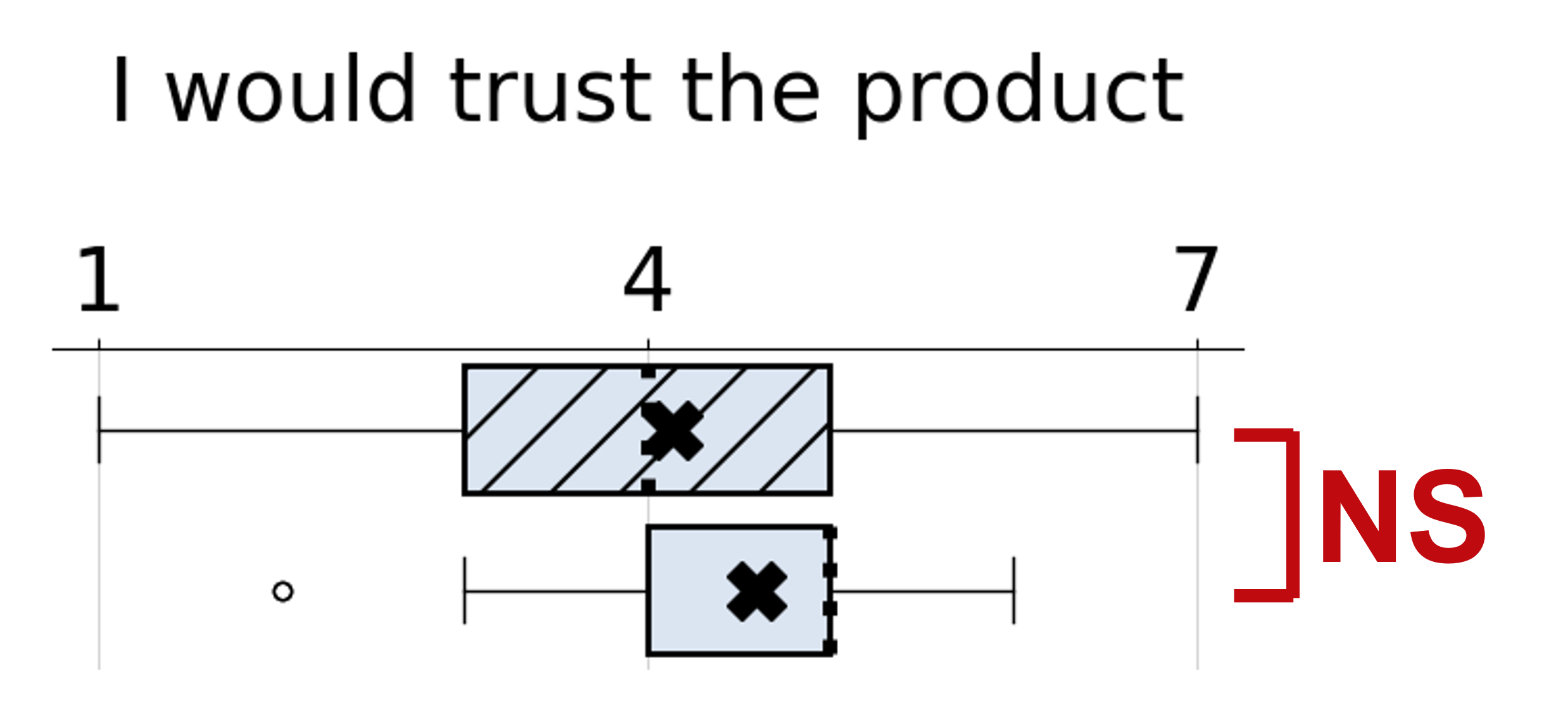} 
         }
         \\
         

         \midrule
         \parbox[c]{\linewidth}{18: Notify users about change}
         &
         \parbox[c]{0.25\linewidth}{
         \includegraphics[width = \linewidth]{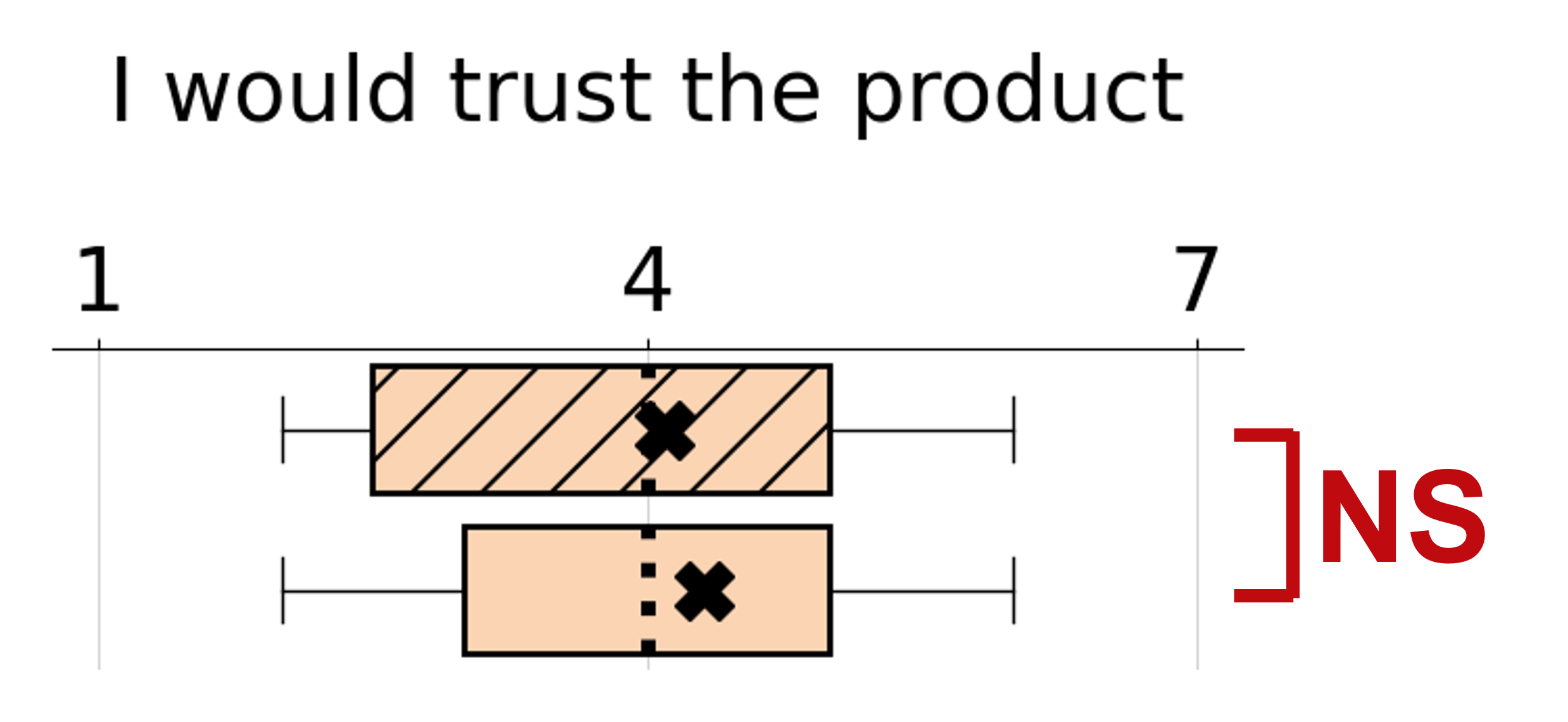}
         }
         & 
         \parbox[c]{0.25\linewidth}{
         \includegraphics[width = \linewidth]{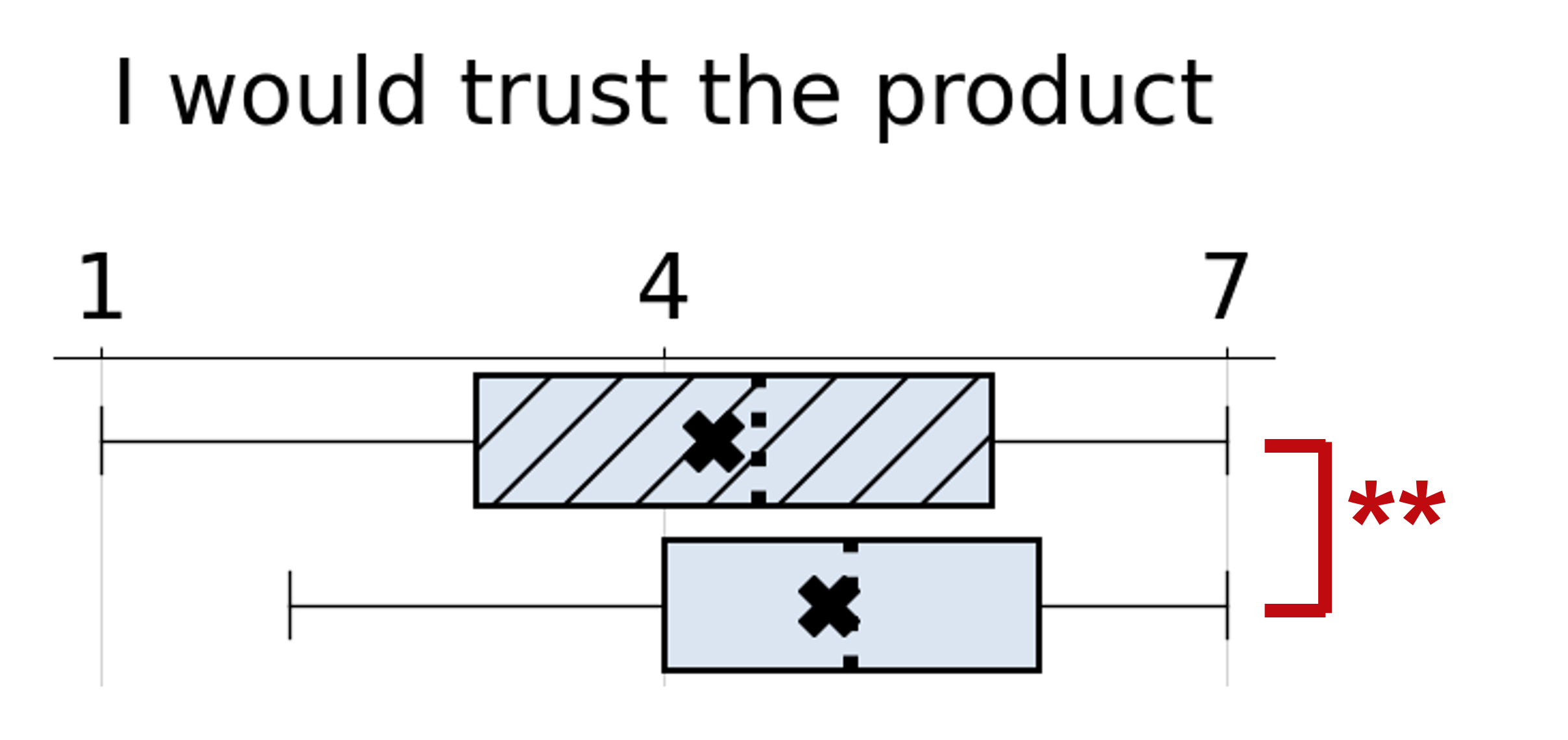} 
         }
         \\

    \end{tabular}
    \caption{Example inclusivity gains by the \colorboxBackgroundForegroundText{AbiOrangeQuote}{black}{risk-averse} only (e.g., the Guideline 14 experiment) or by the \colorboxBackgroundForegroundText{TimBlueQuote}{black}{risk-tolerant} only (e.g., the Guideline 3 and 18 experiments).
    }
    \label{tab:inclusivity-only-one}
\end{table}

%% file: figure/Li-resultThumbnails-forResults.tex
\begin{figure}[h]
    \centering
    \includegraphics[width=0.33\columnwidth]{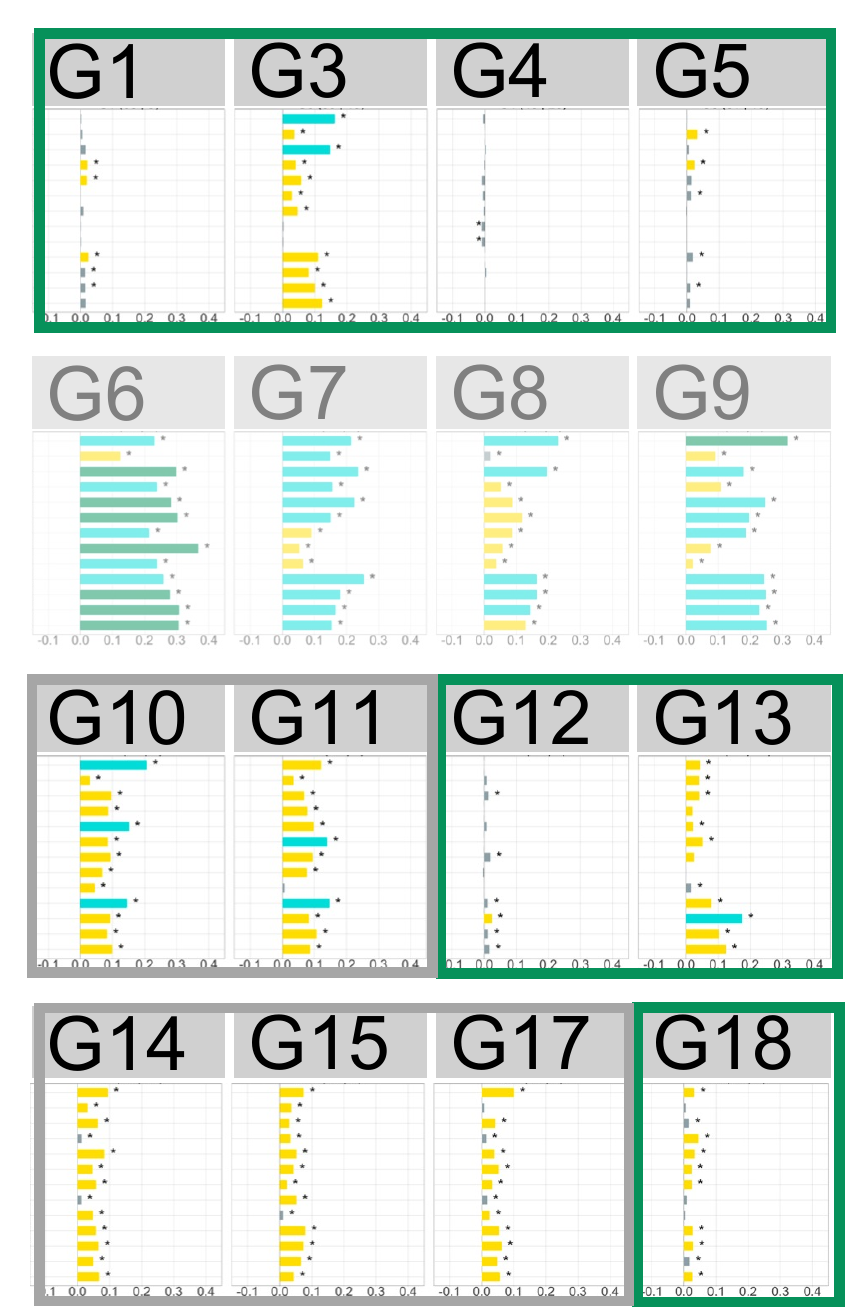}
    \caption{Revisiting Investigation One's results, annotated with Risk implications.  Green boxes: Risk results point to ways to improve those AI products.  Gray boxes: AI products we might like to improve, but Risk results were ``too inclusive'' to help. Unboxed AI products were already very well-received. }
    \label{fig:Li-resultThumbnails-forResults}
\end{figure}

%% file: doc/04-Results-Everything-Else.tex
\section{Results: Beyond risk---the other four problem-solving styles
}
\label{sec:beyond-risk}


\FIXME{\textbf{R4-4}}

Section~\ref{sec:Results-Insights} considered only one type of problem-solving diversity, namely participants' diverse attitudes toward risk.
\FIXED{We now turn to \textbf{RQ2-AllStyles}, which asks ``How inclusive are such products to users with diverse values of GenderMag's other four problem-solving styles?''
Although space does not permit an in-depth analysis of each remaining problem-solving style---motivations, learning style, computer self-efficacy, and information processing style---we summarize in this section whether and how the Risk results of Section~\ref{sec:Results-Insights} generalize to analogous results. }
If they do, we also consider whether those new results add anything to our understanding of the user experiences the AI products offer to diverse problem-solvers. 
Full analyses for all of these styles are in the Appendices.


\boldify{Recall from Table~\ref{table:personas} that someone is ``Abi''-like if they are the following... and ``Tim''-like if they are the following.}

\FIXME{\textbf{R4-3}}
\FIXED{
Recall from Table~\ref{table:personas} that GenderMag uses two personas, ``Abi'' and ``Tim'', to identify the distinguished endpoints of each of GenderMag's five problem-solving style types.
}
As per the table's definitions, we classify participants as more \colorboxBackgroundForegroundText{AbiOrangeQuote}{black}{``Abi''}-like if they had any of the following problem-solving style values: more \colorboxBackgroundForegroundText{AbiOrangeQuote}{black}{risk-averse}, \colorboxBackgroundForegroundText{AbiOrangeQuote}{black}{lower} computer self-efficacy, \colorboxBackgroundForegroundText{AbiOrangeQuote}{black}{task-oriented} motivations for using technology, had a \colorboxBackgroundForegroundText{AbiOrangeQuote}{black}{comprehensive} information processing style, or were a \colorboxBackgroundForegroundText{AbiOrangeQuote}{black}{process-oriented} learner.
Participants nearer the opposite endpoint of these problem-solving spectra are classified to be more \colorboxBackgroundForegroundText{TimBlueQuote}{black}{``Tim''}-like; 
i.e., more \colorboxBackgroundForegroundText{TimBlueQuote}{black}{risk-tolerant}, 
had \colorboxBackgroundForegroundText{TimBlueQuote}{black}{higher} computer self-efficacy, 
had
\colorboxBackgroundForegroundText{TimBlueQuote}{black}{tech-oriented} motivations, 
had a more \colorboxBackgroundForegroundText{TimBlueQuote}{black}{selective} information processing style,
or learned more by \colorboxBackgroundForegroundText{TimBlueQuote}{black}{tinkering}.
\FIXED{As in other persona research~\cite{adlin2010essential}, we use these persona names for two reasons: 1) to provide a vocabulary for an associated collection of traits and collection of traits (recall Section~\ref{sec:background}), and 2) Using the ``Abi'' and ``Tim'' vocabulary helps emphasize which of the ``distinguished endpoints'' of each problem-solving style type tend to co-occur, helping to keep clear which problem-solving value belongs to the underserved population.
}

\boldify{All the problem solving styles mostly showed inclusivity gains.}

As the upcoming Tables~\ref{tab:Motiv-Inclusivity-Frequency}, \ref{tab:Learn-Inclusivity-Frequency}, \ref{tab:SE-Inclusivity-Frequency},~\ref{tab:Info-Inclusivity-Frequency}, and earlier Table~\ref{tab:Risk-Inclusivity-Frequency} show, the first result of these analyses was very good news for most of the \appProduct{}s.
As was also true of risk diversity results, whenever applying an ~\appProduct{} produced a change in inclusivity, it was almost always a \textit{positive} change for at least some <problem-solving value> group of participants---without loss of inclusivity for the other group.
For example, as Table~\ref{tab:Motiv-Inclusivity-Frequency}  shows, whenever applying a guideline produced a gain for either the task-oriented group or the task-oriented group, it almost never produced an inclusivity loss for the other group, with only one exception. 

\Result{7}{\textit{Following the guidelines usually led to inclusivity gains---for \FIXED{every one of these} five problem-solving styles.} Applying these guidelines led to 115, 116, 116, and 116 inclusivity gains, respectively, for motivations-diverse, learning-style-diverse, self-efficacy-diverse, and information-processing-diverse participants; with only 1 or 2 inclusivity losses for any of these types of problem-solving diversity.}

\input{tables/04-Inclusivity-Motivations-Frequency-Table}
\input{tables/04-Inclusivity-Learn-Frequency-Table}
\input{tables/04-Inclusivity-SE-Frequency-Table}
\input{tables/04-Inclusivity-Info-Proc-Frequency-Table}


\boldify{However, the \textit{who} was different, and that's a result.}

One result from \textbf{RQ2-AllStyles} was \textit{who} were advantaged across these 16 product pairs.
Note that the GenderMag assignment of endpoints to ``Abi'' vs. ``Tim'' followed widespread statistical skews of genders toward these particular styles~\cite{burnett2016gendermag};
previous research has shown that ``Abi'' styles have statistical tendencies to cluster, and so do the ``Tim'' styles.
Thus, one might expect color patterns in risk's results (Table~\ref{tab:Risk-Inclusivity-Frequency}) to be similar to the columnar color patterns in, say, the Motivations results (Table~\ref{tab:Motiv-Inclusivity-Frequency}). 

However, this sometimes did not happen.
For example, Table~\ref{tab:Risk-Inclusivity-Frequency} visually contained twice as many \colorboxBackgroundForegroundText{TimBlueQuote}{black}{T cells} for the risk-tolerant participants as there were \colorboxBackgroundForegroundText{AbiOrangeQuote}{black}{A cells} for the risk-averse participants (27 vs. 13 respectively).
The four tables, when considered at a high level, showed that who gained more inclusivity advantages depended on which problem-solving style was considered.
For example, Table~\ref{tab:Motiv-Inclusivity-Frequency} reverses Table~\ref{tab:Risk-Inclusivity-Frequency}'s trend, with more \colorboxBackgroundForegroundText{AbiOrangeQuote}{black}{A cells} for the task-oriented participants than \colorboxBackgroundForegroundText{TimBlueQuote}{black}{T cells} for the tech-oriented.
This suggested that when these guidelines were applied, not only did the \colorboxBackgroundForegroundText{TimBlueQuote}{black}{risk-tolerant} derive more of the HAI-UX inclusivity gains than the \colorboxBackgroundForegroundText{AbiOrangeQuote}{black}{risk-averse} but also that the participants with \colorboxBackgroundForegroundText{AbiOrangeQuote}{black}{task-oriented} motivations derived more inclusivity gains than those with \colorboxBackgroundForegroundText{TimBlueQuote}{black}{tech-oriented} motivations.

\boldify{In fact, this trend of \textit{who} was most advantaged even occurred at the level of granularity of each guideline. Consider G13 for motivations.}

This trend can be found even within each of the experiments, demonstrated in Figure~\ref{tab:G13-Switch-Styles} for Guideline 13's experiment.
Guideline 13's product was a presentation app, and the AI feature was a design helper that recommended designs for alternative layouts.
When participants saw the \vioProduct{}, they were told that ``\textit{...\vioProduct{} has not learned your preferences and blue designs appear in the same place among the suggested designs as the first time you used it,}'' whereas the \appProduct{} ``\textit{...has learned your preferences and now features blue designs prominently.}'' 
Considering participants' attitudes toward risk (first column), the \colorboxBackgroundForegroundText{TimBlueQuote}{black}{risk-tolerant} participants derived the most benefit from the \appProduct{}.
However, the second column shows these participants' same data, but instead considering their motivations.
This provides a nuance to these results;
the \appProduct{} not only benefited the risk-tolerant but also those with \colorboxBackgroundForegroundText{AbiOrangeQuote}{black}{task-oriented} motivations.

\boldify{But again, the question is why did these things happened?}

For HAI practitioners, results like these suggest that different design decisions can appeal to different problem-solving styles for different reasons.
For example, 46\% (13/28) of the participants with \colorboxBackgroundForegroundText{AbiOrangeQuote}{black}{task-oriented} motivations mentioned how efficient they would become with the \appProduct{} or how much time they would save while using it:

\quotateInset{I like to have software that anticipates my needs, because it makes working more efficient. }
{}
{\colorboxBackgroundForegroundText{AbiOrangeQuote}{black}{G13-2178-task-oriented}}
{Whole quote - I like to have software that anticipates my needs, because it makes working more efficient. }

\quotateInset{It is more efficient to see designs similar to those I have used before...it will take me less time to find them.}
{}
{\colorboxBackgroundForegroundText{AbiOrangeQuote}{black}{G13-4099-task-oriented}}
{Whole quote - It is more efficient to see designs similar to those I have used before placed above those I am not interested in because it will take me less time to find them.}

\quotateInset{It [the \appProduct{}] learned my preferences quicker which in time will save me time and trouble.}
{}
{\colorboxBackgroundForegroundText{AbiOrangeQuote}{black}{G13-2740-task-oriented}}
{Whole quote - It learned my preferences quicker which in time will save me time and trouble}

\input{tables/04-Inclusivity-Switch-Styles}

Although the participants with \colorboxBackgroundForegroundText{TimBlueQuote}{black}{tech-oriented} ``Tim''-like motivations also raised efficiency and time savings, they did so less frequently than their \colorboxBackgroundForegroundText{AbiOrangeQuote}{black}{task-oriented} peers (only 17\%---5/29):

\quotateInset{because it saves time than starting from scratch every time I use it [the \appProduct{}].}
{}
{\colorboxBackgroundForegroundText{TimBlueQuote}{black}{G13-662-tech-oriented}}
{Whole quote - because it saves time than starting from the scratch every time i use it.}

\quotateInset{I prefer [\appProduct{}] because of its ability to learn my preferences...thus helping me to work more efficiently.}
{}
{\colorboxBackgroundForegroundText{TimBlueQuote}{black}{G13-662-tech-oriented}}
{Whole quote - I prefer Lone because of its ability to learn my preferences and showing them to me in a prominent way, thus helping me to work more efficiently.}

Comments like these also have ties to the research literature.
In that body of work, people who are more task-oriented prefer to use technologies to accomplish their task, using methods they are already familiar and comfortable with.
Task-oriented people do so in an attempt to focus on the tasks that they care about, which might explain why these task-oriented participants commented so frequently on how the \appProduct{} saved them time;
if the product saves them time, then the task-oriented participants could achieve their task more quickly, devoting more time to what they care about rather than having to spend additional time recreating designs.

\boldify{All-in-all, this leads to a \textit{big} result of this section---the considerations of the HAI practitioners will change based on which problem-solving style they are designing for.}

Section~\ref{sec:beyond-risk} highlighted how providing ways for an AI feature in productivity software to save people time significantly aided those with task-oriented motivations.
This suggests for HAI practitioners that in order to create a more inclusive user experience in human-AI interaction for those with diverse problem-solving styles, one of the best approaches is to consider all five of GenderMag's problem-solving styles.

\FIXME{\textbf{R4-2.5}}

\Result{8}{\FIXED{\textit{The union of these five styles' results revealed more about who was left out---and why---than any one styles' results alone could do.
}
G03, G13 (Figure~\ref{tab:G13-Switch-Styles}), and G18 are cases in point, illustrated through the change in the colors of the cells between \colorboxBackgroundForegroundText{AbiOrangeQuote}{black}{``Abi''} and \colorboxBackgroundForegroundText{TimBlueQuote}{black}{``Tim''} across these five problem-solving styles.
} 
}

%% file: tables/04-Inclusivity-Motivations-Frequency-Table.tex
\begin{table}[h]
    \centering
    \begin{tabular}{c}
         Motivations\\
        \hline
        \includegraphics[width = 0.82\linewidth]{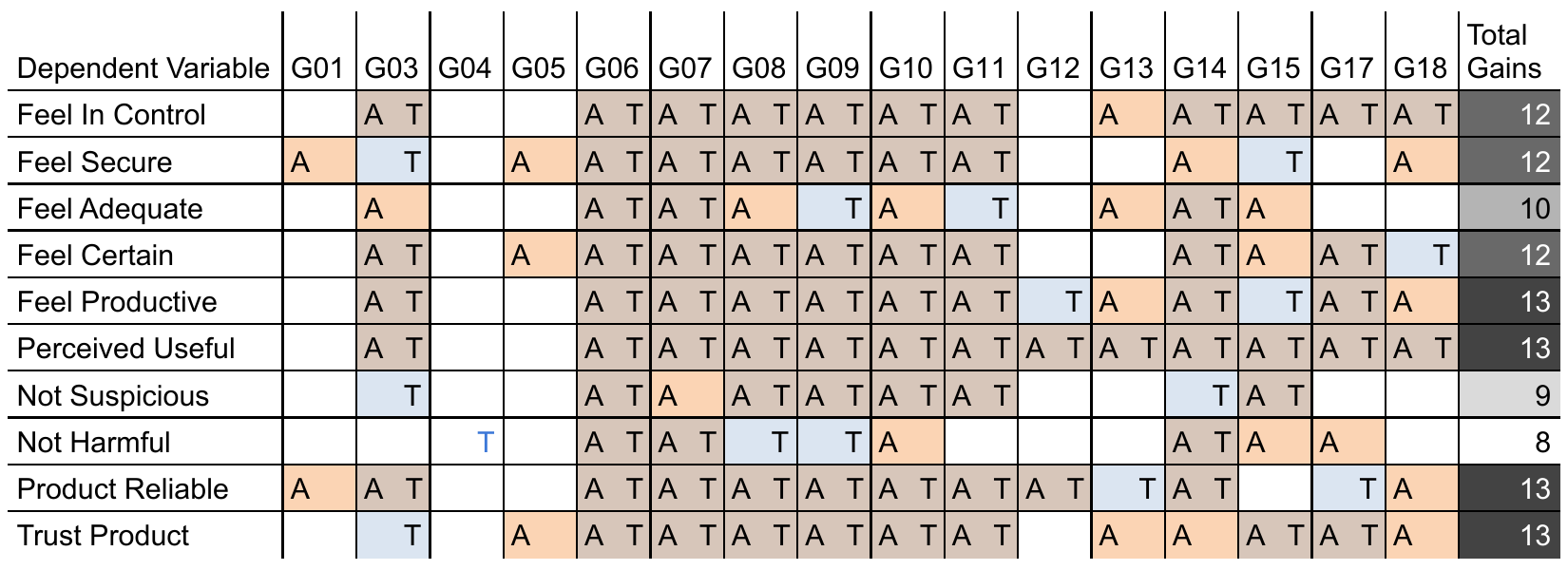} 
    \end{tabular}
    \caption{Motivations results summary, by guideline experiment (columns) and dependent variable (rows), for those with ``\underline{A}bi''-like task-oriented and  ``\underline{T}im''-like tech-oriented motivations.
    Total occurrences: 77 inclusivity gains for \colorboxBackgroundForegroundText{Eggshell}{black}{both motivations (A T)};
    44 without inclusivity gains/losses (blank); 24 inclusivity gains for  \colorboxBackgroundForegroundText{AbiOrangeQuote}{black}{task-oriented only (A)};
     14 inclusivity gains for  \colorboxBackgroundForegroundText{TimBlueQuote}{black}{ tech-oriented only (T)};
     1 inclusivity loss for \textcolor{NegativeTim}{tech-oriented only (T)}. 
     Total possible: 160.
    }
    \label{tab:Motiv-Inclusivity-Frequency}
\end{table}

%% file: tables/04-Inclusivity-Learn-Frequency-Table.tex
\begin{table}[h]
    \centering
    \begin{tabular}{c}
         Learning Style\\
        \hline
        \includegraphics[width = 0.82\linewidth]{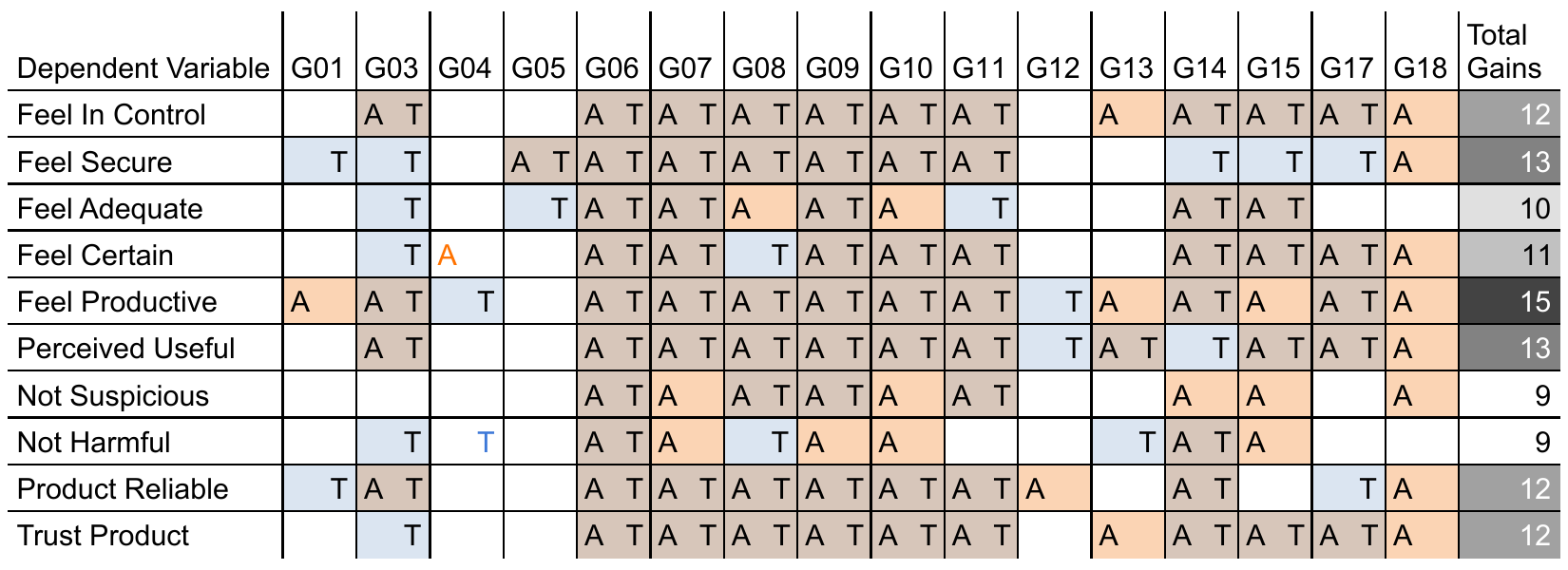} 
    \end{tabular}
    \caption{Learning style results summary, by  guideline experiment (columns) and dependent variable (rows), for the ``\underline{A}bi''-like process-oriented and  ``\underline{T}im''-like tinkering-oriented learners.
    Total occurrences: 72 inclusivity gains for \colorboxBackgroundForegroundText{Eggshell}{black}{both learning styles (A T)};
    42 without inclusivity gains/losses (blank); 13 inclusivity gains for  \colorboxBackgroundForegroundText{AbiOrangeQuote}{black}{process-oriented only (A)};
     27 inclusivity gains for  \colorboxBackgroundForegroundText{TimBlueQuote}{black}{ tinkering-oriented only (T)};
     1 inclusivity loss for \textcolor{NegativeAbi} {process-oriented only (A)}.
     1 inclusivity loss for \textcolor{NegativeTim}{tinkering-oriented only (T)}. 
     Total possible: 160.
    }
    \label{tab:Learn-Inclusivity-Frequency}
\end{table}

%% file: tables/04-Inclusivity-SE-Frequency-Table.tex
\begin{table}[h]
    \centering
    \begin{tabular}{c}
         Computer Self-Efficacy\\
        \hline
        \includegraphics[width = 0.82\linewidth]{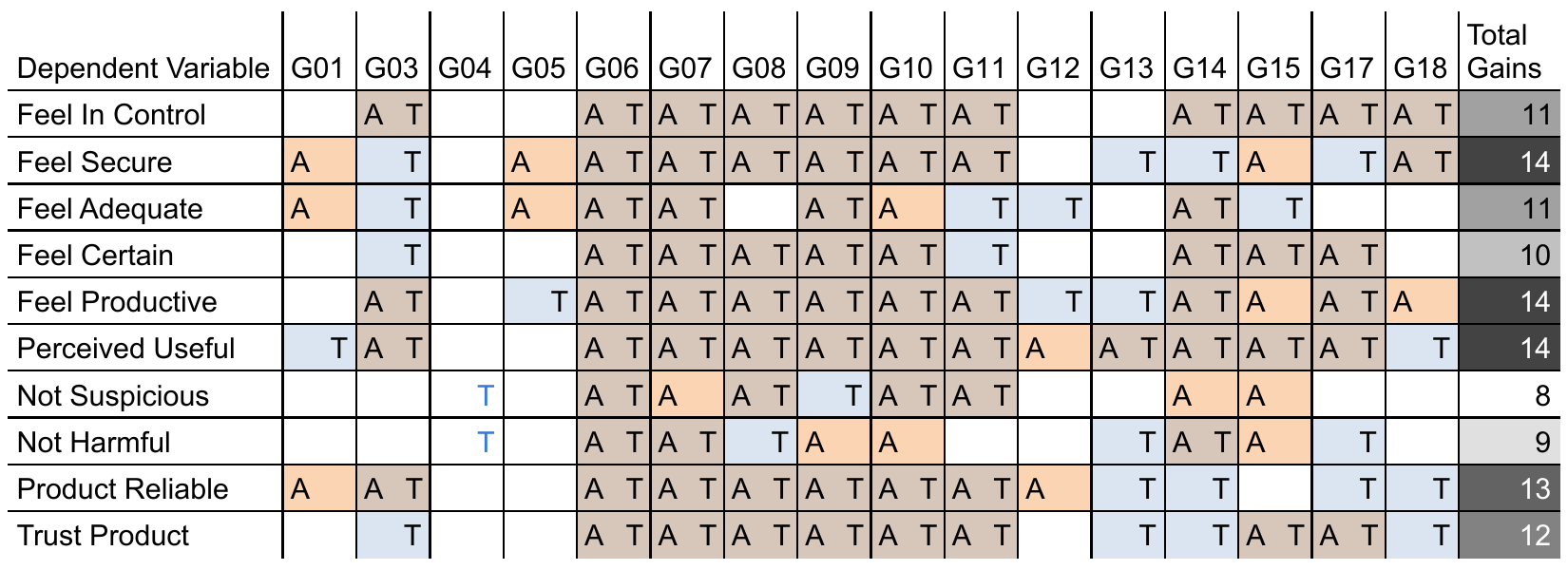} 
    \end{tabular}
    \caption{Computer self-efficacy results summary, by guideline experiment (columns) and dependent variable (rows), for those with ``\underline{A}bi''-like lower and ``\underline{T}im''-like higher computer self-efficacy.
    Total occurrences: 72 inclusivity gains for \colorboxBackgroundForegroundText{Eggshell}{black}{both (A T)};
    42 without inclusivity gains/losses (blank); 
    17 inclusivity gains for  \colorboxBackgroundForegroundText{AbiOrangeQuote}{black}{lower only (A)};
     27 inclusivity gains for  \colorboxBackgroundForegroundText{TimBlueQuote}{black}{ higher only (T)};
     2 inclusivity losses for \textcolor{NegativeTim}{higher only (T)}. 
     Total possible: 160.
    }
    \label{tab:SE-Inclusivity-Frequency}
\end{table}

%% file: tables/04-Inclusivity-Info-Proc-Frequency-Table.tex
\begin{table}[h]
    \centering
    \begin{tabular}{c}
         Information Processing Style\\
        \hline
        \includegraphics[width = 0.82\linewidth]{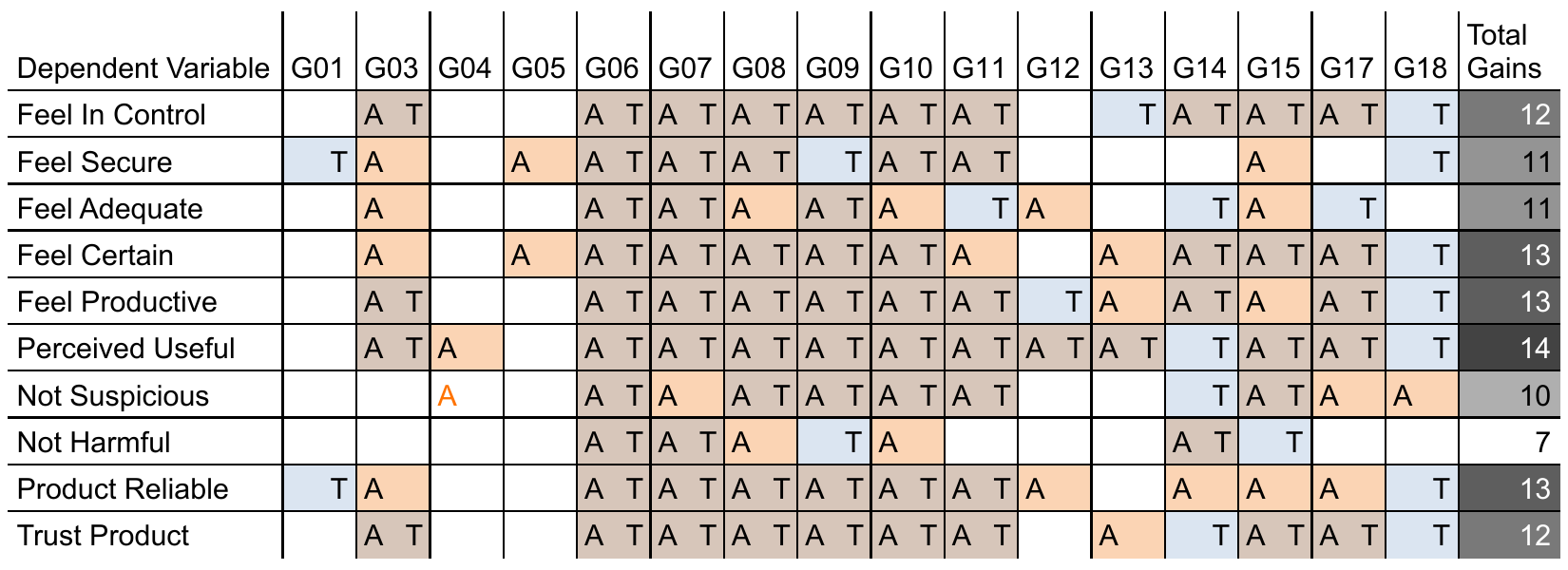} 
    \end{tabular}
    \caption{Information processing style results summary, by  guideline experiment (columns) and dependent variable (rows), for ``\underline{A}bi''-like comprehensive and  ``\underline{T}im''-like selective information processors.
    Total occurrences: 70 inclusivity gains for \colorboxBackgroundForegroundText{Eggshell}{black}{both information processing styles (A T)};
    43 without inclusivity gains/losses (blank); 
    26 inclusivity gains for  \colorboxBackgroundForegroundText{AbiOrangeQuote}{black}{comprehensive only (A)};
     20 inclusivity gains for  \colorboxBackgroundForegroundText{TimBlueQuote}{black}{ selective only (T)};
     1 inclusivity loss for \textcolor{NegativeAbi}{comprehensive only (A)}.
     Total possible: 160.
    }
    \label{tab:Info-Inclusivity-Frequency}
\end{table}

%% file: tables/04-Inclusivity-Switch-Styles.tex
\begin{wrapfigure}{o}{0.4\linewidth}
        \centering
        \begin{tabular}{c}
            \includegraphics[width = \linewidth]{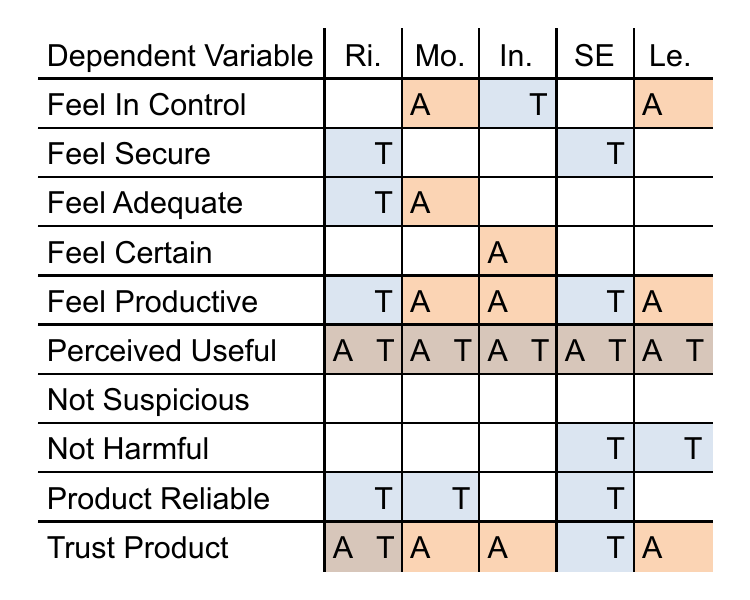} 
        \end{tabular}
        \caption{Guideline 13's inclusivity gains (rows) across the five different problem-solving styles (columns).
        Notice how the values switches between the ``Tim''-like (T) and ``Abi''-like (A) participants, depending on the problem-solving style.
        Ri. = Risk, Mo. = Motivations, In. = Information Processing Style, SE = Self-Efficacy, and Le. = Learning Style.
        }
        \label{tab:G13-Switch-Styles}
        \vspace{-10pt}
\end{wrapfigure}

%% file: doc/06-Gender.tex
\section{Participants' problem-solving styles and their demographics \draftStatus{top is D2.4}
}
\label{sec:gender}

\FIXED{Some HAI research has suggested demographic differences in different AI systems' HAI usability (e.g.,~\cite{zwakman2021usability, kulesza2011bayes, flandorfer2012population}).
Here, we consider whether problem-solving style results like those in Section~\ref{sec:Results-Insights} and Section~\ref{sec:beyond-risk} can shed light on \textit{why} such demographic  differences exist.
}

\boldify{RQ3 seeks to understand the alignment between participants problem-solving diversity and demographic diversity, and we should care about this because ....}

\textbf{RQ3-DemographicDiversity} seeks to understand how participants' problem-solving diversity aligned with their demographic diversity.
The answer to this question will show whether problem-solving disparities in HAI user experiences can help explain demographic disparities in HAI user experiences.

For example, consider the Guideline 18 outcome variable of ``certainty'' and the two genders for whom enough data are present for inferential statistics---women and men.
A statistical peek at the G18 data by gender reveals that the men's inclusivity significantly increased with G18's \appProduct{} over the \vioProduct{} \tTestResult{28}{3.1777}{.004}{590}, whereas the women's did not \tTestResult{34}{1.0359}{.308}{.175}.
This gender disparity seems problematic, but knowing its presence does not suggest a solution%
\footnote{One of the authors of this paper is reminded of the many times she has heard software practitioners say things like, ``what am I supposed to do, paint it pink?''}.



%



\subsection{Problem-solving style diversity, gender, and age \draftStatus{MMBd2.4}
}

If the gender results for the G18 example above show alignment with, for example, the G18 risk results of Section~\ref{subsec:inclusivity-risk}, the risk-oriented solution ideas from that section might help remove the gender disparity.
And indeed, these two results do align: risk analysis showed that G18's \appProduct{}, which added user control to the AI product, did not provide significant inclusivity gains for the \colorboxBackgroundForegroundText{AbiOrangeQuote}{black}{risk-averse} certainty outcome \tTestResult{31}{1.7261}{.094}{.256} but did for the \colorboxBackgroundForegroundText{TimBlueQuote}{black}{risk-tolerant} \tTestResult{34}{2.1884}{.036}{.370}.

Our investigation into RQ3 will enable leveraging this kind of alignment.
For example, if we find that the women participants skewed toward risk aversion, that knowledge would suggest that improving the G18 \appProduct's inclusivity across the risk spectrum could also improve its inclusivity across the gender spectrum.  

Thus, to find out how our participants' problem-solving styles aligned with their genders, 
we counted the number of \colorboxBackgroundForegroundText{AbiOrangeQuote}{black}{``Abi''-like} and \colorboxBackgroundForegroundText{TimBlueQuote}{black}{``Tim''-like} styles of each participant of all 16 experiments, and then compared the counts by gender.
We begin with the two genders for whom enough data are present for inferential statistics---women and men, who provided 98.7\% of the data---and then non-statistically present the data for the participants in the LGBTQIA* community\footnote{LGBTQIA* used based on Scheuerman et al.'s living document~\cite{scheuerman2020hci}.}. 


As Figure~\ref{fig:anderson-facet-gender}(left) shows, the women were split almost equally between having three or more ``Abi''-like styles (first three orange bars, 50.6\%), versus having two or fewer (49.4\%).
For example, the leftmost pair of bars show that 59 women and 24 men had five Abi-like problem-solving style values (0 Tim-like styles).
In contrast, the men skewed heavily toward the right;
only 34.5\% of the men had three or more ``Abi''-like styles (first three blue bars).
As Figure~\ref{fig:anderson-facet-gender}(right) shows, these gender skew differences were statistically significant under Fisher's exact test 
($p < .0001$).%
\footnote{For this test, we used the threshold that \textit{minimized} the chance of showing significance by maximizing the sum of \textit{p}-values~\cite{ramsey2012statistical}.
} 
Vorvoreanu et al.~\cite{vorvoreanu2019gender} found similar gender skew results while investigating an academic search tool.

\boldify{We also analyzed intersectionally}

Adding age demographics into our analysis, an intersectional gender-age analysis showed analogous gender skews in each of the five age groups in our data (Figure~\ref{fig:anderson-facet-gender-proportions}). The results were significant in the three age groups between ages 25---54.

\input{figure/02-Results-Gender-Facets/01-Results-Facets-By-Gender}

\input{figure/01-Gender-Insight/03-Age-Facets-Dist}

We also analyzed the presence of such gender-age intersectional results within each problem-solving style type.
As Figure~\ref{fig:all-facets} suggests, the gender differences did manifest by age in four of the five style types.

\input{figure/6-risk-se-info-trends}

For the four styles shown in Figure~\ref{fig:all-facets}, the gender-by-age differences in these problem-solving attributes are consistent with other gender- and/or age-difference reports in the literature (e.g., \cite{chang2014specialization, dohmen2017risk, dutta2018modeling, flandorfer2012population,  liberatore2022gender,  gunbatar2018gender, lopez2017teacher, schreder2013age}. 
For information processing style, although our participants did not show these demographic differences, others' research has shown both gender differences~\cite{meyers2015infoProcessing,showkat2018identifying} and age differences~\cite{geerligs2018age,guest2015aging,mcintosh2021evaluating,torrens2020lacking}.
Such demographic differences in problem-solving style by gender and by age may help explain demographic differences between people's experiences with AI products (e.g.,~\cite{hulse2018perceptions,gish2017driving,van2019exploring}).

\Result{9}{\textit{Problem-solving styles and gender/age were related.} Participants' problem-solving styles clustered by both gender and age. An implication of this result is that inclusivity gains for certain problem-solving styles, as per the results in Section~\ref{sec:Results-Insights}, should also translate into inclusivity gains for certain genders and/or age groups.}





\subsection{The LGBTQIA* Community \draftStatus{2.4}
}

\boldify{Although most identified as M or W, here's the LGBTQIA* members' problem-solving styles.}

The genders ``woman'' and ``man'' are only two points on the gender spectrum.
Table~\ref{tab:non-binary-results} reports the GenderMag problem-solving style values for the 13 participants who were members of the  LGBTQIA* community.
Although a data set of 13 participants is small, we hope it will add to literature being populated by other researchers with data sets of LGBTQIA* participants (e.g.,~\cite{freeman2022re,acena2021my,hardy2019participatory}), to enable the possibility of future meta-analyses to broaden our understanding of how to inclusively design for users of all gender identities.  
\input{figure/02-Results-Gender-Facets/Results-Non-Binary}

%% file: figure/02-Results-Gender-Facets/01-Results-Facets-By-Gender.tex
\begin{figure}[t]
    \centering
    \begin{subfigure}[h]{0.48\linewidth}
        \includegraphics[width=0.95\columnwidth]{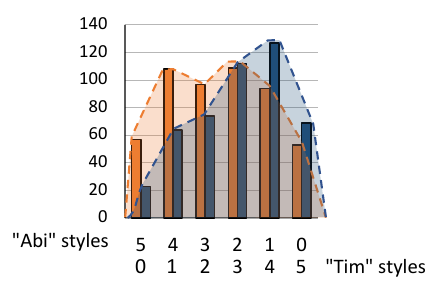}
    \end{subfigure}
    \begin{subtable}[b]{0.48\linewidth}
        \centering
        \footnotesize
        \begin{tabular}{c||cc}
          & 3 or more           &   less than 3 \\
          & ``Abi''-like styles & ``Abi''-like styles \\
         \toprule
         \cellcolor{WomanOrange}{\color{black}Woman}  & \cellcolor{white}277 & 249\\
         
         \cellcolor{ManBlue}{\color{white}Man} & 173 & 302 \\
         \bottomrule
         \cellcolor{white}Estimate (\textit{k}) & \multicolumn{2}{c}{236}\\
         \midrule
         \cellcolor{white}\textit{p}-value & \multicolumn{2}{c}{$< .0001$***}\\
    \end{tabular}
    \end{subtable}  
    \caption{(Left): Counts of women and men (y-axis) in all 16 experiments by the number of Abi-direction or Tim-direction problem-solving styles each participant reported (x-axis). 
    The \colorboxBackgroundForegroundText{ManBlue}{white}{men (blue)} skewed more to the right (i.e., more ``Tim'' styles) than the \colorboxBackgroundForegroundText{WomanOrange}{black}{women (orange)} did.
    (Right): The Fisher's exact test 2x2 contingency table, revealing that the difference was highly significant.
    }
    \label{fig:anderson-facet-gender}
    \Description{FIXME}
\end{figure}

%% file: figure/01-Gender-Insight/03-Age-Facets-Dist.tex
\begin{figure}[b]
    \centering
    \begin{subfigure}[h]{\linewidth}
        \includegraphics[width=\columnwidth]{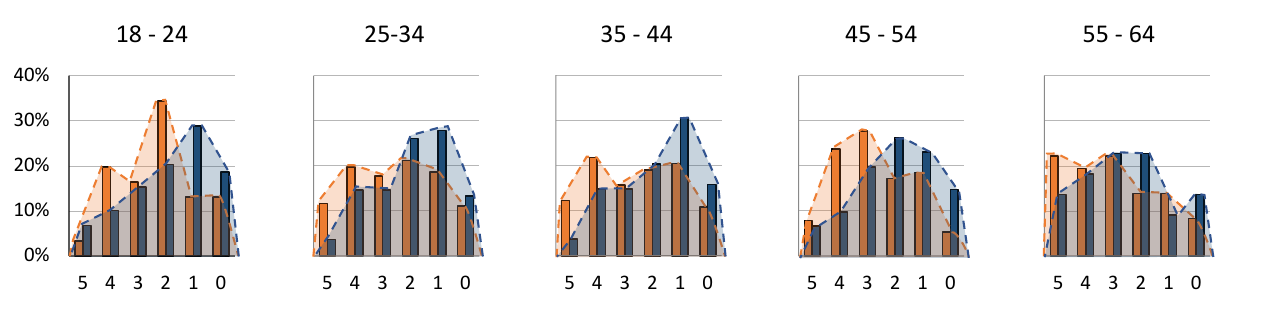}
    \end{subfigure}

    \begin{subtable}[b]{\columnwidth}
    \centering
        \footnotesize
        \begin{tabular}{c||cc|cc|cc|cc|cc}
         & \multicolumn{2}{c|}{18 -- 24} & \multicolumn{2}{c|}{25 -- 34} & \multicolumn{2}{c}{35 -- 44} & \multicolumn{2}{c}{45 -- 54} & \multicolumn{2}{c}{55 -- 64}\\
         
          & $\geq 3$  & $< 3$ & $\geq 3$  & $< 3$ & $\geq 3$  & $< 3$ & $\geq 3$  & $< 3$ & $\geq 3$  & $< 3$ \\
         \toprule
         \cellcolor{WomanOrange}{\color{black}Woman}  & 25 & 36 &  101 & 97 &  77 & 70 & 47 & 29 & 23 & 13 \\
         \cellcolor{ManBlue}{\color{white}Man}  & 19 & 40 & 77 & 142 & 37 & 71 & 23  & 38  & 12 & 10\\
         \midrule
         \cellcolor{white} Estimate (\textit{k}) & \multicolumn{2}{c|}{22} & \multicolumn{2}{c|}{85} & \multicolumn{2}{c|}{66} & \multicolumn{2}{c|}{39} & \multicolumn{2}{c}{22}\\
         \midrule
         \textit{p}-value & \multicolumn{2}{c|}{\cellcolor{white}.3482} & \multicolumn{2}{c|}{.0015**} & \multicolumn{2}{c|}{.0050**} & \multicolumn{2}{c|}{.0061**} & \multicolumn{2}{c}{\cellcolor{white}.5831}\\

    \end{tabular}
    \end{subtable}
        \caption{(Top): Percentage of participants (y-axes) from Figure~\ref{fig:anderson-facet-gender}, divided into age groups.
        \colorboxBackgroundForegroundText{ManBlue}{white}{Men} in all age groups visually skewed towards having fewer ``Abi''-like styles (x-axes) than the \colorboxBackgroundForegroundText{WomanOrange}{black}{women} did.
        (Bottom): The Fisher’s exact test 2x2 contingency tables. The middle three categories had significant gender differences.}
       \label{fig:anderson-facet-gender-proportions}
            \Description{FIXME}

\end{figure}

%% file: figure/6-risk-se-info-trends.tex
\begin{figure}
 \centering
    \begin{subfigure}[t]{0.45\linewidth}
        \centering
        \includegraphics[width = .75\linewidth]{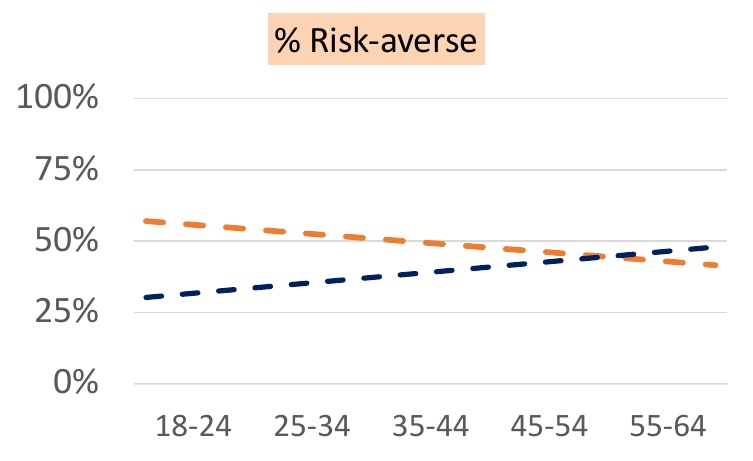}
        \label{fig:risk-by-age}
    \end{subfigure}
    \begin{subfigure}[t]{0.45\linewidth}
        \centering
        \includegraphics[width = .75\linewidth]{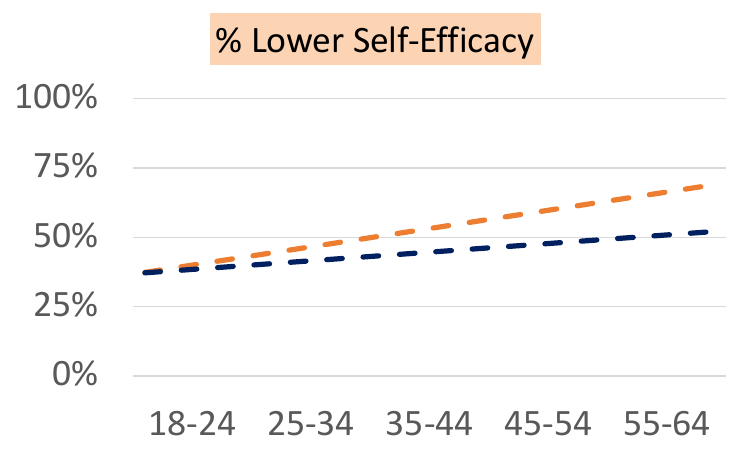}
        \label{fig:self-by-age}
    \end{subfigure}
    
    \begin{subfigure}[t]{0.45\linewidth}
        \centering
        \includegraphics[width = .75\linewidth]{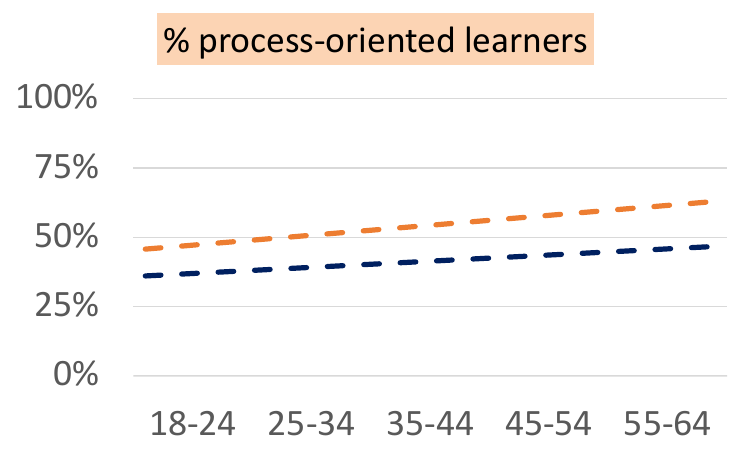}
        \label{fig:learn-by-age}
    \end{subfigure}
    \begin{subfigure}[t]{0.45\linewidth}
        \centering
        \includegraphics[width = .75\linewidth]{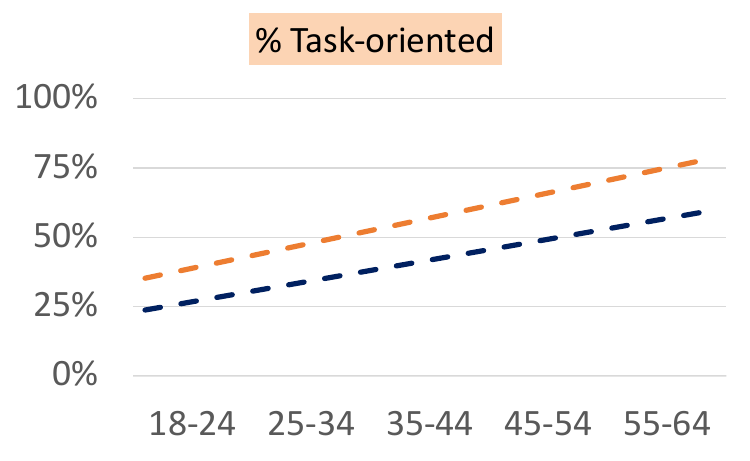}
        \label{fig:mot-by-age}
    \end{subfigure}
    \caption{The percentages (y-axes) across the five  age groups (x-axes) of \colorboxBackgroundForegroundText{WomanOrange}{black}{women (orange)} and \colorboxBackgroundForegroundText{ManBlue}{white}{men (blue)} who exhibited each of four ``Abi''-like problem solving styles---risk-aversion, lower computer self-efficacy compared to peers, process-oriented learning, and task-oriented motivations.
    The information processing style (not shown) trend lines were horizontal at the 50\% mark, indicating no differences between women and men.
    }
    \label{fig:all-facets}
    \vspace{-10pt}
\end{figure}

%% file: figure/02-Results-Gender-Facets/Results-Non-Binary.tex
\begin{table}[h]
	\includegraphics[width =0.8\linewidth]{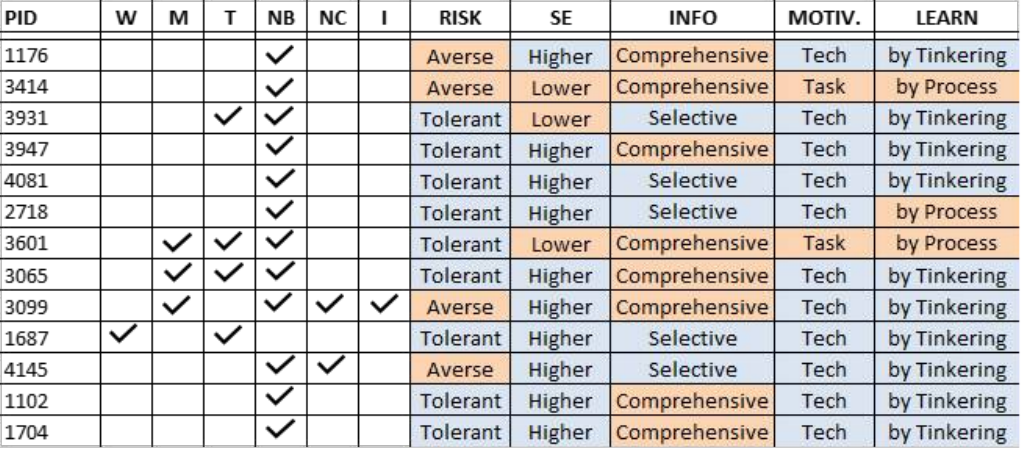}
	
    \caption{LGBTQIA* facet values: Each row shows one LGBTQIA* participant's problem-solving styles. Total LGBTQIA* participants across all experiments: 13.
	W = Woman, M = Man, T = Transgender, NB = Non-Binary, NC = Gender Non-Conforming, I = Intersex.
    }
	\label{tab:non-binary-results}
\end{table}



%% file: doc/05-Discussion-Inclusiveness-Equity.tex
\section{Discussion}

\subsection{Inclusivity and equity: Complements in HAI-UX fairness \draftStatus{MMBd2.4}}
\label{subsec:equity-inclusiveness-discussion}

\boldify{In AI, fairness has received attention. HAI should get fairness attention too too. Fairness has inclusivity and equity.}

Ideas about fairness in AI, what it is, and how to achieve it, have recently received substantial attention (e.g., ~\cite{buolamwini2018gender,foulds2020intersectional,green2019disparate,harrison2020empirical}).
Research and conversations in this area usually refer to algorithmic or data fairness---but the ideas are also relevant to HAI-UX fairness.

\boldify{Fairness includes inclusivity and equity.}

In considering any type of fairness, two concepts often drive the discussion---inclusivity and equity.
This paper has considered inclusivity, but not equity.

A way to think about inclusivity in HAI is as an ``outcome-oriented'' concept that applies \textit{within} a specific group of people.
As shown in earlier sections, when an AI product somehow led to disadvantageous outcomes for some particular group of participants (e.g., \colorboxBackgroundForegroundText{AbiOrangeQuote}{black}{risk-averse} participants), then that product was not inclusive to \textit{that} group.

\boldify{The above section measured change in inclusivity -- but how much is enough?  For this, we need equity.}

Although this paper's inclusivity results revealed who the guidelines were helping the most and who was being left out, they do not answer how much more inclusivity progress an AI product still needs to make and for whom.
Measuring equity can help to answer this question. Like inclusivity, equity in HAI is also an ``outcome-oriented'' concept---but it applies to between-group comparisons.
For example, if an AI product's user experiences for two groups (e.g., \colorboxBackgroundForegroundText{AbiOrangeQuote}{black}{risk-averse} participants and \colorboxBackgroundForegroundText{TimBlueQuote}{black}{risk-tolerant} participants) were of the same high---or low---quality, then the product was equitable.

\boldify{Here's what  equity showed for Risk.}

Ideally, one would like the inclusivity gains \appProduct{} achieved to result in a final outcome that is equitable to the two groups.
To explore how useful a measure of equity would be to our investigation's results, we measured equity of a dependent variable's outcome for a given product as the absence of a significant difference between the two participant groups. 
Table~\ref{fig:Risk-EquityResults+InclusivityColors} shows risk-group equity outcomes by this measure, superimposed on the risk-group inclusivity outcomes.

\input{figure/Risk-EquityResults+InclusivityColors}

For example, Table~\ref{fig:Risk-EquityResults+InclusivityColors}'s G15 column shows that the G15 \appProduct{} achieved inclusivity gains two times for \colorboxBackgroundForegroundText{AbiOrangeQuote}{black}{risk-averse} participants (orange cells) only, and three times for \colorboxBackgroundForegroundText{TimBlueQuote}{black}{risk-averse} participants (blue cells) only.
The G15 column further shows that those five targeted gains, along with the inclusivity gains experienced by everyone, ultimately led to fully equitable outcomes the risk spectrum ("=" markings).  
Thus, applying the G15 guideline ended up targeting exactly who it should have targeted in order to bring everyone up to an equitable state.

In total, Table~\ref{fig:Risk-EquityResults+InclusivityColors} shows that the guidelines' resulting \appProduct{}s almost always produced equitable outcomes across the risk spectrum.
Specifically, 129/160 outcomes (81\%) were equitable, marked by ``='' in the table.
Of the 31/160 outcomes that were \textit{in}equitable, only 3/160 (2\%) favored the \colorboxBackgroundForegroundText{AbiOrangeQuote}{black}{risk-averse} (``A''), and 28/160 (17\%) favored the \colorboxBackgroundForegroundText{TimBlueQuote}{black}{risk-tolerant} (``T'').

Of course, equity does not always mean success. 
The G15 \appProduct{} produced entirely equitable Table~\ref{fig:Risk-EquityResults+InclusivityColors}, but only moderately positive HAI user experiences (revisit Figure~\ref{fig:Li-resultThumbnails-forResults}). 
In contrast, the G1 \appProduct{} produced mostly equitable but extremely low HAI user experiences, and the G6 \appProduct{} produced entirely equitable and very positive HAI user experiences. 

Due to space limitations, we do not present equity results in detail for Risk or for the other four problem-solving styles.
Still, our limited exploration here shows the additional value measuring equity can bring, so we advocate for measuring equity as well as inclusivity as a way to fully understand user experiences.
Measures such as these of inclusivity and equity can provide complementary information on the quality of user experiences an AI product is offering.

\FIXME{PV dealing with R4-5: ``R4-5: There are some paragraphing issues in the discussion, including several very short and sometimes choppy paragraphs (1-2 sentences).  More cohesive and developed discussion paragraphs would improve the flow and readability of this section." - I adjusted a few paragraphs by combining sentences if there was relevance. The main issue the reviewer had was the succint nature of the paragraphs but I think extending the pargaraphs is not necessary therefore I merged paragraphs together that flowed naturally. I wasn't sure if I should remove the bolded titles, so I left those there}

\subsection{Practical implications for HAI practitioners \draftStatus{MMBd2.3}}

Measuring inclusivity and equity can reveal insights that HAI practitioners can use to understand and improve their AI products' HAI user experiences.
As we pointed out in Section~\ref{sec:gender}, although it is common practice to use empirical studies with human participants to measure user experience success, investigating inclusivity by anything other than demographics (e.g., gender, age, etc.) is relatively rare. 

\FIXME{MMB@anyone: "As we pointed out in Section x" is a promise, so chk just before fix to see if we kept that promise.}

As our results show, incorporating users' problem-solving into AI products' HAI-UX work can sometimes point out where and why mismatches are arising between a group of users and an AI product.
Some particularly actionable examples were given in Section~\ref{subsec:riskActionability}.
A way to gather participants' problem-solving styles would be to incorporate the validated  survey~\cite{hamid-2023} we used into user testing.

Armed with this new information, HAI practitioners could gain actionable insights in use-cases like the following:

~\textit{HAI Practice Use-Case 1}: For an AI product with a problematic HAI-UX, measure \textit{equity state}, to see which problem-solving groups of users are being left behind on a problematic product. 
To do so, for each problem-solving style, HAI practitioners could compare equity outcomes that are significantly different between the ``A'' participant group and the ``T'' participant group.

~\textit{HAI Practice Use-Case 2}: For an AI product that has just been changed, measure ~\textit{inclusivity changes} to see who a particular product change/new feature has benefited. 
To do so, as we did in our Inclusivity Results section, HAI practitioners could compare ``A'' participants before the change versus after the change, and likewise for ``T'' participants.

~\textit{HAI Practice Use-Case 3}: After an AI product has changed, complement a measure of its \textit{equity state} (as per Use-Case 1) with a measure of \textit{dependent variable final outcomes} (e.g., as in Figure~\ref{fig:Li-resultThumbnails-forResults}).
This combination shows not only final equity state, but also how successful the AI product's HAI-UX is for each group of participants.

Our results suggest that doing measures like these can provide new, valuable information on who is being included, who is being left out, and how a product can improve.

\subsection{Threats to Validity \& Limitations
\draftStatus{1.9-D2.0??}
}
\boldify{Every study has threats to validity...}

As with every empirical study~\cite{Wohlin-2012, ko2015practicalempirical}, our investigation has limitations and threats to validity.

\boldify{These UX metrics might not have captured all information that people wanted. For example, some participants mentioned security in Section~\ref{sec:Risk}, but there was no direct question about their sense of security.}

In any study, researchers cannot ask participants every possible question, having to balance research goals with participant fatigue.
As such, the dependent variables we analyzed may not have captured all information about people's reactions.
For example, some participants' free-text remarks suggested outcomes that our Likert-style questionnaires did not cover; one example was participants' mentions of privacy concerns while interacting with certain products.
Because the study was not designed with a dependent variable about privacy,  we cannot be certain if remarks such as these indicated only isolated cases or more prevalent phenomena.

\boldify{We also did not actively deal with ``idk'' answers, even though methods like imputation exist.}

Another threat was how to handle missing data.
Since participants had the option to say ``I don't know'' for any of the questions, we had to decide whether to 1) impute the data or 2) drop the ``I don't know'' values, costing degrees of freedom in our statistical tests.
We chose the latter, because although there are many imputation methods to leverage (e.g., hot-deck, cold-deck, regression), any inferences are then limited to the imputed data, rather than the original data.



\boldify{One  threat for us is that there were a lot of statistical tests reported in this paper. We don't correct in the paper but do in the appendices because some reviewers may not agree with our decision.}

Another threat was how to handle the number of statistical tests we ran.
As mentioned in Section~\ref{subsec:methods-phase-two-analysis}, we did not report statistically corrected results in this paper because every test corresponded to a pre-planned hypothesis~\cite{armstrong2011statistical,armstrong2014use}.
That said, we recognize that some readers may not agree with this decision, so we also provide all Holm-Bonferroni corrected results in Appendix D.


\FIXME{\textbf{R9-4}: Participants didn't use an HCAI system, but rather Vignettes which are arguably less realistic...\\ \textbf{R4-7:} I would like to see the paper discuss the pros and cons of the vignette-approach ...there is likely a breadth vs. depth trade-off in terms of data collection. A discussion of this trade-off would make for a useful addition to the paper.\\
\textcolor{red}{MMM@AAA: DONE if you like what I wrote (thx for finding those old words)}}

\FIXED{
\boldify{One possible threat to validity is using vignettes instead of real systems.}

Also, we chose to use vignettes vs. a real system.
Each approach has its own advantage:
Using vignettes allows enough control to genuinely isolate the experimental variation to vary ONLY the independent variable, and this isolation was critical to our statistical power.
In contrast, a real system's strength is realism in the external world, but at the cost of controls.
Because this was a set of controlled experiments, we chose control, leaving  to other studies to investigate  external validity questions (faithfulness to real world conditions).    
} 

\FIXME{\textbf{R9-5: MMB on 1/28/24 says DONE} familiarity with systems could be an issue... }
\boldify{Other threats could have come from how we assigned vignettes or how participants understood vignettes.}

\FIXED{Other threats to validity could arise from the particular pairing of vignette to product, and/or from participants associating a vignette with a specific real product with which they had familiarity. 
We attempted to avert the latter by randomly assigning generic names (Ione and Kelso) instead of real product names, but participants may have still imagined their favorite productivity software.
If this occurred, it would contribute an extra source of variation in these data.}

\boldify{One limitation of this investigation was that the study was limited to people living in the USA.
In order to broaden inclusivity, we require the diverse voices of people in more countries.}

\FIXME{\textbf{R9-6:} \textcolor{red}{MMB@AAA, 1/29/24: I pronounce DONE if you like it. }Sample only people in USA, does GM even apply beyond USA}
\FIXED{
Although the productivity software and GenderMag problem-solving styles have been shown to be viable/useful in countries around the world, the participants in our study were restricted to those who lived in the USA at the time of the study.
As such, the results in this paper cannot be generalized to other countries around the world.
However, since the methodology is not U.S.-specific, replicating the study with participants from additional countries should be straightforward.
}

\boldify{There are always limitations to any study, and one limitation was that there is always a balance between the number of d.v's you can ask participants. For example, some participants commented on safety}

One limitation of this investigation is that its results cannot be generalized to AI-powered systems outside of productivity software.
This suggests the need to investigate HAI-UX impacts on diverse problem-solvers  across a spectrum of domains, from low-stakes domains (e.g., music recommender systems) to high-stakes domains (e.g., automated healthcare or autonomous vehicles).

Threats and limitations like these can only be addressed through additional studies across a spectrum of empirical methods and situations, in order to isolate different independent variables of study and establish generality of findings across different AI applications, measurements, and populations.

%% file: figure/Risk-EquityResults+InclusivityColors.tex
\begin{table}[h]
    \centering
    \includegraphics[width=.9\columnwidth]{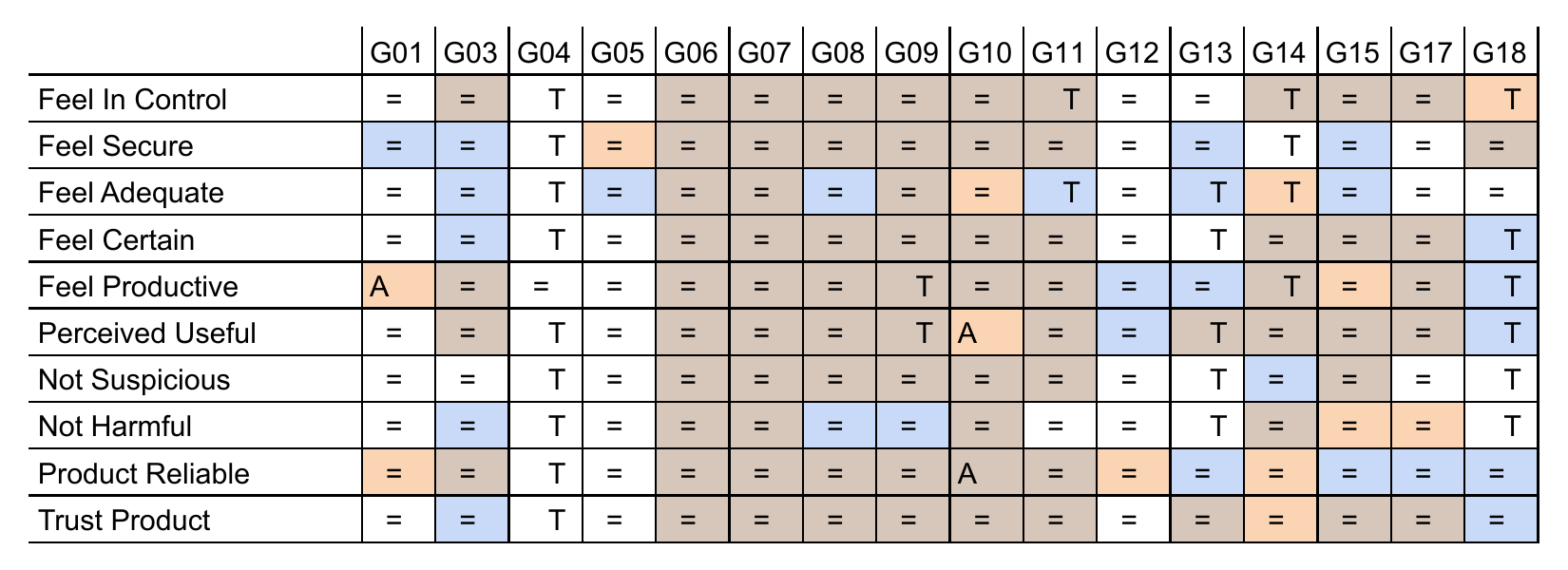}
    \caption{Risk equity of all experiments' \appProduct{}s. Equity symbols are superimposed on the Risk inclusivity result colors.\\ 
    \textbf{Equity symbols}: equitable (\textbf{=}); inequitable favoring risk-Averse (\textbf{A}) or risk-Tolerant (\textbf{T}).\\
     \textbf{Inclusivity colors}: 
     Inclusivity gains for \colorboxBackgroundForegroundText{Eggshell}{black}{both risk-Averse and risk-Tolerant};
     inclusivity gains for 
    \colorboxBackgroundForegroundText{AbiOrangeQuote}{black}{risk-Averse only};
    inclusivity gains for 
    \colorboxBackgroundForegroundText{TimBlueQuote}{black}{risk-Tolerant only};
     inclusivity gains for nobody (no color).}
    \label{fig:Risk-EquityResults+InclusivityColors}
 
\end{table}

%% file: doc/08-Conclusion.tex
\section{Conclusion 
\draftStatus{0.0}
}

\boldify{In this paper, we... and our results suggest...}

We investigated whose user experiences 16 AI products included by considering a set of five problem-solving styles.
We measured beyond \textit{whether} 16 AI-powered products improved HAI-UX to \textit{who} was included in such improvements and who was left out.
We recruited 1,016 participants, who saw two products, which violated and applied guidelines, measuring their user experience through 10 dependent variables.
Among our results, we found:

\begin{itemize}
    \item \textit{Problem-solving style inclusivity in HAI-UX}: Not only did most of the 16 product changes inclusively improve the user experience of participants with diverse attitudes toward risk, they inclusively improved diverse values in all five GenderMag problem-solving style spectra (\textbf{Results \#1 \& 7 })
    \item \textit{Actionable inclusivity in HAI-UX}: these results also suggest actionable steps that HAI practitioners can take to make an AI product more inclusive (\textbf{Results \#2, 3, 4, 5, 6, \& 8})
    \item \textit{Problem-solving diversity and demographic diversity}: The relationships between participants' problem-solving styles and their intersectional gender-and-age demographic diversity uncovered ways for HAI practitioners to bring actionable results from problem-solving diversity investigations to bear on demographic disparities (\textbf{Result \#9}).
\end{itemize}


These inclusivity results were prevalent across all 16 products---of 160 possible UX measures across the 16 experiments, user experience inclusively improved at least 115 times (72\%) for all five of the participants' GenderMag problem-solving styles.
Not only that, but which kind of participant was most advantaged depended on both the problem-solving style and the product.

Although prior literature has reported results like these for non-AI-powered technologies~\cite{burnett2010gender,burnett2016finding,guizani2022debug,padala2020gender,stumpf2020gender,vorvoreanu2019gender}, this paper revealed the importance of investigating people's problem-solving style to understanding how inclusively AI-powered technologies are serving their many diverse users.
Our work directly relates to one of Shneiderman's three ideas for Human-Centered Artificial Intelligence~\cite{shneiderman2020three}, the ``shift from emulating humans to empowering people''.
We believe that our work provides an actionable way for HAI practitioners to address the needs of people with diverse problem-solving styles in human-AI interaction, which helps move current efforts beyond \textit{who} needs empowering to also consider \textit{how} to empower people with AI products.    



\section*{Acknowledgments}

We thank Rupika Dikkala, Catherine Hu, Jeramie Kim, Elizabeth Li, Caleb Matthews, Christopher Perdriau, Sai Raja, and Prisha Velhal for their help with this paper.
We are grateful to the editors and reviewers for their encouragement and constructive engagement, which greatly helped to improve this paper.
This work was supported in part 
by Microsoft, 
by NSF \#1901031 and \#2042324; and by USDA-NIFA/NSF  \#2021-67021-35344. 
Any opinions, findings, conclusions, or recommendations expressed are the authors’ and do not necessarily reflect the views of the sponsors.

\FIXME{**just before ship** Check that we acknowledge everyone (alphabetically) that we need to.}